\RequirePackage{fix-cm}
\documentclass[natbib]{svjour3}              % onecolumn (standard format)
\smartqed  % flush right qed marks, e.g. at end of proof

\usepackage{xcolor}
\usepackage{graphicx}
\usepackage{mathptmx}      % use Times fonts if available on your TeX system
\usepackage{aps-bibstyle}  % use this style if you upload to .tex file only a part of Bibtex created bbl.
\usepackage{microtype}
\usepackage{textcomp}
\usepackage{amssymb, mathrsfs,color}

 \journalname{Space Science Reviews}
\begin{document}

\title{Electron Power-Law Spectra in Solar and Space Plasmas}
%\title{Particle Acceleration in \\Solar Flares and Terrestrial Substorms}%\thanks{Grants or other notes


%\titlerunning{Electron Power-Law Spectra in Solar and Space Plasmas}        % if too long for running head

%\author{ISSI Team 347}
\author{M. Oka \and J. Birn \and M. Battaglia \and C. C. Chaston \and S. M.  Hatch \and G. Livadiotis \and S. Imada \and Y. Miyoshi \and M. Kuhar \and F. Effenberger  \and E. Eriksson \and Y. V. Khotyaintsev  \and A. Retin\`{o}}

%        \and
%        Second Author %etc.
%}

%\authorrunning{Oka et al. } % if too long for running head

\institute{
	M. Oka \and C. C. Chaston \at
	Space Sciences Laboratory, University of California Berkeley \\
	7 Gauss Way, Berkeley, CA 94720 \\
	Tel.: +1-510-642-1350 \\
	\email{moka@ssl.berkeley.edu}
	\and
 	J. Birn \at 
 	Space Science Institute, Boulder, Colorado, USA \\
	Los Alamos National Laboratory, Los Alamos, NewMexico, USA 
	\and
	M. Battaglia \and M. Kuhar \at
	Institute of 4D Technologies, School of Engineering, University of Applied Sciences and Arts Northwestern Switzerland, CH-5210 Windisch,Switzerland  
	\and
	S. M. Hatch \at
	Department of Physics and Astronomy, Dartmouth College, Hanover, New Hampshire, USA
	\and
	G. Livadiotis \at
	Southwest Research Institute, San Antonio, TX-78238, USA
	\and
	S. Imada \and Y. Miyoshi \at
	Institute for Space-Earth Environmental Research, Nagoya University, Furo-cho, Chikusa-ku, Nagoya, 464-8601 Aichi, Japan
	\and
	F. Effenberger \at
	Helmholtz Centre Potsdam, GFZ German Research Centre for Geosciences, Potsdam, Germany \\
	Bay Area Environmental Research Institute, NASA Research Park, Moffett Field, CA, USA \\
	\and
	E. Eriksson and Y. V. Khotyaintsev \at
	Swedish Institute of Space Physics, Uppsala, Sweden
	\and 
	A. Retin\`{o} \at
	Laboratoire de Physique des Plasmas, Palaiseau, France
}
\date{Received: date / Accepted: date}

\maketitle

\begin{abstract}

Particles are accelerated to very high, non-thermal energies in solar and space plasma environments. While energy spectra of accelerated electrons often exhibit a power law, it remains unclear how electrons are accelerated to high energies and what processes determine the power-law index $\delta$. Here, we review previous observations of the power-law index $\delta$ in a variety of different plasma environments with a particular focus on sub-relativistic electrons. It appears that in regions more closely related to magnetic reconnection (such as the `above-the-looptop' solar hard X-ray source and the plasma sheet in Earth's magnetotail), the spectra are typically soft ($\delta \gtrsim$ 4). This is in contrast to the typically hard spectra ($\delta \lesssim$ 4) that are observed in coincidence with shocks. The difference implies that shocks are more efficient in producing a larger non-thermal fraction of electron energies when compared to magnetic reconnection. A caveat is that during active times in Earth's magnetotail, $\delta$ values seem spatially uniform in the plasma sheet, while power-law distributions still exist even in quiet times. The role of magnetotail reconnection in the electron power-law formation could therefore be confounded with these background conditions. Because different regions have been studied with different instrumentations and methodologies, we point out a need for more systematic and coordinated studies of power-law distributions for a better understanding of possible scaling laws in particle acceleration as well as their universality.

%Insert your abstract here. Include keywords, PACS and mathematical subject classification numbers as needed.
\end{abstract}

\newpage
\tableofcontents

%#####################################################################
%#####################################################################
\section{Introduction}
\label{sec:intro}
\addtocontents{toc}{\setcounter{tocdepth}{3}}% Allow \section in ToC
%#####################################################################
%#####################################################################

%----------------------------------------
\subsection{Why Electron Power Laws?}
%----------------------------------------

A solar flare is an explosive energy-release phenomenon on the sun, and non-thermal (power-law) electrons alone appear to carry up to 50\% of the released energy \citep[e.g.][and references therein]{Lin1971, Lin2003, Emslie2012, Aschwanden2014}. While it has been established that magnetic reconnection -- a plasma process that converts magnetic energy into particle energy -- plays an important role during flares, the precise mechanism for producing such energetic electrons is still unclear.

A substorm is an explosive energy-release phenomenon in Earth's magnetotail that also produces non-thermal ions and electrons reaching energies of hundreds of keV\footnote{A `substorm' occurs over the time scale of up to a few hours, mostly due to phenomena in the magnetotail leading to expansion of aurora in the polar ionosphere. Electron acceleration occurs on a time scale of minutes. There is also a `storm', which is a disturbance of the magnetic field and associated plasmas mostly in the inner magnetosphere. It occurs over a time scale of hours or days and generally includes multiple substorms.}. As in solar flares, magnetic reconnection is thought to play a major role in the energy release and conversion.  However, ion fluxes have been shown to dominate the energy outflow from tail reconnection sites \citep[e.g.][]{Oieroset2002, Eastwood2013} but the relative contributions of thermal and suprathermal ions were not investigated. Also, thermal and non-thermal electrons do not provide a comparable or even dominant contribution to the energy outflow, due to their lower average energy (typically less than 1 keV compared to 5 -- 10 keV mean ion energies).\footnote{In the absence of a clearly identifiable Maxwellian in the magnetotail, the term `thermal' here may denote energies comparable to or less than the mean energy, rather than a Maxwellian component (See Section \ref{sec:definition} for more details). The separation of thermal from non-thermal populations appears more arbitrary than in the solar case, also because of the absence of collisions that would cause the particle populations to relax toward a Maxwellian.}

The similarities and differences between the energetic particle properties in solar and terrestrial environments thus were a strong motivation for this paper. A possible key to understanding and describing the suprathermal populations, here specifically of electrons, is the power-law index, which may be used to measure the amount of energization and provide insight into acceleration mechanisms. Thus, by comparing power-law indices in solar flares (Section \ref{sec:flares}) and Earth's magnetotail (Section \ref{sec:emt}), we may be able to find similarities and differences of electron acceleration mechanisms, although both the measurement techniques and the plasma parameters are very different in each environment (Section \ref{sec:cav}, Appendix \ref{sec:param}). From this perspective, it is instructive to compare these power-law indices with those of other regions such as shocks and the solar wind because electrons in these regions are also known to exhibit a non-thermal tail (Section \ref{sec:oth}). Through detailed comparisons and discussions of power-law spectra in different environments, we aim to search for and discuss a possible scaling law or universality of electron acceleration that may exist, at least, in the sub-relativistic plasmas in the heliosphere (Section \ref{sec:dsc}). It should be noted that particle acceleration occurs everywhere in the universe, not only in solar and terrestrial settings but also in astrophysical settings. Thus, we hope this review of solar and space plasmas will ultimately contribute toward a better understanding of astrophysical plasmas as well.

%----------------------------------------
\subsection{Caveats of Comparing Solar and Space Plasmas}
\label{sec:cav}
%----------------------------------------

The diagnostics and methodologies of high-energy electrons are very different in solar and space environments, i.e. remote-sensing solar X-rays and {\it in situ} measurements of particles in space. In principle, an electron distribution $f(t,\mathbf{v},\mathbf{r})$ is intrinsically a function of time $t$, particle velocity $\mathbf{v}$ and position $\mathbf{r}$, where $f$ is the differential density. However, in our observations of solar flares via X-rays, we can obtain photon distributions as a function of energy $E$ and time $t$ only, with limited spatial/time resolutions and no information along the line of sight. Upon interpreting the observed spectrum from a hard X-ray source (such as footpoint, looptop and above-the-looptop sources, as will be described in Section \ref{sec:flares}), we have to consider it as the mean distribution of the source at a given time,
\begin{equation}
\langle nVf \rangle = \int_V n(\mathbf{r}) f(E, \mathbf{r}) dV, 
\end{equation} 
where $n$ and $V$ are the density and volume of the hard X-ray source, respectively. In reality, a hard X-ray source is not necessarily uniform and there could be a variety of different spectral forms at various positions within the source. It is also difficult to estimate the size and shape of the source volume $V$. In some cases, a stereoscopic observation can be achieved with multiple telescopes, but the error bars of the $V$ estimates are still large with current instruments. In addition, we need to convert a photon spectrum to an electron spectrum for comparison, which involves an assumption on how electrons lose their energy. 

On the other hand, {\it in-situ} measurement in space by a spacecraft provides full 3D distributions of particle velocities $f(\mathbf{v})$ at a given time $t$ and position $\mathbf{r}$, where $f$ is the phase space density. While anisotropies and other kinetic and/or non-thermal features can be inferred from $f(\mathbf{v})$, it is often difficult to distinguish between spatial and temporal variations from the $f$ data alone. For example, when a spacecraft detected an increase of $f(\mathbf{v})$, it could be a passage of a plasmoid filled with a larger number of particles (i.e., spatial variation) or an arrival of particles from a remote structure that released the particles (i.e., temporal variation). A contextual data from different plasma parameters (such as magnetic fields) and/or multi-point measurements (with multiple spacecraft) are used to discuss such spatial/temporal variations.

Because of these differences, we have not focused on the details of, for example,  pitch angle distributions. If a magnetotail study reported the power-law index of spectra for both  parallel and perpendicular directions with respect to the magnetic field, then we have taken the values from the spectrum that showed the largest flux to represent the power-law index of the features/phenomena. Also, {\it in situ} measurements often show time variations within particular features/phenomena. In such cases, we focused on an average value and the standard deviation.

%----------------------------------------
\subsection{Definitions of Power-Law Index}
\label{sec:definition}
%----------------------------------------

While there are different definitions of the power-law index, we will use $\delta$ as measured in electron differential flux throughout this paper. In this section,  we summarize some commonly used definitions so that we can more easily compare power-law indices reported in the literature. Table \ref{tab:index} provides a conversion table.

\begin{table*}
\caption{A conversion table for various power-law indices in the sub-relativistic regime (See text for values in the ultra-relativistic regime).  From left to right are phase space density ($f(p), f(E)$), differential density $N(E)$, differential flux (or flux density) $\mathscr{F}(E)$, the kappa distribution $f_\kappa(v)$, thin- and thick-target X-ray emission $I_{thin}(\varepsilon)$ and $I_{thick}(\varepsilon)$. Note $\delta = \kappa$ and $\Gamma = \gamma_{thin}$. $f(E)$ is listed in addition to $f(p)$ because it has been used in magnetospheric observations. }
\label{tab:index}       % Give a unique label
\renewcommand{\arraystretch}{1.7}
\setlength{\tabcolsep}{3pt}
\begin{tabular}{lllllll}
	\hline\noalign{\smallskip}

	% 1st ROW
	$f(p)$ & 
	$f(E)$ &
	$N(E)$ &
	$\mathscr{F}(E)$&
	$f_\kappa(v)$ &
	$I_{thin}(\varepsilon)$ &
	$I_{thick}(\varepsilon)$ \\

	% 2nd ROW
	$\;\; \propto p^{-s}$ & 
	$\;\; \propto E^{-\Gamma}$ &
	$\;\; \propto E^{-\delta'}$ &
	$\;\; \propto E^{-\delta}$ &
	$\;\; \propto v^{-2(\kappa+1)}$ &
	$\;\; \propto \varepsilon^{-\gamma_{thin}}$ &
	$\;\; \propto \varepsilon^{-\gamma_{thick}}$ \\

	\noalign{\smallskip}\hline\noalign{\smallskip}
	$s$ & 
	$\frac{s}{2}$ & 
	$\frac{s-1}{2}$ & 
	$\frac{s-2}{2}$ & 
	$\frac{s-2}{2}$ & 
	$\frac{s}{2}$   & 
	$\frac{s-4}{2}$ \\

	$2\Gamma$ & 
	$\Gamma$           & 
	$\Gamma-\frac{1}{2}$ & 
	$\Gamma-1$ & 
	$\Gamma-1$ & 
	$\Gamma$ & 
	$\Gamma-2$ \\

	$2\delta'+1$ & 
	$\delta'+\frac{1}{2}$ & 
	$\delta'$ & 
	$\delta'-\frac{1}{2}$ & 
	$\delta'-\frac{1}{2}$ & 
	$\delta'+\frac{1}{2}$ & 
	$\delta'-\frac{3}{2}$ \\

	$2(\delta+1)$ & 
	$\delta+1$ & 
	$\delta+\frac{1}{2}$ & 
	$\delta$ & 
	$\delta$ & 
	$\delta+1$ & 
	$\delta-1$ \\

	$2(\kappa+1)$ & 
	$\kappa+1$ & 
	$\kappa+\frac{1}{2}$ & 
	$\kappa$ & 
	$\kappa$ & 
	$\kappa+1$ & 
	$\kappa-1$ \\

	$2\gamma_{thin}$ & 
	$\gamma_{thin}$           & 
	$\gamma_{thin}-\frac{1}{2}$ & 
	$\gamma_{thin}-1$ & 
	$\gamma_{thin}-1$ & 
	$\gamma_{thin}$ & 
	$\gamma_{thin}-2$ \\

	$2\gamma_{thick}+4$ & 
	$\gamma_{thick} + 2$ & 
	$\gamma_{thick} + \frac{3}{2}$ & 
	$\gamma_{thick} + 1$ & 
	$\gamma_{thick} + 1$ & 
	$\gamma_{thick} + 2$ & 
	$\gamma_{thick} $ \\

	\noalign{\smallskip}\hline
\end{tabular}
\end{table*}

In an isotropic, three-dimensional (3D) form, the \textbf{phase space density}  of the power-law distribution can be written as
\begin{equation}
f(p) \propto p^{-s}
\end{equation}
where $p$ is the particle momentum and $s$ is the power-law index.  Of course, for many cases of solar and space plasma, the particle distributions are not necessarily isotropic, and the data obtained by {\it in-situ} measurements is often analyzed in 3D velocity space $f(\mathbf{v})$ where $\mathbf{v}$ is the particle velocity. The phase space density is also used for analyzing energy spectra (especially in magnetospheric studies), as a function of particle energy $E$, i.e. $f(E)$, after taking an average over all or a part of the pitch-angles. In this form the power-law index becomes
\begin{equation}
f(E) \propto E^{-\Gamma}, 
\end{equation}
where $\Gamma = s/2$ for sub-relativistic particles  ($E=p^2/2m$) and $\Gamma = s$ for ultra-relativistic particles (i.e., $E=pc$).

The \textbf{differential density} (cm$^{-3}$ keV$^{-1}$) is defined as $N(E)dE = 4 \pi p^2f(p)dp$ and the power law can be written as
\begin{equation}
N(E) \propto E^{-\delta'},
\end{equation}
where $\delta' \equiv (s-1)/2$ for sub-relativistic particles and $\delta'\equiv s-2$ for ultra-relativistic particles. Many studies of particle simulations use this form, because an energy spectrum can be directly obtained from a histogram of energies of particles in the simulation box.

The \textbf{differential flux} or flux density (cm$^{-2}$s$^{-1}$keV$^{-1}$) is defined as $\mathscr{F}(E) dE = v \, N(E)dE$ and the power law can be written as
\begin{equation}
%\mathscr{F}(E) \equiv \frac{dJ}{dE}
\mathscr{F}(E) \propto E^{-\delta}
\end{equation}
where $\delta = (s-2)/2$ for sub-relativistic particles. For ultra-relativistic particles, $v \sim c$ so that $\mathscr{F} \sim c\, N$ and $\delta$ remains the same as in the differential density case, $\delta = s-2$.

To diagnose accelerated electrons in the solar corona, hard X-ray (HXR) observations of electron bremsstrahlung emission are used \citep[e.g.][]{Brown1971a, Tandberg1988, Holman2011a} as summarized below. A caveat is that the initial distribution $\mathscr{F}(E_0)$
%$\mathscr{F}_0(E)$ 
may evolve into a different distribution $\mathscr{F}(E)$ by energy loss processes (primarily Coulomb collisions) within the duration of the X-ray measurement. These distributions are often described as `injected' and `instantaneous' distributions in solar flare studies, respectively. Two extreme models can be considered for the evolution.  

In the {\it thin-target} model, non-thermal electrons do not lose much energy and preserve their distribution so that $\mathscr{F}(E) = \mathscr{F}(E_0)$. The \textbf{differential photon flux} (or flux density) $I(\varepsilon)$ (photons cm$^{-2}$s$^{-1}$keV$^{-1}$)  can be derived by
\begin{equation}
I(\varepsilon) = \frac{S\Delta N}{4\pi R^2} \int_\varepsilon^{\infty} \mathscr{F}(E_0) \, \sigma_B(\varepsilon, E_0) dE_0, 
\label{eq:thintarget}
\end{equation}
where $S$ is the flare area, $\sigma_B$ is the Bremsstrahlung cross section (differential in photon energy $\varepsilon$, cm$^2$ keV$^{-1}$), $R$ is the distance between the observer and the X-ray source, $\Delta N = \int_{\rm source}n_p(s)ds$ is the column density of the source observed, and $n_p(s)$ is the ambient proton density as a function of the distance along the injected electron path \citep[e.g.][]{Tandberg1988}. It has been shown that, for a single power-law electron population $\mathscr{F}(E_0) \propto E_0^{-\delta}$, $I(\varepsilon)$ is also a power-law of the form
\begin{equation}
I(\varepsilon) \propto \varepsilon^{-\gamma_{thin}}, 
\label{eq:convertthin}
\end{equation}
where $\gamma_{thin} = \delta + 1$.

In this conversion, the nonrelativistic Bethe-Heitler (NRBH) bremsstrahlung cross section is used for $\sigma_B$. Thus, there would be an error in this conversion if there were significant flux at and above (near-)relativistic energies. Some studies report electron power-law index $\delta$ obtained by numerical computations with the relativistic Bethe-Heitler cross-section incorporated into the RHESSI spectral analysis software (OSPEX) \citep{Holman2011a}, removing our need to convert from $\gamma$ to $\delta$. However, some other studies report photon power-law index $\gamma$ only. Furthermore, in addition to the assumption in the cross-section, there are other important sources of errors and uncertainties in spectral analysis (See Section 3 of \cite{Holman2011a}). It is beyond the scope of this paper to correctly and accurately derive the electron power-law index $\delta$ for all published values of $\gamma$. Thus, for the sake of comparison with the $\delta$ values in space obtained via {\it in-situ} measurements, we simply keep using this conversion as needed (i.e., Eq. (\ref{eq:convertthin})), while focusing on the values obtained in the sub-relativistic energy range (typically $\lesssim$ 100 keV). 

In the {\it thick-target} model,  non-thermal electrons lose all of their non-thermal energies and thermalize. By considering the energy loss (from $E_0$ to $E$), the photon spectrum can be described as
\begin{equation}
I(\varepsilon) = \frac{S}{4\pi R^2} \frac{1}{C} \int_{E_0=\varepsilon}^{\infty} \mathscr{F}(E_0) \, \int_{\varepsilon}^{E_0} E \sigma_B(\varepsilon, E) dE dE_0, 
%I(\varepsilon) = \frac{S\Delta N_{\rm eff}}{4\pi R^2} \int_{E_0=\varepsilon}^{\infty} \mathscr{F}(E_0) \, \sigma_B(\varepsilon, E) dE_0, 
\label{eq:thicktarget}
\end{equation}
where $C \equiv 2 \pi e^4 \ln{\Lambda}$ and $\Lambda$ is the Coulomb logarithm. Comparing this expression with Equation (\ref{eq:thintarget}), we can define the effective column density $\Delta N_{\rm eff}$ as
\begin{equation}
\Delta N_{\rm eff} \equiv \frac{1}{C\sigma_B(\varepsilon, E_0)} \int_\varepsilon^{E_0} E\,\sigma_B(\varepsilon,E) dE.
\end{equation}
This model considers the case where all non-thermal electrons thermalize due to Coulomb collisions, and the X-ray emission does not explicitly depend on the source density. However, the effective column density $N_{\rm eff}$  arises from a consideration of the evolution from $E_0$ to $E$. This thick-target model is valid if the actual (observed) column density  is sufficiently large, i.e., $\Delta N \gtrsim \Delta N_{\rm eff}$. Otherwise ($\Delta N \ll \Delta N_{\rm eff})$, the model should not be applied.  It has been shown that, for a single power-law electron population $\mathscr{F}(E_0) \propto E_0^{-\delta}$, $I(\varepsilon)$ is also a power law of the form
\begin{equation}
I(\varepsilon) \propto \varepsilon^{-\gamma_{thick}}, 
\end{equation}
where $\gamma_{thick} = \delta -1$. (Again, the nonrelativistic Bethe-Heitler (NRBH) bremsstrahlung cross section is used for $\sigma_B$.) Because higher energy electrons lose energy and emit more photons than in the case of thin-target model, the resultant thick-target photon spectrum becomes harder, i.e., $\gamma_{thick} < \gamma_{thin}$.

%XXXXXXXXXXXXXXXXXXXXXXXXXXX
\begin{figure*}
\includegraphics[width=1.\textwidth]{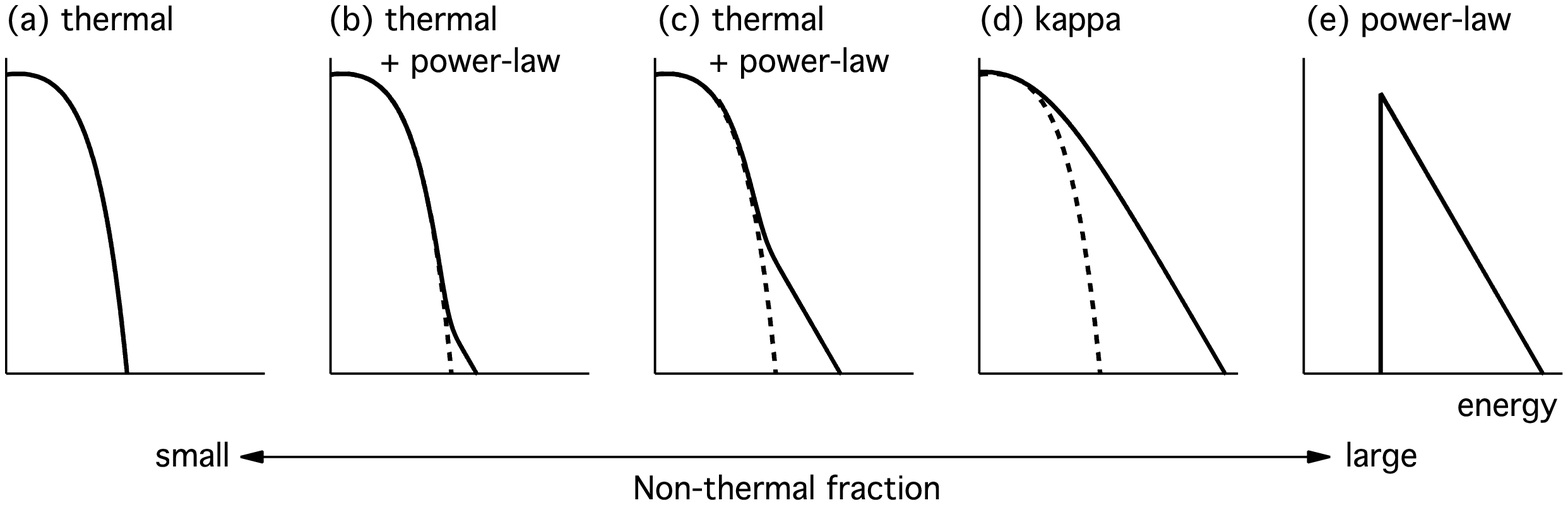}
\caption{Schematic illustrations of basic non-thermal distributions, adapted from \cite{Oka2015}. The horizontal and vertical axes (both logarithmic) represent particle energy and differential density, respectively. The core distribution and the power-law slope are fixed while the intensity of the power-law component is varied. To quantify non-thermal fraction of particle densities/energies, a lower-energy cutoff $E_{\rm c}$ of the power-law tail is required. For the kappa distribution, $E_{\rm c}$ is no longer needed and a formula for estimating the non-thermal fraction is proposed in \cite{Oka2015}.}
\label{fig:specmodels}
\end{figure*}
%XXXXXXXXXXXXXXXXXXXXXXXXXXX

While a single power-law is already useful when characterizing particle energy spectra with a clear distinction between thermal and non-thermal components (Figure \ref{fig:specmodels}(b,c)), the kappa distribution is also useful when there is no clear spectral break between the lower and higher energy components (Figure \ref{fig:specmodels}(d)). The isotropic, three-dimensional (3D) form of the \textbf{kappa distribution} $f_\kappa(v)$ (s$^3$ cm$^{-6}$) is written as
\begin{equation}
f_\kappa(v) = \frac{N_\kappa}{(\pi\kappa\theta^2)^{3/2}} \frac{\Gamma(\kappa+1)}{\Gamma(\kappa-1/2)} 
\left(1+\frac{v^2}{\kappa \theta^2} \right)^{-(\kappa+1)}, 
\label{eq:kappa}
\end{equation}
where $v$ is the particle speed, $\kappa$ is the power-law index, $\Gamma$ is the Gamma function, $N_\kappa$ is the number density, and $\theta$ is the most probable particle speed at which the differential flux becomes maximum\footnote{This kappa distribution is equivalent to the modified kappa distribution formulated by \cite{Leubner2004a} and used by, for example, \cite{Bian2014a}. The transformation can be expressed as $\kappa \mbox{*} = \kappa+1$ and $\theta \mbox{*} = \theta \sqrt{\kappa/\kappa+1}$ where $\kappa \mbox{*}$ and $\theta \mbox{*}$ are the spectral index and the most probable speed of the modified kappa distribution, respectively \citep[][]{Livadiotis2009}.}.

Note that the thermal speed (of a particle with mass $m$) depends only on the temperature $T$ of the system, $v_{th}=\sqrt{2k_BT/m}$, and it actually represents the temperature in speed units ($k_B$ is the Boltzmann constant). The auxiliary quantity $\theta \equiv v_{th}\cdot\sqrt{(\kappa-3/2)/\kappa}$  is used in 3-D kappa distributions, as it coincides with the most probable speed. Nevertheless, care is needed because (i) the coincidence of $\theta$ with the most probable speed holds only for the 3-D case, and (ii) the temperature and not the most probable speed constitutes a fundamental thermodynamic parameter, as it is determined from the equipartition theorem and the thermodynamic definition of temperature; therefore, the speed $\theta$ depends on $\kappa$, and should not be used for demonstrating the kappa distributions for various kappa indices, while the temperature, $v_{th}$ or $T$, is independent of $\kappa$, and is  preferred when considering variations of $\kappa$ (see \cite{Livadiotis2015b} and Chapter 1 of \cite{Livadiotis2017} for further details; see also \cite{Lazar2016a} for different perspectives of the definition of the temperature).

The kappa distribution is suitable for characterizing an observed energy spectrum with no clear distinction at the interface between the thermal core and the power-law tail (Figure \ref{fig:specmodels}(d)). For such a distribution, a fit with the combined `thermal+power-law’ distribution would lead to systematically higher temperatures and lower densities due to an artificial, lower-energy cutoff of the power-law \citep{Oka2015}. A combined `thermal+power-law' model is suitable if there was a clear spectral transition, from soft (steep) to hard (flat), at the interface between the two component (Figure \ref{fig:specmodels}(b,c)). It should also be noted that the kappa distribution maximizes the entropy of nonextensive statistical mechanics, providing an important context for our studies of power laws (Appendix \ref{sec:nonextensive}). 

In space, a particle velocity distribution can often have a non-thermal tail extending from a flattop core distribution \citep[e.g.][]{Feldman1982, Feldman1983,Thomsen1983a, Chateau1989}. Such a \textbf{flattop distribution} is sometime expressed empirically as 
\begin{equation}
f_{L}(v) = \frac{N_L \kappa \sin{(\pi/2\kappa)}}{\pi^2 v_{\perp L}^2 v_{||L}} \left[1+\left(\frac{v_{\perp}}{v_{\perp L}}\right)^{2\kappa}+\left(\frac{v_{||}}{v_{||L}}\right)^{2\kappa} \right]^{-\frac{\kappa+1}{\kappa}}, 
\end{equation}
where $N_L$ is the density, $\kappa$ is the spectral index and $v_L$ is the location and sharpness of the spectral break (or `shoulder'). In the higher energy limit ($v\gg v_L$), the distribution approaches a power-law $f \propto v^{-2(\kappa+1)}$ as is the case with the kappa distribution \citep[e.g.][]{Vasyliunas1968a}. In the lower energy limit ($v\ll v_L$), the distribution becomes flat at $f = N_L \kappa \sin{(\pi/2\kappa)}/(\pi^2 v_{\perp L}^2 v_{||L})$. Also, \cite{Stverak2009} proposed a similar but slightly different form of the flattop distribution.

The flattop distribution has been observed in a kinetic-scale region where a potential drop develops. Examples include the shock transition layer \citep[e.g.][]{Feldman1982, Feldman1983,Thomsen1983a} and the vicinity of magnetic reconnection diffusion region \citep[e.g.][]{Asano2008, Chen2009, Egedal2010a, Wang2010b, Teh2012a, Nagai2013, Oka2016}. We note that there have been some physical explanations to the flattop core component \citep[e.g.][]{Dum1978, Karlicky2012a, Fujimoto2014, Egedal2015}. For the entire spectral shape including both the flattop and non-thermal tail features, there has been a discussion from the non-Euclidean-normed statistical mechanics \citep[Eq.(66) of][]{Livadiotis2016}.

%----------------------------------------
\subsection{General Theories of Power-Law Formation}
\label{sec:gen}
%----------------------------------------

Where does a power law come from? There is a wide variety of theories of particle energization in both solar and space physics contexts, and readers are referred to comprehensive reviews for the details of those theories \citep[e.g.][]{Miller1997a, Aschwanden2005, Zharkova2011, Birn2012, Petrosian2012}. When it comes to the formation of power laws, however, many of those theories resort to stochasticity (i.e., random motion of particles). Fermi's approach (Section \ref{sec:stochastic}) is the classical way to treat stochastic acceleration. A Fokker-Planck approach is a convenient way to study stochastic acceleration in detail, and it even leads to the kappa distribution as a solution under certain conditions (Section \ref{sec:fokker}). %We also note that a power-law form of particle energy spectrum is often viewed as a summation of many thermal distributions with  different temperatures, from both data analysis \citep[e.g.][]{Krucker2008} as well as theoretical \citep[e.g.][]{Nishizuka2013} points of view. However, as far as we are aware, the physical origin of such a multi-thermal plasma has not been discussed in detail. 

%XXXXXXXXXXXXXXXXXXXXXXXXXXX
\begin{figure}
\includegraphics[width=\textwidth]{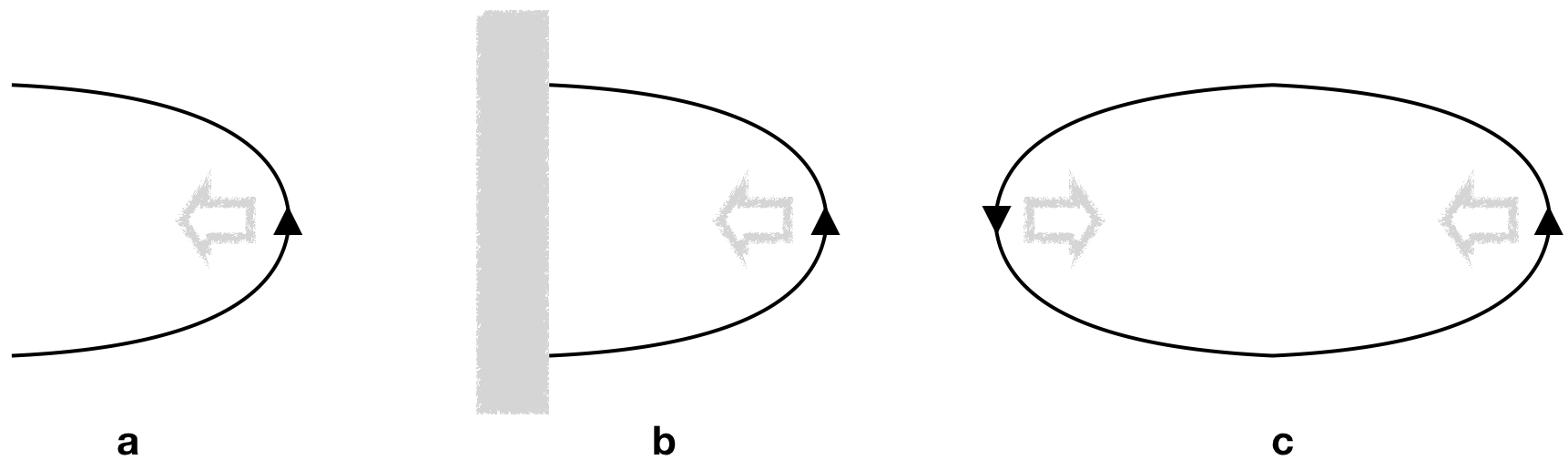}
\caption{Schematic illustrations of a magnetic field line favorable for Fermi Type B acceleration: (a) a curved magnetic field line as in the original illustration by \cite{Fermi1949}, (b) a line-tied version, and (c) a magnetic island version. In the line-tied version, particles can be trapped due to mirror reflection at the footpoints. An energization `kick' can occur at the curved tip of the field line and/or the mirror points.  See Appendix \ref{sec:fermitheory} for more details.}
\label{fig:fermitypeB}
\end{figure}
%XXXXXXXXXXXXXXXXXXXXXXXXXXX

\subsubsection{Fermi's approach}
\label{sec:stochastic}

In a stochastic process, particles bounce back and forth between scattering agents (such as plasma waves and turbulence). A net energy gain is achieved when the frequency of head-on collision with scattering agents is larger than the frequency of head-tail collision. If there is a probabilistic escape process in the system,  then a power-law energy spectrum can be formed and the power-law index depends on the timescale of the particle energization process as well as the timescale of the probabilistic escape process. In the original theory by \cite{Fermi1949} who pioneered the stochastic process in the context of the origin of galactic cosmic rays, the resultant energy spectrum is derived as
\begin{equation}
N(E) \propto E^{-\left(1+ \frac{\tau_{\rm acc}}{\tau_{\rm esc}} \right)}
\label{eq:stochastic}
\end{equation}
where $E$ is the particle energy, $\tau_{\rm acc}$ is the acceleration time scale, and $\tau_{\rm esc}$ is the escape time scale (The quantity $\tau_{\rm acc}/\tau_{\rm esc}$ is assumed to be energy independent).  See Appendix \ref{sec:fermitheory} for the derivation and variations of the theory, i.e.,  the first and second-order Fermi acceleration for relativistic and non-relativistic conditions. It should be noted that, even if the acceleration region was sufficiently large or there was no escape process such that $\tau_{\rm esc} \rightarrow \infty$, we can still expect a power law.

Usually, waves and turbulence are invoked in many applications of the stochastic acceleration process. The importance of turbulence in a stochastic process is more fully described in \cite{Petrosian2012}. The widely-known application of the stochastic process with waves and turbulence is the diffusive shock acceleration (DSA) \citep[e.g.][]{Blandford1987}. In DSA, particles experience stochastic first-order Fermi acceleration by moving back and forth across the shock front. They can be confined at and around the shock front by waves and turbulence. 

However, we emphasize that stochasticity is not always required for particle energization, although it appears necessary for a power-law formation. For example, during shortening of a magnetic flux tube (as schematically illustrated in Figure \ref{fig:fermitypeB}b), all particles moving along the flux tube (with small pitch angles) may experience multiple Fermi acceleration (via, for example, mirror reflection and `slingshot' effect (See Appendeix \ref{sec:fermitheory})) but not necessarily escape or scattering (if waves/turbulence were absent).   Here we simply define \textbf{Fermi acceleration} as an energization `kick' by a moving structure (such as a curved magnetic field line, waves, turbulence). Fermi acceleration usually works statistically through multiple kicks.  A \textbf{stochastic acceleration} is statistical Fermi acceleration combined with stochasticity with the form of, for example, probabilistic escape process. In the Fokker-Planck approach (also described in the following subsection), an escape process may not be included explicitly, but the stochasticity is represented by diffusion. It can lead to a power-law energy spectrum when a power-law form of turbulence spectrum is assumed. Sometimes, the two terms (i.e., Fermi acceleration and stochastic acceleration) are used interchangeably, probably because the original work by \cite{Fermi1949} considered stochastic process to derive a power law. However, it is customary to distinguish the two terms (especially in the mangetospheric community) and we also follow the same definitions throughout this paper.

Another example that illustrates the difference between Fermi and stochastic acceleration is the particle energization process in magnetic islands (or fluxropes in 3D). Particles trapped within an island flanked by active X-lines at both ends (as illustrated in Figure \ref{fig:fermitypeB}c) are energized through Fermi Type B acceleration \citep{Kliem1994}. The process is named `contracting island mechanism' and can be expanded to a stochastic acceleration process in  a system with many islands \citep{Drake2006}. The system with many islands may act like turbulence and produce high-energy particles, especially in a three-dimensional system \citep[e.g.][]{Dahlin2015}. A prediction of the power-law index, however, requires a phenomenological escape or diffusion process. In some cases, a sufficiently large escape time scale ($\tau_{\rm esc} \rightarrow \infty$) is considered so as to produce the hardest power law (and the smallest power-law index) \citep[e.g.][]{Drake2010, Guo2014} (See Eq. \ref{eq:stochastic}).  It is also to be noted that, if the distribution of the island sizes followed a power law, i.e., fractal distribution, then the resultant particle energy spectrum may become a power law even if each island did not produce a power law \citep{Nishizuka2013}. Such an idea may be connected to the nonextensive statistical mechanics as it was introduced to describe multifractals (Appendix \ref{sec:nonextensive}).

\subsubsection{Fokker-Planck approach}
\label{sec:fokker}

A stochastic process can be more conveniently described by a Fokker-Planck approach, and it has been used in various problems of particle acceleration. In collisionless plasmas, pitch angle scattering (by waves and turbulence) and associated diffusion are considered for the effective collisional term. This leads to the particle transport equation, which has been used to explain power-law spectra at shocks \citep[e.g.][]{Blandford1987}.  In such a treatment, particle distributions are typically a function of time $t$, velocity $\mathbf{v}$ and space $\mathbf{x}$, where $\mathbf{x}$ is needed to describe the shock structure. 

However, a Fokker-Planck approach shows that turbulence alone (without a shock) can still lead to a non-thermal power-law tail,  although the turbulence spectrum must be in a power-law form \citep[e.g.][]{Miller1996, Yoon2006a, Zhdankin2017}. Furthermore, the Fokker-Planck equation can also yield the kappa distribution as an analytical solution, as shown by e.g., \cite{Hasegawa1985} and \cite{Ma1998a}. These authors included collisional friction in addition to diffusion by a specific type of waves  such as high-intensity radiation field \citep{Hasegawa1985} and whistler waves \citep{Ma1998a}. 

Interestingly, \cite{Bian2014a} followed the same derivation but made it clearer that the power-law index $\kappa \mbox{*}$ is the ratio between the acceleration time scale and collisional friction/deceleration time scale. (As noted in the footnote in Section \ref{sec:definition}, their definition of kappa ($\kappa \mbox{*}$) is slightly different from the conventional definition of kappa ($\kappa$) and $\kappa \mbox{*} = \kappa + 1$.) Their argument implies that any type of waves/turbulence can lead to a power-law  as long as the diffusion coefficient $D$ has the form of $D \propto 1/v$. Their result, i.e., $\kappa \mbox{*} = \tau_{\rm acc}/2\tau_{\rm c}$ where $\tau_{\rm acc}$ and $\tau_{\rm c}$ are the acceleration and collisional friction time scales, respectively, reminds us of the power-law index given by Fermi (Eq. (\ref{eq:stochastic})). A caveat is that this formula and the observed values of $\kappa$ (as reviewed in this paper) implies the collision time   to be only an order of magnitude smaller than the acceleration time scale. In contrast, the collision time-scale can be many orders of magnitude larger in Earth's magnetotail and in solar flares where a thin-target model of the hard X-ray emission can be assumed. And yet, the electron kappa distribution has been directly observed in Earth's magnetotail (Section \ref{sec:emt}) and it may also exist in a thin-target hard X-ray source of flares (Section \ref{sec:alt}).  

Nevertheless, the possible involvement of Coulomb collisions in the generation of kappa distributions is not surprising if we consider the energy dependence of the collision frequency. The cross-section for the Coulomb collisions among particles varies with particle energy $E$ as $E^{-2}$ and the collision frequency scales as $E^{-3/2}$ (See textbooks by, for example, \cite{Benz2002} and \cite{Kulsrud2005}). Thus, it takes longer for higher-energy particles to equilibrate \citep[e.g.][]{Dudik2017}. Such an effect can be considered in an environment with density gradient (such as the solar atmosphere) to explain the origin of the non-thermal tail without resorting to wave-particle interactions \citep[e.g.][]{Scudder1979}. (A detailed review of non-thermal particles in a collisionally dominated non-equilibrium plasma can be found in \cite{Dudik2017}.) 

On the other hand, many of the theories of power-law formation by waves and turbulence (as reviewed above) consider the test-particle limit in which there is no feedback to particles from the waves and turbulence, although recent particle simulations of relativistic plasmas produce power laws \citep[e.g.][]{Guo2014, Zhdankin2017}. In this regard, \cite{Yoon2006a} considered the self-consistent acceleration of electrons to suprathermal energies by weak turbulence processes which involve the Langmuir/ion-sound turbulence and the beam-plasma interaction. In their theory, the spontaneously emitted thermal fluctuations act like the collisional drag in the Fokker-Plank equation. Further developments of the theory can be found in their recent work \citep[e.g.][]{Yoon2012,Yoon2014,Yoon2016}. Because they argue that the kappa distribution represents what they call `turbulent equilibrium' between electrons and enhanced Langmuir turbulence, their theory may have a connection to the nonextensive statistical mechanics in which the kappa distribution is illustrated as the state of maximum entropy of non-equilibrium plasmas as we discuss in Appendix \ref{sec:nonextensive}.

%#####################################################################
%#####################################################################
\section{Power Laws in Solar Flares}
\label{sec:flares}
\addtocontents{toc}{\setcounter{tocdepth}{3}}% Allow \section in ToC
%#####################################################################
%#####################################################################

\begin{figure}[t]
\includegraphics[width=\textwidth]{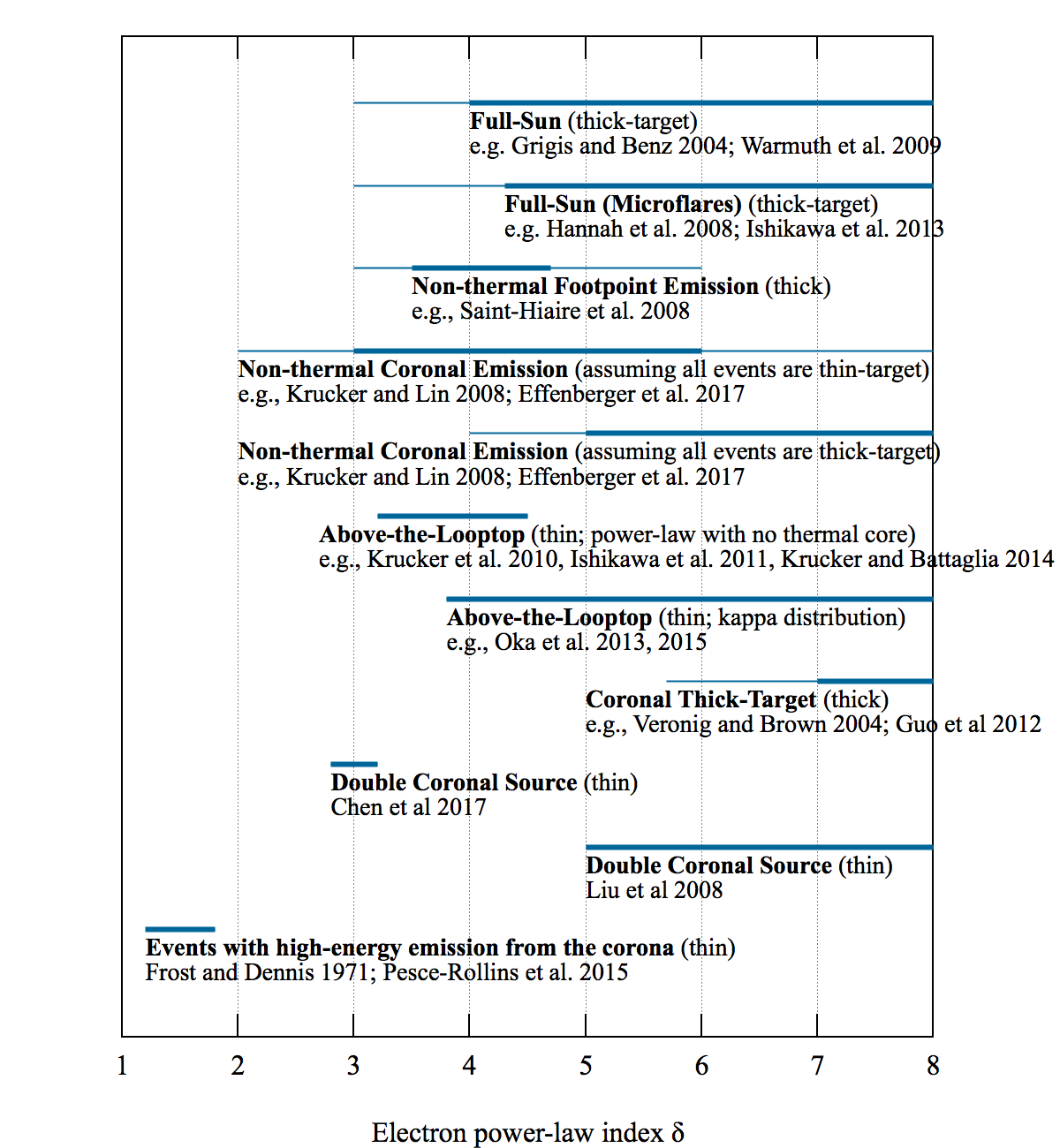}
\caption{An overview of electron power-law index $\delta$ reported by different studies of solar hard X-ray measurements in the $<$ 100 keV energy range. Various assumptions and/or different degrees of uncertainties could be involved in the conversion from the photon power-law index $\gamma$ to the electron power-law index $\delta$ (See Sections \ref{sec:definition} and \ref{sec:overview}). The assumption of thin- or thick-target emission model is indicated in the parenthesis for each category. For statistical studies, typical values are shown by thick lines. For non-statistical studies, only the thick lines are used to display reported values. See texts in this section for more details and caveats of each category. }
\label{fig:flr_index}
\end{figure}

There are many reviews of solar flares and associated particle acceleration \citep[e.g.][]{Miller1997a, Aschwanden2005, Krucker2008a, Lin2011a, Kontar2011a, Fletcher2011, Zharkova2011, Holman2011a,  Petrosian2012, Benz2017, Dudik2017}. Here we focus on the  power-law index measured  in the sub-relativistic energy range (typically $<$ 100 keV), as summarized in Figure \ref{fig:flr_index}. 

%XXXXXXXXXXXXXXXXXXXXXXXXXXX
%XXXXXXXXXXXXXXXXXXXXXXXXXXX

%----------------------------------------
\subsection{Overview}
\label{sec:overview}
%----------------------------------------

Solar flares are the most energetic phenomena in our solar system. They release energy, previously stored in the magnetic field, of the order $10^{32}-10^{33}$ erg in only seconds to minutes, converting it to accelerated particles and heating all the layers of the solar atmosphere. 

\begin{figure}
\includegraphics[width=0.6\textwidth]{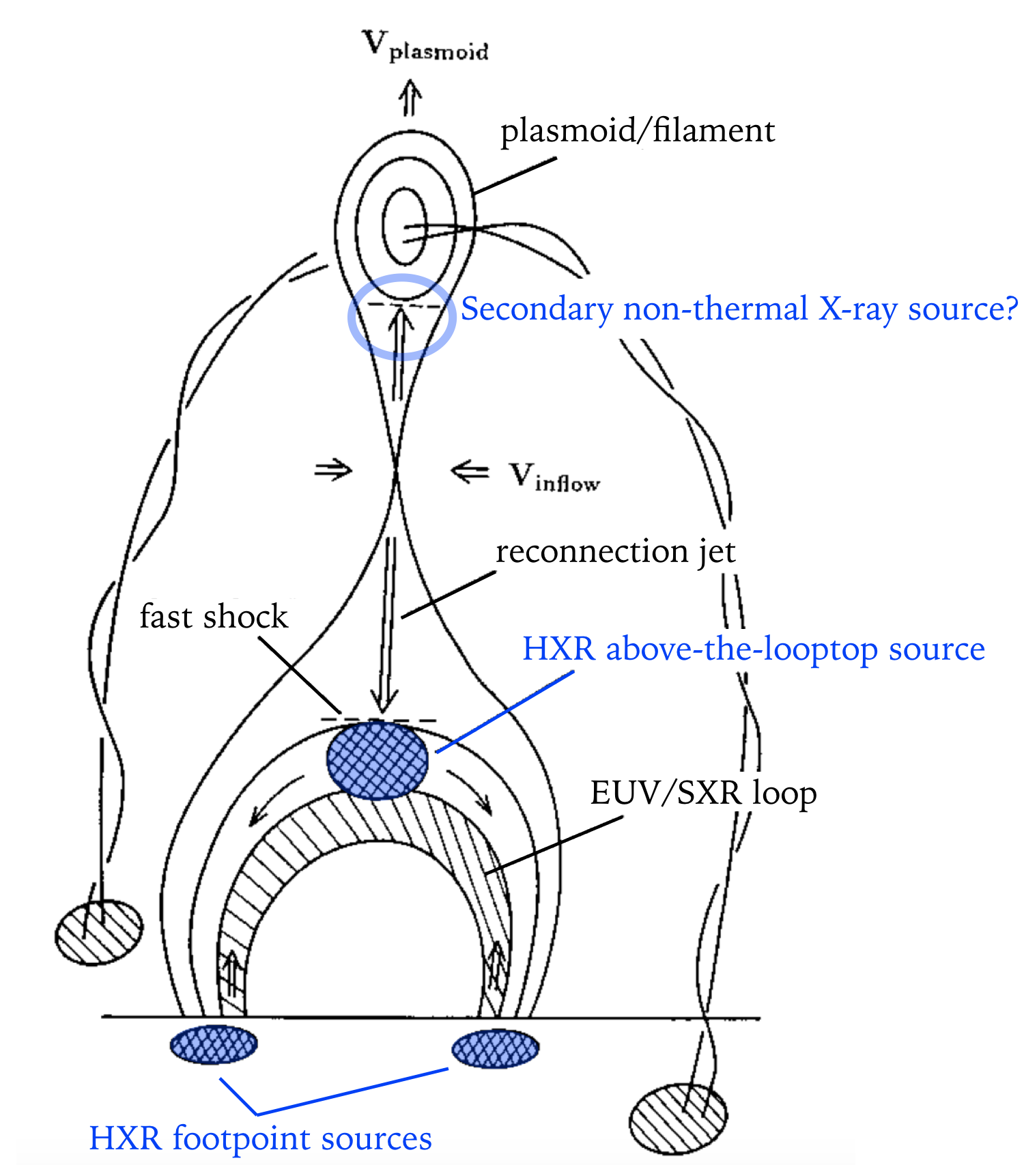}
\caption{Cartoon of the standard solar flare scenario, as adapted from \cite{Shibata1995}. This cartoon is based on the discovery of the Masuda flare in which the coronal hard X-ray (HXR) source was located `above' the EUV/SXR loop (See also Section \ref{sec:alt} and Figure \ref{fig:flr_data}).  A power-law spectrum has been observed in the blue-shaded regions at the footpoints and the above-the-looptop region. In many cases, however, the flare size is smaller than that of the Masuda flare, and the coronal HXR source appears cospatial with the lower-energy X-ray source and closer to the apex of the EUV/SXR loop. }
\label{fig:flr_model}
\end{figure}

The standard solar flare scenario envisages energy release via reconnection near or above the top of magnetic loops that are rooted to the solar surface. As a result, electrons, protons, and ions are accelerated up to relativistic speeds.  The accelerated particles are then either ejected into interplanetary space along open field lines (and may be observed {\it in-situ} as solar energetic particles (SEPs)) or get trapped in closed magnetic loops where they propagated downward toward the dense chromosphere and photosphere. In the latter case, the accelerated particles deposit their energy into the chromosphere. This results in heating and expansion of chromospheric plasma upwards into the magnetic loop (termed chromospheric evaporation). Figure \ref{fig:flr_model} illustrates this model. See also \cite{Holman2016} and references therein for the standard flare model. 

%XXXXXXXXXXXXXXXXXXXXXXXXXXX
%XXXXXXXXXXXXXXXXXXXXXXXXXXX

The signatures of accelerated electrons and heated plasma are readily observed in X-rays at energies of a few keV up to a few hundred keV. They are bremsstrahlung emission caused by electrons suffering Coulomb collisions with the ambient plasma \citep[e.g.][and references therein]{Holman2011a}. Their intensity depends on the plasma temperature, amount of heated plasma (emission measure), ambient density, and electron number, but not on the magnetic field strength\footnote{In addition to X-rays, radio emissions can also be used to diagnose energetic electrons during flares \citep[e.g.][]{White2011, Nita2015}. The radio emission can depend on the magnetic field strength.} (Eqs. (\ref{eq:thintarget}) and (\ref{eq:thicktarget}) in Section \ref{sec:definition}). If a photon energy spectrum exhibits a power-law distribution (with the power-law index $\gamma$), then the spectrum of the electrons that emitted the X-rays must also have a power-law component (with the power-law index $\delta$). In converting $\gamma$ to $\delta$, a caveat is that one has to make an assumption on the density of the {\it target} (the ambient plasma) with which X-ray-emitting electrons collide. For large column densities, such as in the dense chromosphere, in which the electrons lose their energy completely in collisions, the thick target model for bremsstrahlung emission is applicable \citep[e.g.][]{Brown1971a,Kontar2011a}. For smaller column densities, the electron spectrum will not be affected significantly and a thin target approximation is used. Whether a target acts as thin or thick depends on the energy of the accelerated electron and the ambient density.  Hence, in many cases, an `intermediate thin-thick target' model \citep[e.g.][]{Wheatland1995,Battaglia2007a} may be needed to describe the full spectrum. See Section \ref{sec:definition} for more details of the conversion between $\gamma$ and $\delta$ and the target-assumptions.

%XXXXXXXXXXXXXXXXXXXXXXXXXXX
\begin{figure}
\includegraphics[width=0.6\textwidth]{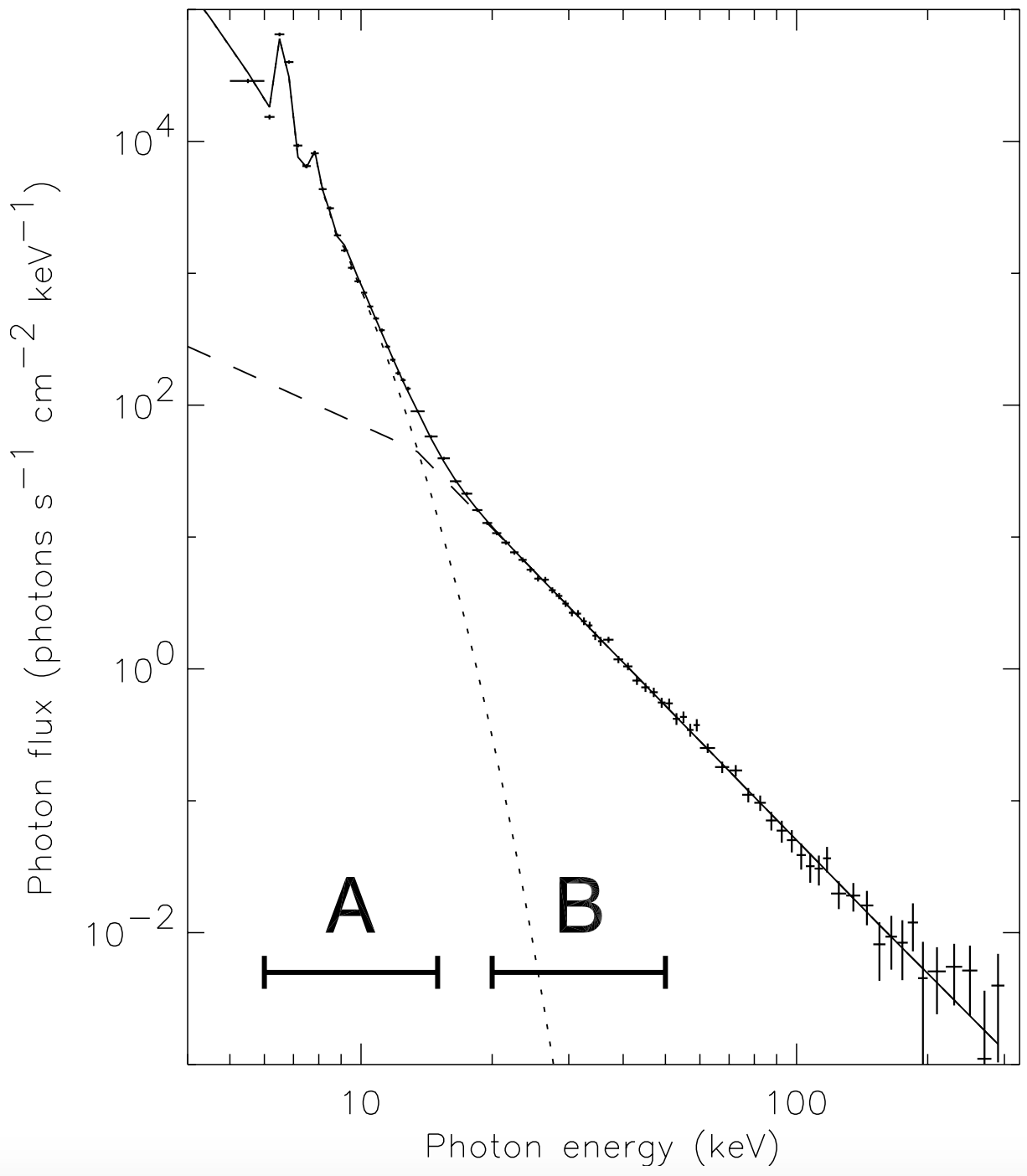}
\caption{A typical `full-sun' (spatially integrated) X-ray spectrum during flares, adapted from Figure 1 of \cite{Grigis2004}. Reproduced with permission \textcopyright ESO. The dotted line is a thermal distribution model, representing the lower-energy component. The dashed line is a power-law model with a lower-energy cutoff at $E_{\rm cutoff}$ = 13.4 keV, representing the higher-energy component. Note the extension of the power-law below the cutoff energy. This part of the spectrum is unobservable due to the intense low-energy component. However, under the assumption that the electron spectrum has a sharp cut-off at $E_{\rm cutoff}$, the resulting photon spectrum would have the shape of the dashed line, as every electron of a given energy (say 10 keV), can emit many photons below this energy.
%Because an electron can emit photons with different energies (for example, a 10 keV electron can emit many photons in the $<$ 10 keV range), the lower-energy cutoff in the electron spectrum appears as a turnover in the photon spectrum, and the power-law spectrum extends toward lower energies with a harder (flatter) spectral slope. 
The narrow peaks in the lower energy range are the Fe ($\sim$6.7 keV) and Fe - Ni ($\sim$8 keV) excitation line complexes. } 
\label{fig:flr_fullsun}
\end{figure}
%XXXXXXXXXXXXXXXXXXXXXXXXXXX

While hard X-ray (HXR) imaging capability was already available by the 1980s \citep[e.g.][]{vanBeek1980}, the observationally easiest and most often applied analysis of flare spectra  is the fitting of `full-Sun,' i.e. spatially integrated, spectra. Figure \ref{fig:flr_fullsun} shows an example of a full-Sun X-ray spectrum with no spatial information, as obtained by Reuven Ramaty High Energy Solar Spectroscopic Imager \citep[RHESSI, ][]{Lin2002}\footnote{RHESSI observes the full Sun through 9 rotating collimators, allowing to make images of flaring sources and infer spectra from individual sources \citep[e.g.][]{Battaglia2006,Krucker2014}. However, in Figure \ref{fig:flr_fullsun}, the data is spatially-integrated to produce the full-sun spectrum. Similar spectral forms have been already obtained by earlier works \citep[e.g.][]{Lin1981a}}. There is clearly a spectral break at around 18 keV in this case. (Typically, the break energy is in the 10 -- 20 keV range.) Thus, when characterizing and fitting spectral data,  thermal and non-thermal power-law distributions are used for the lower and higher energy components, respectively. For the lower energy range (typically $<$ 10 keV or `soft X-rays (SXRs)'), this iso-thermal (single Maxwellian) model usually fits the data well, although flaring plasma is known to be multi-thermal (However, RHESSI's temperature sensitivity range (upward of $\sim 8$ MK) does not allow for constraining cooler temperatures). For the higher energy range (typically $>$ 20 keV or `hard X-rays (HXRs)'), the power-law model also fits the data well, although it requires a somewhat unphysical, lower-cutoff energy $E_{\rm cutoff}$.  There is a wide variety of discussions on $E_{\rm cutoff}$ as reviewed by \cite{Holman2011a} (See also \cite{Holman2016} for more details on the terminology). 

\begin{figure}
\includegraphics[width=\textwidth]{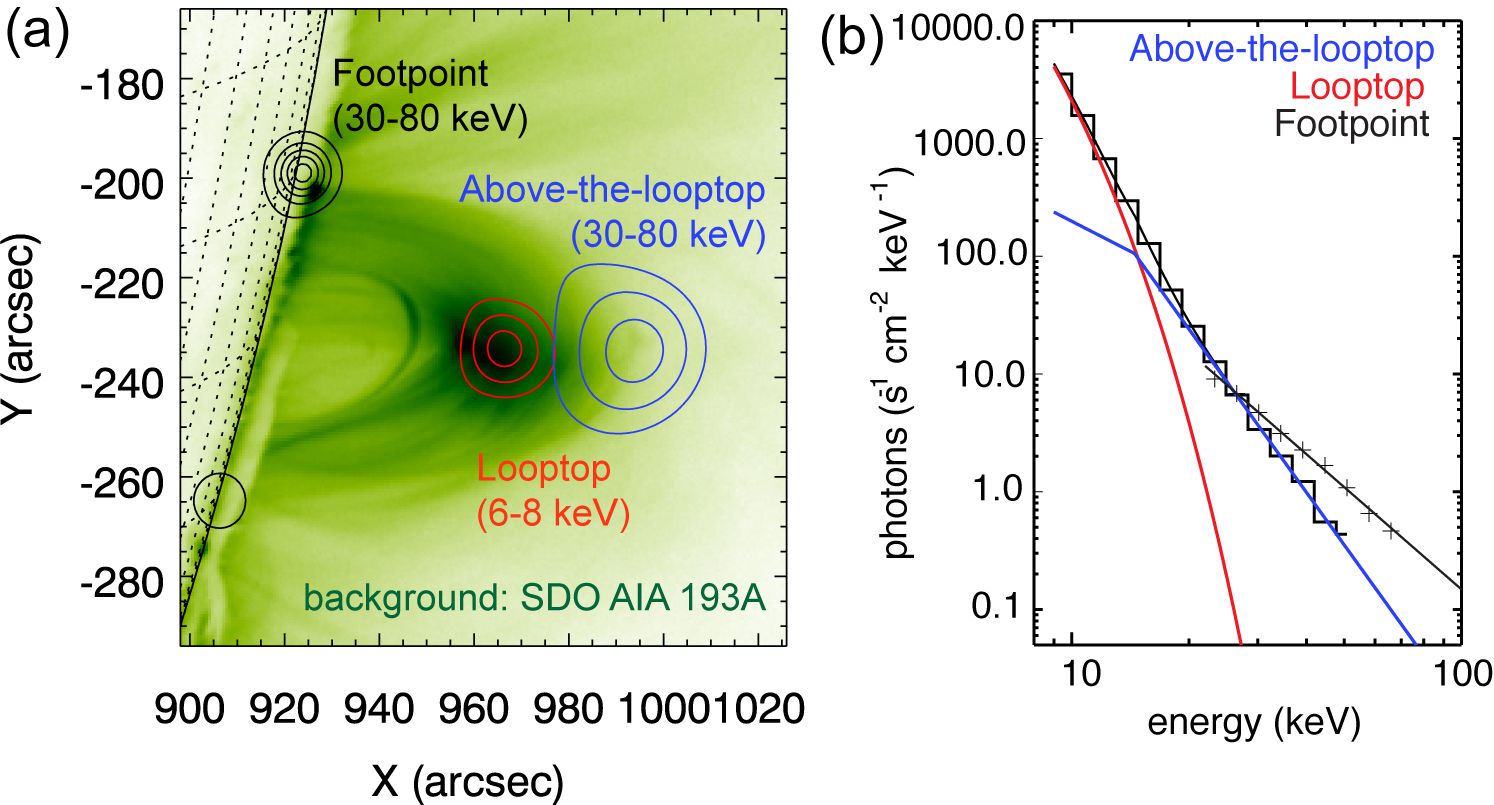}
\caption{Imaging spectroscopy from the solar limb flare SOL2012-07-19T05:58. Left: X-ray images as measured by RHESSI (contours) superposed on an extreme ultra-violet (EUV) image  as measured by SDO/AIA. Right: Photon energy spectra from the chromospheric footpoint sources (crosses) and the combined coronal sources (histogram). The crosses are fitted with a power-law (black) and the histogram is fitted with a combined thermal (red; representing the looptop source) and non-thermal (blue; representing the above-the-looptop source) spectra. Both the right and left panels are adapted from Fig. 1 of \cite{Krucker2014} but the color scheme is altered so that the same color indicates the same source in both panels. The left panel is actually a reproduction by \cite{Oka2015}.}
\label{fig:flr_data}
\end{figure}

X-ray imaging revealed that the higher-energy, non-thermal component comes primarily from the chromosphere at the footpoints of the flaring loop, while the lower-energy, thermal component  comes from the corona at and around the top of the flaring loop, as illustrated in Figures \ref{fig:flr_model} and \ref{fig:flr_data}.  The higher-energy non-thermal emissions can also come from the corona at and around the looptop region, although the intensity is much lower than that of the footpoint emissions. In rare cases, such faint and non-thermal emissions are originating from somewhat `above' the mostly thermal looptop region \citep[e.g.][ See also Figures \ref{fig:flr_model} and \ref{fig:flr_data}]{Masuda1994}. The separation distance between the thermal `looptop' region and non-thermal `above-the-looptop' region can be roughly 10 - 20 Mm \citep[e.g.][]{Krucker2010a, Oka2015}. This does not necessarily mean that thermal component does not exist  in the `above-the-looptop' region. Due to observational constraints (in particular, the limited dynamic range) and the presence of the very bright emission from the looptop region, it has been difficult to measure the thermal component that exists locally in the above-the-looptop region. (With the similar reason, the thermal components from the footpoint sources are difficult to observe.) 

Below, we focus only on the non-thermal, power-law component observed in full-sun spectra, chromospheric footpoint sources, and in the corona. 

%XXXXXXXXXXXXXXXXXXXXXXXXXXX
%XXXXXXXXXXXXXXXXXXXXXXXXXXX

%----------------------------------------
\subsection{Full-Sun Spectra}
%----------------------------------------

Full-sun spectra allow for investigating the time-evolution of single events and for statistical studies of larger sets of flares of different sizes \citep[e.g.][]{Holman2003a, Grigis2004,Emslie2004, Warmuth2009}. It has been shown that, in many flares, the time-evolution of the non-thermal tail follows the so-called soft-hard-soft pattern \citep[e.g.][]{Grigis2004}, i.e. the photon spectrum starts out rather soft, becomes harder (flatter) as the X-ray flux increases, and becomes softer (steeper) again as the emission decreases. This pattern has even been observed for individual peaks of the same flare \citep{Grigis2005a}. One interpretation of this is that the acceleration mechanism itself undergoes a change in efficiency as the flare progresses \citep[e.g.][]{Grigis2006,Bykov2009a}. Full-sun spectra have also been used in various studies of correlations between flare parameters, such as temperature, emission measure, X-ray flux and power-law index of flares of different sizes (as measured by GOES class). A typical finding is that larger flares tend to be hotter and have more non-thermal flux \citep[e.g.][]{Ryan2012,Warmuth2016}. However, \cite{Battaglia2005} showed that there seems to be no correlation between the photon power-law index $\gamma$ and the GOES class, i.e. even some smaller flares exhibit rather hard spectra. In fact, there have been statistical studies of power-law spectra in microflares \citep[e.g.][]{Christe2008, Hannah2008}. Using Suzaku data, \cite{Ishikawa2013} showed that the power-law spectrum can reach up to at least 100 keV even in smaller events. 

In the compilation of the reported numbers of the power-law index $\delta$ in Figure \ref{fig:flr_index}, we assumed the thick-target model for full-sun spectra. This is because, in many cases, non-thermal X-ray emission originates primarily from the footpoint sources located in the dense chromosphere, which acts as a thick-target on all electrons.

%----------------------------------------
\subsection{Non-thermal Emission in Footpoints}
\label{sec:footpoints}
%----------------------------------------

As outlined above, hard X-ray (HXR) sources at the footpoints of magnetic loops are non-thermal, thick target emission of accelerated electrons as they are stopped in the dense chromosphere \citep[e.g.][]{Hoyng1981, Hoyng1981a, Sakao1994, Sakao1996, Aschwanden1999, Krucker2002, Emslie2003a, Saint-Hilaire2008, Yang2012}. Due to the high chromospheric density they are typically bright and thus easy to observe, often up to energies of more than 100 keV. Most flares show two footpoints \citep[e.g.][]{Saint-Hilaire2008}, although events with three or more footpoints have been reported \citep[e.g.][]{Emslie2003a}. Because of their brightness, HXR footpoints have invaluable diagnostic potential for the locations and amount of energy deposition into the chromosphere the acceleration mechanism and even chromospheric properties such as density \citep[e.g.][]{Kontar2008, Battaglia2011, Chen2013a}. 

According to a statistical analysis by \cite{Saint-Hilaire2008}, the typical photon power-law index $\gamma$  is 2.5 - 4.5. These authors also reported that,  in double-footpoint structures, the power-law index differences $\Delta \gamma$ is not large, ranging mostly between 0 and 0.6. Such an asymmetry can be attributed to (a) different magnetic field intensities and associated mirror effect \citep[e.g.][]{Sakao1994, Sakao1996, Aschwanden1999, Saint-Hilaire2008, Liu2009} or (b) different column densities and associated acceleration and/or transport effect along the loops \citep[e.g.][]{Emslie2003a, Liu2009}. Recently, \cite{Daou2016} used 1D Fokker-Planck simulations to investigate the relationship between the footpoint X-ray emission and the magnetic field ratio. They reported that the asymmetry $A$ of the X-ray intensity $I$, where $A \equiv (I_1-I_2)/(I_1+I_2)$ \citep[See also][]{Alexander2003, Liu2009}, reaches its saturation value close to unity when the footpoint magnetic field ratio reaches $\sim$4. As the bremsstrahlung emission itself does not depend on the magnetic field intensity, these studies indicate that the X-ray emitting electrons behave differently in the two end of the loop.

%----------------------------------------
\subsection{Non-thermal Emission in the Corona}
%----------------------------------------
As outlined above,  non-thermal X-ray emission from the corona is significantly fainter when compared to that from the footpoint sources. Thus, characterizing the spectral properties of non-thermal coronal emission remains challenging as described below. 

If the standard flare configuration was viewed from the top by an earth-orbiting satellite, it would be difficult to distinguish between emission originating from the chromosphere and emission from the corona, due to projection effects. Hence, a large number of studies has focused on flares that occurred typically a few degrees or more behind the solar limb \citep[e.g.][]{Tomczak2001, Krucker2008, Tomczak2009, Effenberger2016}. In such `occulted' events, footpoint sources are masked by the solar disk, and high-energy emissions in the corona can be observed without being limited by the dynamic range of the instrument.  \cite{Krucker2008} studied 55 occulted events in the RHESSI database and found higher-energy emissions that could be fitted with a power law in 50 events, establishing the presence of non-thermal plasmas in the corona. They also reported that the centroid positions of such  non-thermal sources are close to the location of thermal emission observed at lower energies to  within $\sim$2 Mm, consistent with an earlier finding by Yohkoh \citep{Tomczak2001}. More recently, \cite{Effenberger2016} added 61 occulted flares and, from the total of 116 occulted flares, confirmed the results by \cite{Krucker2008}.

Typical electron power-law indices $\delta$ in these events range from 3 to 6 (compare Figure \ref{fig:flr_index}). These were inferred under the thin-target assumption, as the density in these sources can be expected to be sufficiently low for the target to act as thin. However, without a rigorous determination of the density and the column depths, one cannot exclude the possibility that, at least some of these events act as a thick target on most electron energies.

In addition to these frequently observed emissions, there is a number of special types of events, as described in the following subsections.

\subsubsection{Above-the-looptop events}
\label{sec:alt}

A special case of coronal hard X-ray (HXR) source is sometimes observed above the top of the EUV loop. Such above-the-looptop (ALT) sources are also called `Masuda-type' sources, after the author who first reported them \citep{Masuda1994}.  Only a few cases of above-the-looptop sources have been reported by RHESSI  \citep[e.g.][]{Krucker2010a,  Ishikawa2011a, Oka2015, Liu2013, Liu2013a}. Unlike the aforementioned, typical cases in which the non-thermal source height $H$ (from the solar surface)  ranges between roughly 4 - 20 Mm, the above-the-looptop source is found in large flares with $H$ roughly 20 - 60 Mm.  The separation distance $d$ between the ALT source and the thermal X-ray source around the looptop (LT) region is also large, 10 - 20 Mm, in contrast to the values in the typical cases,  i.e., $|d|<$ 2 Mm \citep{Krucker2008, Effenberger2016}.

The X-ray spectrum from the above-the-looptop source exhibits a non-thermal power-law component.  It was argued that a standard scenario, with hard X-rays produced by a beam comprising the tail of a dominant thermal core plasma, does not work \citep{Krucker2010a}. Instead, it was proposed that all electrons in the ALT source are `bulk-energized' to form a power law with no thermal particles left \citep{Krucker2010a, Krucker2014, Ishikawa2011a} and that precipitating electrons are emptying out of the ALT source to produce the footpoint emissions \citep{Ishikawa2011a}. 

What leads to such a `bulk-energization' of electrons remains unclear to date. Radio emission from the above-the-looptop region  can exhibit spectral features similar to solar Type II radio bursts (which are associated with propagating shocks in the outer corona), suggesting that there could be a fast-mode shock in the above-the-looptop source as a consequence of collision between the downward reconnection outflow and the pre-existing flaring loop \citep[e.g.][]{Aurass2002, Aurass2004, Mann2009}. A recent study provided  more direct evidence of dynamically evolving (instead of stationary) fast-mode shock \citep{Chen2015a}. On the other hand, another study of microwave emissions indicated the hardest power law at and around the reconnection point, suggesting that the reconnection region (instead of the above-the-looptop region) is the key location of electron acceleration \citep{Narukage2014a}. 

The spectral analysis of the above-the-looptop (ALT) source is challenging because of the limited dynamic range (and hence limited energy coverage, say, $\lesssim$80 keV). The lower-energy end of the ALT spectrum is masked by the intense emission from the adjacent (mostly thermal) source at the looptop (LT) and, in fact, it is difficult to completely separate the spectra from LT and ALT regions. Thus, the combined spectrum from both LT and ALT regions is fitted with a combined, thermal and power-law model where the thermal and power-law components represent the LT and ALT sources, respectively \citep[e.g.][]{Krucker2010a}.   Such a model implicitly assumes (and can lead to a conclusion) that there is no thermal electron in the ALT region \citep[e.g.][]{Krucker2014}. The possibility of a presence of a cold thermal core within the ALT region is rejected by a constraint derived separately from imaging analysis. An alternative (and yet simple) approach is to assume a hot thermal core in the ALT region while maximizing the intensity of the power-law component \citep{Oka2013, Oka2015}. Such a model can be represented by the kappa distribution as illustrated in Fig. \ref{fig:specmodels} and used in other flare studies \citep[e.g.][]{Kasparova2009a, Hannah2010, Battaglia2015}. We emphasize that the kappa distribution model is still consistent with the idea of `bulk energization' in a sense that all electrons in the ALT region have experienced energization.  It should also be noted that a super-hot component has been identified in full-sun spectra in `on-disk' events  in addition to the brightest thermal component originating from the loop  \citep[e.g.][]{Caspi2010, Longcope2010, Caspi2014}.  Such a super-hot plasma may be related to the hot thermal core component in the ALT source as inferred by the kappa distribution model.

Regarding the power-law index,  the introduction of the hot thermal core (and hence the extension of the thermal component toward the higher energy range) in the ALT region leads to a slightly softer power-law tail, although the boundary between thermal and non-thermal components is not explicitly visible in the case of the kappa distribution. For example, in the limb flare of SOL2012-07-19T05:58, \cite{Krucker2014} used a single power-law with no thermal core to represent the ALT source and obtained  $\delta \sim$ 3.6 while \cite{Oka2015} used the kappa distribution and obtained $\delta \sim$ 4.1.  Of course, in principle, both models should return the same power-law index if the power-law tail extended toward much higher energies ($\gg$ 80 keV) and if the dynamic range of the measurement was large enough to cover such higher energy ranges. After all, the spectral form of the ALT source is not well constrained leading to the somewhat different values of the estimates of the power-law index.

\subsubsection{Coronal thick-target events}
\label{sec:thick}

While many solar flares display the typical morphology with two footpoints,  there are cases for which footpoint sources are absent (even though they occur `on-disk' where footpoint sources would not be masked) and flare emission at all energies originates from the corona \citep[e.g.][]{Veronig2004, Sui2004}. It has been described that the flaring loop was dense enough to become collisionally thick at observed energies and that all electrons deposit most of their energy before reaching the footpoints.  Thus, such cases are referred to as coronal thick-target events and have been studied by different approaches  \citep[e.g.][]{Xu2008a, Guo2012b, Guo2012c, Lee2013, Fleishman2016}. Based on analysis and modeling of the source properties (such as the source size and number density distributions), it was argued that the coronal thick-target sources can be interpreted as sites of electron acceleration \citep[e.g.][]{Xu2008a, Fleishman2016}. 

As for the energy spectra, the power-law slope is steep with the photon power-law index $\gamma >$6 \citep[e.g.][]{Veronig2004, Krucker2008a}. \cite{Guo2012b} presented 22 cases of spectral analyses obtained from 11 flare events. It was reported that the lowest value of the electron power-law index $\delta$ (with the assumption of thick-target emission) is 5.7 but the typical value is in the range 7 - 9. They pointed out that, while the steepness could be a property of the electron acceleration process, it is also consistent with the absence of footpoint emission that the higher-energy electrons would produce in less-dense cases.

\subsubsection{Other notable events}
\label{sec:double}

Magnetic reconnection generates bi-directional jets. In the standard flare model (Fig. \ref{fig:flr_model}), the `above-the-looptop' source is found at or near the location where the downward jet would be obstructed by the flaring loop. However, in rare cases,  a secondary, non-thermal emission can be found on the opposite (i.e., non-obstructed) side of the presumed location of magnetic reconnection (See the open circle in Figure \ref{fig:flr_model}) \citep[e.g.][]{Sui2003, Sui2004, Liu2007, Liu2013, Chen2017}. The dependence of source height on energy (temperature) was positive and negative in the primary (lower) and secondary (upper) source, respectively, suggesting a presence of highest temperature electrons in between the two sources. \cite{Sui2003} interpreted this finding as evidence for a large-scale current sheet. Because of the dependence of source height on energy, the lower source can be viewed as the above-the-looptop (ALT) source. Furthermore, in the events reported by \cite{Sui2003} and \cite{Sui2004}, the lower source was actually considered thick-target, as described by \cite{Veronig2004} and reviewed in the previous subsection.

\cite{Liu2007} reported a case in which both sources are non-thermal and the photon power-law index $\gamma$ was 6 - 9. While the dependence of source height on energy was similar to that of \cite{Sui2003}, they argued that the inner emissions (i.e., those closer to the center of the primary (lower) and secondary (upper) source) showed a harder power-law spectrum than the outer emissions. \cite{Chen2017} argued, in another case study, that the upper source could be where side lobes (two different arcades) are rapidly approaching toward each other as a result of breakout eruption. The photon index was relatively small $\sim$4. 

While coronal emission is generally faint and can be detected typically up to $\sim$ 100 keV, some cases show coronal emissions that extend up to $\gamma$-ray ranges.  For example, \cite{Krucker2008c} reported very hard spectra with the photon power-law index $\gamma$ between 1.5 and 2 in the 200 -- 800 keV range. These events are not shown in Figure \ref{fig:flr_index} because of the very different energy range. 

\cite{Frost1971} also reported a high-energy coronal emission up to at least 250 keV in a behind-the-limb event. While the power-law spectrum was soft in the $\gtrsim$ 100 keV range, it was very hard in the $\lesssim$ 100 keV range   with the photon power-law index $\gamma \sim$ 2.3.  With the thin-target emission model, the electron power-law index is $\delta \sim$1.3. This event  was associated with Type II radio bursts and solar energetic particles (SEPs, see Section \ref{sec:sep}),  indicating that a coronal mass ejection (CME) occurred and a CME-driven shock had formed in the solar corona. 

While the precise origin of such a hard power law from the corona remains unclear, similar behind-the-limb events with CME/SEP signatures were reported \citep{Pesce-Rollins2015}. The main focus of this new study was a detection of $>$100 MeV gamma-rays for $\sim$ 30 min by the {\it Fermi} satellite, but it also showed a relatively hard power-law in the X-ray range ($\lesssim$ 100 keV). The authors reported that the photon power-law index $\gamma$ was $\sim$3.8. This corresponds to the electron power-law index $\delta$ of $\sim$ 2.8 with the thin-target emission model. Possible connections between the $>$ 100 MeV gamma-rays, hard X-ray flare, and CME/SEPs are still debated \citep[e.g.][]{Ackermann2017}. 

%It should be noted again that the simple power-law relationship is not valid when there is a higher-energy (relativistic) emission and/or a spectral break close to and above the photon energies of interest \citep{Holman2011a}. Thus, while \cite{Frost1971} and \cite{Pesce-Rollins2015} reported hard power-law spectra (i.e., small $\gamma$), we have not shown their $\delta$ values in Fig. \ref{fig:flr_index}. 

%----------------------------------------
\subsection{Relation Between Coronal and Footpoint Sources}
\label{sec:relation}
%----------------------------------------
If the loop size is sufficiently large, the non-thermal coronal source can be well separated from the footpoint sources even in `on-disk' events. In such a case, imaging spectroscopy (i.e., using multiple image to construct an energy spectrum for individual sources) allows us to compare spectral relation between the two sources. In the simplest flare scenario, where the corona acts as a thin target on the same accelerate electron beam that, after propagation and neglecting transport,  results in the observed thick target footpoint emission, the inferred spectral indices $\delta$ have to agree. In terms of the photon power-law index $\gamma$, it must be somewhere between the two extreme assumptions, i.e., $\gamma_{thin} = \delta + 1$ and $\gamma_{thick} = \delta - 1$. Thus, the difference between $\gamma_{c}$ in the coronal source and $\gamma_{f}$ in the footpoint sources must be no larger than 2. 

However, \cite{Battaglia2006} reported that, in 2 events out of 5, $\gamma_{c} - \gamma_{f}$ was larger than 2, indicating there was additional process that would make the footpoint spectra harder. They proposed transport effects due to return current, causing lower-energy electrons to preferentially lose their energies. On the other hand, \cite{Simoes2013} reported 4 events that showed the electron power-law index $\delta_{f}$ in the footpoint sources significantly softer than that of the non-thermal coronal source $\delta_{c}$. They suggested a presence of mechanisms that would keep a fraction of  non-thermal electrons trapped inside the coronal loop. It is also to be noted that \cite{Krucker2008c} reported 3 cases of coronal $\gamma$-ray emission. While the energy range was much higher (200 - 800 keV), the photon power-law index $\gamma$ (between 1.5 and 2) was substantially smaller than that of footpoint sources (between 3 and 4). Thus, they suggested that flare-accelerated high-energy ($\sim$MeV) electrons stay long enough in the corona to lose their energy by collisions producing $\gamma$-ray emission, while lower energetic electrons precipitate more rapidly to the footpoints. Regarding the theoretical interpretation of data, \cite{Chen2013a} examined two different flare events to discuss the energy dependences of basic characteristics in the framework of a stochastic acceleration mechanism.

%#####################################################################
%#####################################################################
\section{Power Laws in Earth's Magnetotail}
\label{sec:emt}
\addtocontents{toc}{\setcounter{tocdepth}{3}}% Allow \section in ToC
%#####################################################################
%#####################################################################

A comprehensive review of particle acceleration mechanisms in Earth's magnetotail and auroral region can be found in \cite{Birn2012}. Here we focus on the power-law index measured in the magneotail in the typical energy range of less than a few hundreds of keV.

%XXXXXXXXXXXXXXXXXXXXXXXXXXX
\begin{figure}
\includegraphics[width=0.7\textwidth]{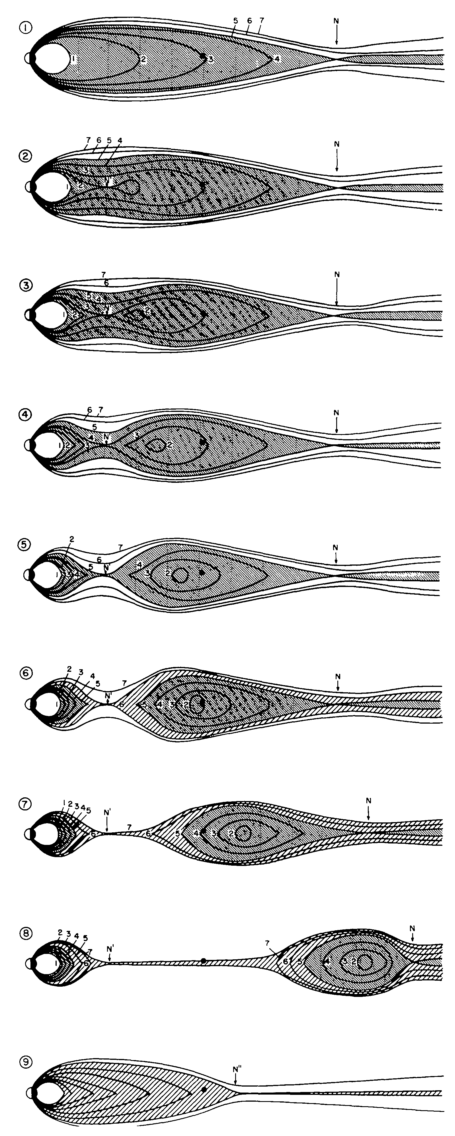}
\caption{Schematic illustration of an evolution of Earth's magnetotail, illustrating how and where magetic reconnection can take place, adaptes from \cite{Hones1977}. Each line represents a magnetic field line and the numbers (1 - 7) follow the same field lines. The near-Earth reconnection site (as illustrated in panels 3 - 8) is typically located at 20 - 30 $R_{\rm E}$ away from Earth, where $R_{\rm E}$ is Earth's radii $\sim$6378 km. The distant reconnection site (as illustrated in all panels) are typically located at 100 - 200 $R_{\rm E}$ away from Earth. }
\label{fig:emt_hones}
\end{figure}
%XXXXXXXXXXXXXXXXXXXXXXXXXXX

%XXXXXXXXXXXXXXXXXXXXXXXXXXX
\begin{figure}[t]
\includegraphics[width=\textwidth]{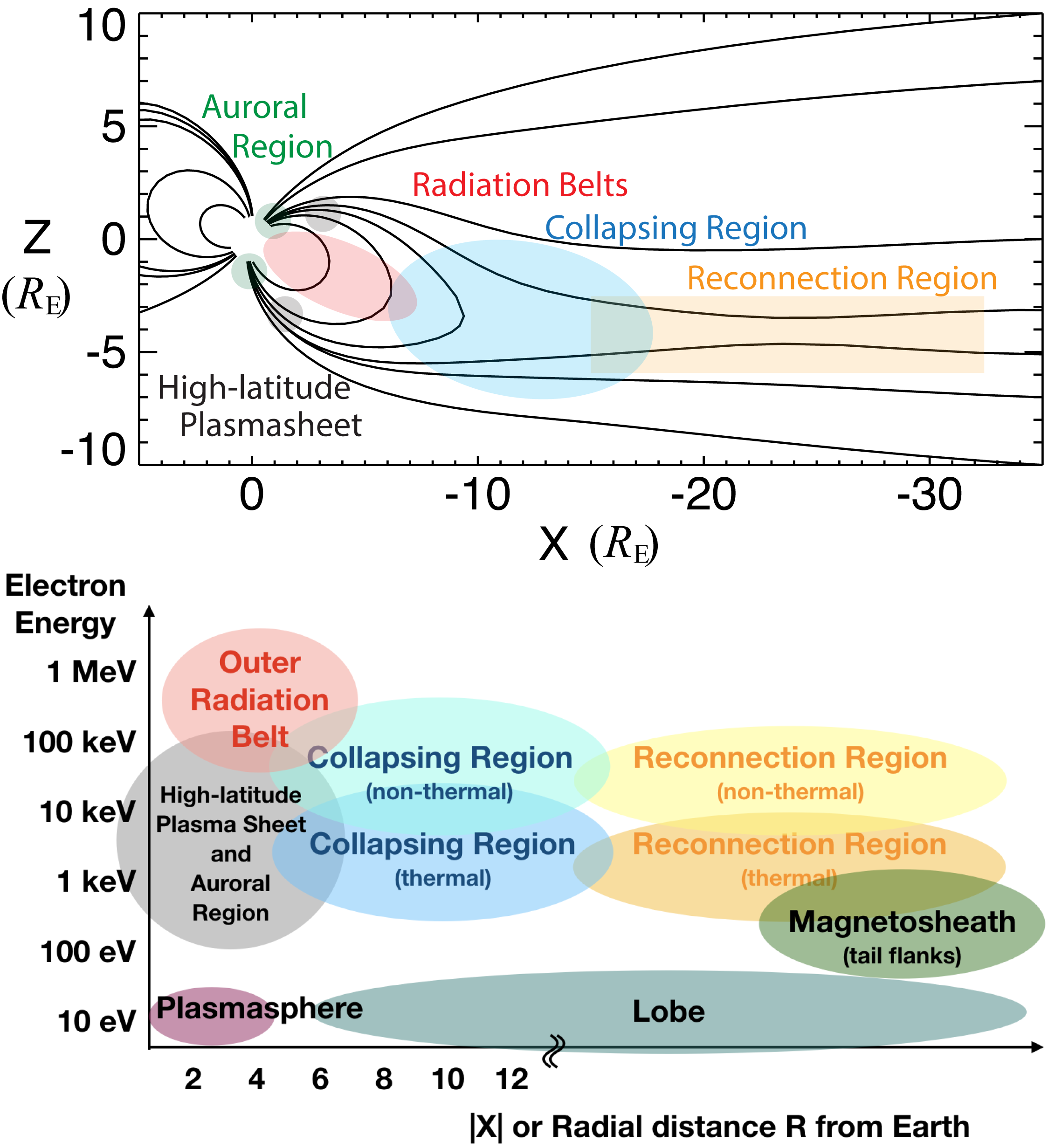}
\caption{Schematic illustrations of the regions referred to in this paper. {\it Upper panel}: The regions with respect to a semi-empirical model of the magnetic field lines (solid lines) \citep{Tsyganenko1995a}. The date of the model, i.e., 2009 February 7, is chosen arbitrarily for illustrative purpose. The Geocentric Solar Ecliptic (GSE) coordinate is used with the unit of Earth's radii R$_{\rm E}, \sim$6378 km. The GSE coordinate system is defined by $X$ toward the sun, $Z$ perpendicular to the plane of Earth's orbit around the Sun (positive North), and $Y$ completing the right-hand system (approximately toward dusk). {\it Lower panel}: Key regions in the magnetotail and the typical electron energy ranges in those regions.}
\label{fig:emt_model}
\end{figure}
%XXXXXXXXXXXXXXXXXXXXXXXXXXX

%----------------------------------------
\subsection{Overview}
\label{sec:emt_overview}
%----------------------------------------

It has been reported that Earth's magnetotail extends more than 1000 R$_{\rm E}$ (or 6000 Mm) from Earth \citep[e.g.][]{Ness1967, Scarf1987}. Some studies reported possible detections of the magnetotail at farther locations such as 3000 R$_{\rm E}$ \citep{Intriligator1979} and 15,000 R$_{\rm E}$ (or 0.63 AU) \citep{Ashford1998} from Earth. Based on previous observations at various distances from Earth, it is now generally considered that, while a reconnection X-line forms sporadically at near-Earth, typically 20 -- 30 R$_{\rm E}$ from Earth, there also exists a quasi-stationary X-line at the typical distance of 100 -- 200 R$_{\rm E}$.  

Figure \ref{fig:emt_hones} shows a schematic illustration of an evolution of the magnetotail, involving both `near-Earth neutral line' (NENL) and `distant neutral line' (DNL). This illustration, based on observations in 1970s, indicates a formation and tailward (i.e., anti-sunward) motion of a plasmoid in the distant magnetotail. Previous missions such as {\it International Sun-Earth Explorer-3} (ISEE3, later renamed as {\it International Cometary Explorer}), {\it Geotail}, {\it Wind} and {\it ARTEMIS} explored the distant tail. These missions contributed to our understanding of the basic structure and convective motions of the distant magnetotail including slow shocks and plasmoids \citep[e.g.][and references therein]{Nishida1998}. Energetic electrons are also observed in the distant tail as a detectable signature of the Hall effect of magnetic reconnection \citep[e.g.][]{Fujimoto1997a, Manapat2006}.

Figure \ref{fig:emt_model} shows a typical structure of Earth's magnetotail more focused on the near-Earth region.  Some of the regions discussed in this paper are schematically highlighted with colors. The \textbf{plasma sheet} is centered around the (magnetic equatorial) plane at which $B_x$ reverses its sign and contains a relatively hot and dense plasma. It starts to be bifurcated in the collapsing region and extends along the inner magnetospheric field lines. When a polar-orbit spacecraft crosses such a bifurcated plasma sheet at high latitude (for example, at the locations highlighted by the gray-shaded regions), it is referred to as `high-latitude plasma sheet'.  The typical parameters of the plasma sheet at the location of $X \sim$ 8 -- 11 R$_{\rm E}$ can be found in Table \ref{tab:parameters} of Appendix \ref{sec:param}. 

In the plasma sheet, a spacecraft can detect additional features that indicate a passage of the \textbf{reconnection region}. During magnetic reconnection, magnetic field lines of opposite directions `break' and `reconnect' in the diffusion region. Within the ion diffusion region (or more precisely the Hall region), a quadrupolar structure of the magnetic field in the out-of-plane direction (i.e., $Y$ direction) can be identified. A passage of the X-line can be identified by a correlated reversal of ion bulk flow $V_x$ and magnetic field north-south component $B_z$. Such features are found typically between X = -15 and -30 $R_{\rm E}$ (but may occur even farther out). Reconnection in this location occurs sporadically and is closely related to magnetotail activity including the so-called `substorms'.

On the earthward side of the near-tail reconnection site, magnetic field lines quickly collapse toward Earth's intrinsic dipole (\textbf{collapsing region}). This is typically associated with magnetic field dipolarization, manifested primarily by an increase of the north-south component of the magnetic field ($B_z$). Reconnection also results in fast plasma flows, which may be grouped into a $\sim$10-min sequence of multiple $\sim$1-min bursts (`Bursty Bulk Flows' (BBFs), \cite{Angelopoulos1992}). Such events need not be associated with substorms, but are generally correlated with activity.

Considering dipolarizations, one has to distinguish transient events, which are characterized by a short, temporary increase in $B_{\rm z}$, from more permanent increases, which are generally observed closer to Earth \citep{Nakamura2009}. The transient events propagate earthward in association with earthward flows. Their earthward propagation can often also be demonstrated by multiple, radially spaced, satellites, specifically THEMIS \citep[e.g.][]{Runov2009}.
In contrast, the permanent dipolarizations typically do not show fast flows \citep[e.g.][]{Nakamura2002a}. While they are sometimes considered separate events, a more likely explanation, strongly supported also by MHD simulations \citep[e.g.][]{Birn2011a}, is that they result from the braking and diversion of the fast flows \citep[e.g.][]{Shiokawa1997a, Nakamura2013}.  This causes an accumulation and pileup of the intensified $B_z$ field, which then may lead to bouncing \citep{Panov2010a, Birn2011a, Nakamura2013} and an expansion outward, both azimuthally \citep[e.g.]{Nagai1982a, Miyashita2009} and radially \citep[e.g.][]{Jacquey1991, Hesse1991, Miyashita2009}. 

The region, or time interval, of a transient dipolarization is sometimes denoted Dipolarizing Flux Bundle (DFB) \citep{Liu2013b} or Flux Pileup Region (FPR) \citep{Khotyaintsev2011b}. It should be noted, however, that this region is not the piled-up flux of the surrounding medium but rather the low-density material of an underpopulated flux tube \citep[e.g.][]{Pontius1990, Birn2004, Forsyth2008a, Runov2011}, generated by reconnection.

The leading edge of an earthward DFB (or jet front) typically exhibits a distinct and localized structure of increasing $B_{\rm z}$, the thickness of which is of the order of the ion inertia length $d_{\rm i}$ \citep[e.g.][]{Nakamura2002a, Runov2009, Sergeev2009}. Thus, the leading edge of the DFB has been termed `dipolarization front'. %Although the role of the entropy reduction in the propagation and earthward penetration of seems well documented, its necessity for the formation of dipolarization fronts is not. Sharp DFs are found in PIC simulations of 2D reconnection.   Also, tailward jets with a similar structure (but with reversed $B_{\rm z}$ profile) have been observed in the distant magnetotail (See, for example, a Geotail observation presented  in Figure 1 of \cite{Fujimoto1997a}), as expected from the 2D picture of magnetic reconnection. 

The electric field associated with dipolarization can penetrate even to within geosynchronous orbit. Otherwise one would not see dispersionless injection at the geosynchronous orbit \citep[e.g.][]{Gabrielse2014}. This can make a large difference in allowing the energetic particles to get onto closed drift paths in the inner magnetosphere (from which they are excluded if the dipolarization stops farther out) and thereby increase the ring current population.

In the following sub-sections, we first describe some caveats that arise from the fact that there is a variety of different methodologies, datasets, and instrumentations. Then, we describe previous studies of power-law energy spectra focused on the plasma sheet without any identification of specific features (Section \ref{sec:activequiet}), the reconnection region (Section \ref{sec:mrx}), the collapsing region (Section \ref{sec:collapse}),  radiation belts (Section \ref{sec:belts}), and the auroral region (Section \ref{sec:aurora}). We note that the classifications of previous studies into the reconnection and collapsing region is only for convenience.  In Section \ref{sec:mrx}, we primarily focus on studies of specific features in the Hall region and/or the diffusion region. In Section \ref{sec:collapse}, we primarily focus on studies of the reconnection downstream region including reconnection fast flows. These studies do not focus on the kinetic features of the Hall region or the diffusion region, but may be presenting some data that were obtained close to (or even within) the reconnection Hall region. 

%----------------------------------------
\subsection{Caveats of Comparing Energy Spectra}
%----------------------------------------

\begin{table}
\begin{minipage}{\textwidth}
\caption{Technical details of selected studies of electron power-law energy spectrum, demonstrating the variety of different methodologies, datasets, and instrumentations. This is not an exhausive list of all the studies mentioned in the text.}
\label{tab:instr}       % Give a unique label
\begin{tabular}{lllllcll}
\hline\noalign{\smallskip}
& & & Energy & & PL & &  \\
Reference
& R $(R_{\rm E})$\footnote{Location of observation indicated by the radial distance R from Earth's center}
& Rgn.\footnote{The region or phenomena observed. The following acronyms are used: PS (plasma sheet without specific features of reconnection and collapsing), RX (reconnection),  DF (dipolarization front), Inj (injection event), DFB (dipolarizing flux bundles), HL (high-latitude plasma sheet) and Arr (below/above auroral region).}     
& range, keV\footnote{The energy range (in keV) used to fit the data, as shown in the text or figures in each reference.  The ranges do not necessarily represent the coverage by the instruments. $\phi$ is the potential in the auroral acceleration region and is typically 1 -- 10 keV.}
& Mdl.\footnote{Spectral models used to measure the power-law indices. PL,  $\kappa$, and FT denote power-law, kappa, and flattop distributions, respectively. Note also that $\delta = \kappa$. }
& Index \footnote{Power-law index $\delta$. We show the largest range of values, taken from each study, and it does not necessarily indicate the typical values.}   
& Proj.\footnote{Project (or spacecraft mission). The following acronyms and abbreviations are used: GTL (Geotail), CL (Cluster), THM (THEMIS) and SR (sounding rocket).}  
& Type\footnote{Type of methodology. The following abbreviations are used: Cs (specific values from up to a few cases), St (values obtained by a statistical study of many events), TV (time variations of certain time periods), and SE (averaged values from superposed epoch analyses).}       \\
\noalign{\smallskip}\hline\noalign{\smallskip}
PLASMA SHEET  & & &  & & & \\
\cite{WuMingyu2015}\footnote{Their Event 2; Event 1 had the energy range of 8 -- 30 keV.}
						& 11 -- 22& PS & 2 -- 30 & PL & 3.5 -- 5 & THM & TV \\
\cite{Imada2011}       & 17 -- 96& RX &0.07 -- ($>$38)& $\kappa$ & 3 -- 4.5 & GTL & St \\
\cite{Bame1967}\footnote{The electrostatic analyzer covered the lower energy range up to $\sim$ 20 keV. An additional data point was added from the integrated flux in the $>$45 keV range. We converted their power-law index based on integrated flux (2.7 -- 3.5) to that of differential flux, $\delta$. \cite{Montgomery1965} also used integrated flux measured by the Vela satellite and obtained similar power-law indices.}        
& 16 -- 21& PS & 0.4 -- ($>$45)& PL & 3.7 -- 4.5 & Vela & Cs. \\			\cite{Oka2015}\footnote{Although unpublished, they performed additional analysis using data from the solid state detector and found $\kappa$ = 4 -- 5 at and around the EDR in the 30 -- 300 keV range.}         & 30      & RX & 0.03 -- 28 & FT & 4 -- 5 & THM & Cs \\
\cite{WuMingyu2015}\footnote{Their main event on 2008 Feb. 26}    
						& 17 -- 21& RX & 30 -- 200 & PL & 3.3 -- 5.2 & THM & TV \\
\cite{Imada2007}\footnote{See also \cite{Chen2009, Huang2012}}          
                       & 16      & RX & 40 -- 110 & PL & 4 -- 5.7   & CL & Cs \\
\cite{Nakamura2013}    & 13 -- 15& DF & 40 -- 200 & PL & 3.5 -- 4.6 & CL & Cs \\
\cite{Zhou2016}        & unclear & RX & 40 -- 244 & PL & 1 -- 6     & CL &    \\
\cite{Asnes2008}\footnote{See also \cite{BurindesRoziers2009}}
						& $>$ 15.5& PS & 40 -- 400 & PL & 0 -- 10    & CL & TV \\
\cite{Gabrielse2014}\footnote{Their $\kappa$ values were defined in differential energy flux. The values are thus converted to our $\delta$ here.}
					   & 6 -- 30 & Inj & 41 -- 140 & PL & 3.4 -- 4.1 & THM & SE\\
\cite{Runov2015}\footnote{Their $\kappa$ values appear to be defined in differential energy flux and are converted to our $\delta$ here.}       
						& 6 -- 25 & DFB & 0.1 -- 200 & $\kappa$ & 3.4 -- 5.5 & THM & SE \\
\cite{Oieroset2002}    & 60      & RX & 0.02 -- 300 & PL & 3.8 -- 4.7 & Wind & Cs \\
S \& A (2015)\footnote{\cite{Stepanova2015}}
                       & 7 -- 30 & PS & 0.05 -- 500 & $\kappa$ & 1.7 -- 5.5 & THM & St \\
\cite{Christon1991a}    & 12 -- 23& PS & 0.03 -- 1000& $\kappa$ & 3 -- 16 & ISEE1 & St \\
& & &  & & & \\
AURORAL REGION & & &  & & & \\
\cite{Kletzing2003}\footnote{They used the pitch angle range of 0 -- 30$^{\rm o}$ because they interpret that these are the electrons which map to the auroral acceleration region.}   & 5 -- 6 & HL & 0.01 -- 4 & $\kappa$ & 2 -- 10 & Polar & St \\
O \& J (1998)\footnote{\cite{Olsson1998}; While they had data points above 20 eV, they used data points with energies above the potential $\phi$.}      
						& 0.3 & Arr & $\phi$ -- 100 & $\kappa$ & 2 -- 9 & Freja & St \\
M \& A (2014)\footnote{\cite{McIntosh2014}} & 800 km & Arr & 0.03 -- 30 & $\kappa$ & 3 - $\infty$ & DMSP & St \\
\cite{Kaeppler2014}\footnote{See also \cite{Ogasawara2017}} & 130 km & Arr & $\phi$ -- 20 & $\kappa$ & 2 -- 11 & SR & Cs \\
%\cite{Ogasawara2017} & $>$100\footnote{100 -- 350 km} km& Arr & 10 -- 40 & $\kappa$ & 3 -- 30 & SR & Cs \\
%\cite{BurindesRoziers2009} & 4 -- 19 & PS & 40 -- 400 & PL & 1.5 -- 7.5 & CL \\
\noalign{\smallskip}\hline
\end{tabular}
\end{minipage}
\end{table}

There is a wide variety of different instrumentations, dataset, and methodologies used in observational studies of the magnetotail, as demonstrated in Table \ref{tab:instr}.  For example, different energy ranges and different spectral models are used to derive a power-law index, leading to various concerns. When a study derives a power-law index from a very narrow energy range (say, 40 -- 100 keV) using a power-law, there is a concern that the data could be fitted with a Maxwellian with a very high temperature. When a study derives a power-law index from a wide energy range (say, 0.04 -- 400 keV) using a kappa distribution, there is a concern how well the spectrum was fitted, because an observed spectrum usually show complexity originating from, for example, a cold thermal component in the lower energy range and a higher-energy (cosmic-ray-like) background. Only some (and not all) studies use a combination of different spectral models. 

Instrumentations can be different in different studies. Typically, up to a few tens of keV are covered by electrostatic analyzers. The higher energy ranges are covered by solid state detectors. These data are analyzed separately in some studies, while these data are combined to make one spectrum for fitting in other studies. 

Also, methodologies can be different. Some studies take a few samples of measurements at the highest time resolution (say, 3s) for fitting and so their results may not represent average properties. Some other studies performs a superposed epoch analysis, smoothing out the fine structures in the spectrum. 

Ideally, a coordinated and systematic study should be carried out with the same instrumentation and methodology, but it is beyond the scope of this paper. By reviewing previous reports of power laws, we hope to obtain some hint for a better understanding of particle acceleration and for possible future studies. The reported values of the power-law index in the magnetotail are summarized in two separate figures, Figures \ref{fig:emt_index_a} and \ref{fig:emt_index_b}. Some technical details of selected studies are compiled in Table \ref{tab:instr}. 

%----------------------------------------
\subsection{Plasma Sheet}
%----------------------------------------

\subsubsection{Active vs Quiet Times}
\label{sec:activequiet}

%XXXXXXXXXXXXXXXXXXXXXXXXXXX
\begin{figure}[t]
\includegraphics[width=0.9\textwidth]{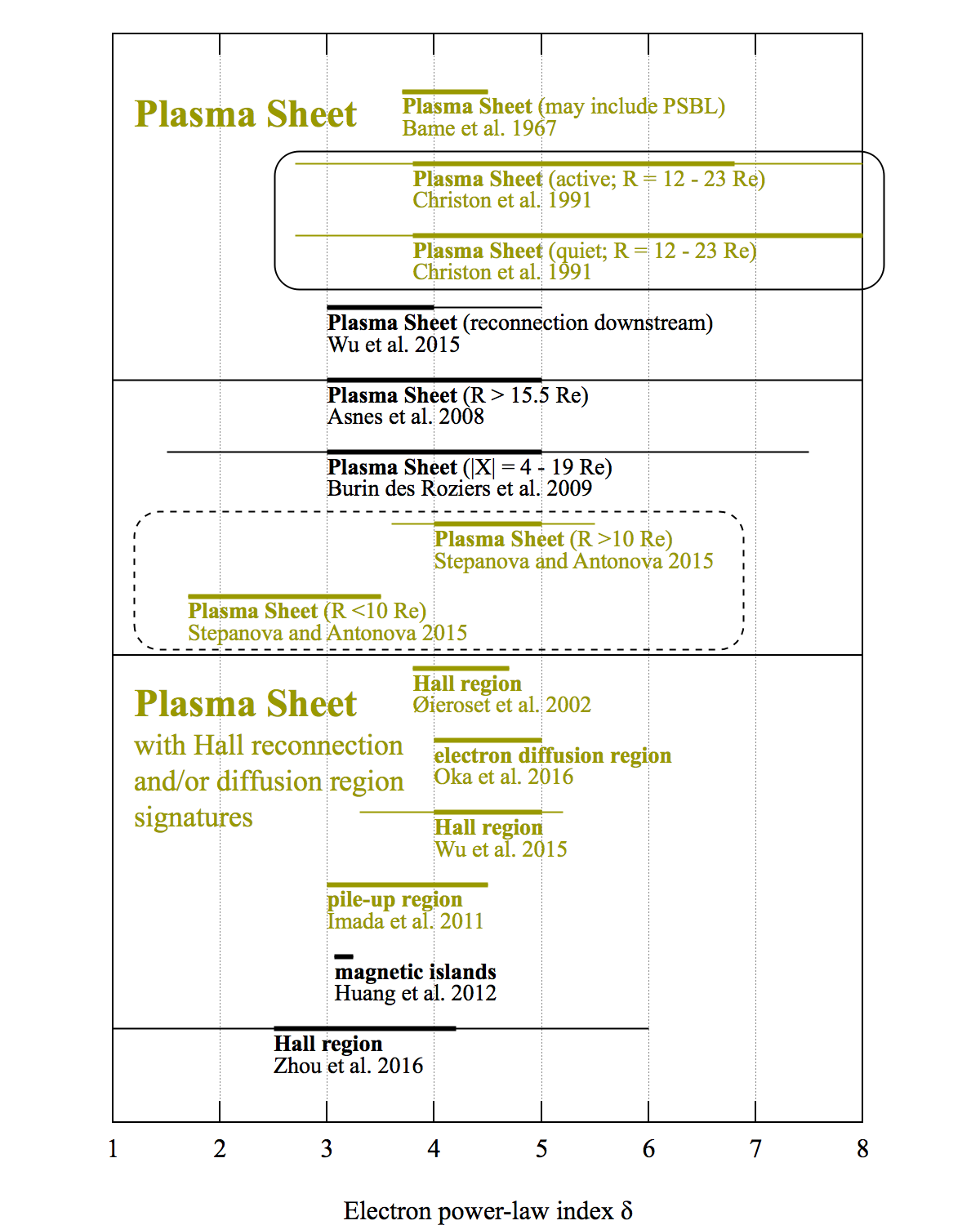}
\caption{An overview of electron power-law index $\delta$ reported by different studies of electron energy spectra in Earth's plasma sheet (See also Fig. \ref{fig:emt_index_b}). For statistical studies, typical values are shown by thick lines. For non-statistical studies, only  thick lines are used to display reported values. Studies (spectral analysis) over an energy range larger than two-orders of magnitudes are shown in ocher color. The other items with narrower energy ranges are shown in black. Note that different studies may be using not only different energy ranges but also different methodologies and instrumentations. The round-corner rectangles highlight the studies that compare power-law indices in different conditions but with the same methodology and instrumentation.}
\label{fig:emt_index_a}
\end{figure}
%XXXXXXXXXXXXXXXXXXXXXXXXXXX

Pioneering observations revealed the presence of energetic electrons up to several MeV in the magnetotail \citep[e.g.][]{Frank1965a, Anderson1965, Montgomery1965, Bame1966,Bame1967}. It was reported that the energetic electrons appear as bursts or `isolated patches' in the magnetotail plasma sheet \citep[e.g.][]{Frank1965a, Anderson1965} and that they form a power-law energy spectrum \citep{Montgomery1965, Bame1966, Bame1967}. In some cases, the power-law index in the magnetotail plasma sheet between $X$ = -15 R$_{\rm E}$ and -20 R$_{\rm E}$ were reported to be $\delta$ = 3.7 -- 4.5 \citep{Bame1967}, where $X$ is in the geocentric solar magnetospheric (GSM) coordinate and $|X|$ is roughly the radial distance from Earth's center along the sun-Earth line. Not only energetic electrons but also energetic ions were reported in later years \citep[e.g.][]{Fan1975, Hones1976, Sarris1976a, Sarris1976b, Baker1979}. In particular, \cite{Baker1979} reported that protons form a power-law energy spectrum and that the power-law index $\delta$ was 5.1 -- 6.5 at geosynchronous orbit. 

\cite{Anderson1965} already suggested that the energetic electron fluxes are sensitive to geomagnetic activity as inferred by the $K_p$ index\footnote{The activity level of Earth's magnetosphere has been inferred from different indices derived from different sets of magnetic field measurements on the ground. There are many geomagnetic observatories (stations) across the globe. Some of the widely used geomagnetic indices are the $AE$ (Auroral Electrojet) index, $K_p$ index  and $Dst$ index, which are derived from geomagnetic measurements in the high-latitudes, mid-latitudes, and low-altitudes, respectively.  The $AE$ index better represents substorms and associated auroral activities. The $Dst$ index better represents storm activities.}. Later, a more direct association between energetic electrons and reconnection-related features such as magnetic neutral line and plasmoids (or `fireballs') was identified \citep{Terasawa1976a, Baker1976}, indicating that magnetic reconnection plays an important role in producing  energetic electrons (see also a review by, for example, \cite{Hones1979}). Similar associations were also discussed for energetic ions \citep[e.g.][]{Mobius1983}.

A puzzle is that energetic particles exist in the magnetotail even during geomagnetically quiet periods \citep[e.g.][]{Anderson1965}. A  systematic analysis using International Sun-Earth Space Explorer (ISEE) data was carried out by \citet{Christon1988, Christon1989a, Christon1991a}. \cite{Christon1989a} reported that, during quiet times AE $<$ 100, electron energy spectra exhibit a kappa distribution (i.e., a thermal component extending smoothly to a power-law component in the higher energy range).  During active times (AE $>$ 100), the spectra become more complex and cannot be represented, in general, by a single functional form although the power-law tail persists. The spectra often show different forms of roll-off and/or excess fluxes in lower energy ranges when compared to a simple Maxwellian or Kappa distribution \citep{Christon1991a}. More recently, \cite{Asnes2008} reported that the power-law index is independent of geomagnetic activity ($K_p$), while there is a significant local time dependence, with harder spectra observed at dawn compared to the dusk side. Also, \cite{DesRoziers2009} studied the energetic electron fluxes ($>$ 40 keV) instead of the power-law index. They reported that the energetic electron fluxes inside the plasma sheet can still undergo rapid variations when the solar wind is calm and geomagnetic activity is low. 

It should be noted that  the spatial variation of the power-law index also remains unclear.  \cite{WuMingyu2015}  reported that the electron power-law index did not fluctuate very much ($\delta \sim$ 3.5 -- 5; converted from their values in differential energy flux) when magnetic reconnection and associated phenomena were observed by multiple probes of the Time History of Events and Macroscale Interactions during Substorms (THEMIS) mission \citep{Angelopoulos2008}, distributed over large distances in the magnetotail (in the $X$-direction). \cite{Stepanova2015} studied 5 cases of multi-probe THEMIS observations of the magnetotail.  They found that the power-law index $\delta$ is $\sim$4 or larger (softer) at X $\lesssim$ -10 R$_{\rm E}$ but the power-law becomes harder at the inner locations (-5 $<$ X $<$ -10 R$_{\rm E}$).

%----------------------------------------
\subsubsection{Reconnection Region}
%----------------------------------------
\label{sec:mrx}

During an encounter with the plasma sheet, a spacecraft often detects signatures of the reconnection diffusion region at the ion-scale (or more precisely the Hall region).  Some of the studies of such reconnection signatures argue that certain features such as magnetic island and magnetic field pile-up, embedded in the Hall region, could be important for electron acceleration. 

\cite{Oieroset2002} reported the first observation of energetic particles inside the diffusion region (the Hall region), in a fortuitous encounter by the Wind spacecraft in Earth's distant magnetotail (see also \cite{Egedal2005} for interpretations) . The diffusion region was identified based on (i) an uninterrupted transition of flow reversal without leaving the reconnection layer, (ii) quadrupolar Hall magnetic field signatures, (iii) an electron beam with direction consistent with being the Hall current carrier. They demonstrated that the fluxes become more intense and the power law becomes harder with decreasing distance from the diffusion region center. Also, they fitted the lower and higher part of the observed energy spectrum by a Maxwellian and power-law distributions, respectively.  From their analysis, the power-law index ($\delta$) inside the diffusion region was 3.8 -- 4.7. 

An encounter with the electron diffusion region (EDR) is more challenging as its size is much smaller than the diffusion region at the ion scale (the Hall region). Nevertheless, before the Magnetospheric MultiScale (MMS) mission  achieved electron-scale measurements \citep[e.g.][]{Burch2016}, there were observational reports of EDR detection based on (1) decoupling of ion and electron bulk flow velocities \citep{Nagai2011, Nagai2013}, (2) a higher-order scalar measure derived from particle data \citep{Scudder2012, Zenitani2012a, Tang2013} and (3) non-gyrotropic distribution of electron velocities \citep{Oka2016}. \cite{Oka2016} reported more than an order of magnitude energization across the EDR, from immediate upstream to immediate downstream. In their analysis, the power-law indices ($\delta$) at the EDR and its immediate downstream region were 4.0 -- 5.0.

Based on full-particle simulations and Geotail \citep{Nishida1994} observations, \citet{Hoshino2001a} proposed that electrons are first accelerated at the $X$-line where electrons can be demagnetized and then further accelerated through gradient B and curvature drift at the `piled-up' magnetic field lines  in the immediate downstream region. This idea spurred  interpretations of data with a similar scenario \citep{Imada2005, Imada2007, Asano2008, WuMingyu2015}. While a statistical picture of the locations of energetic electrons relative to the X-line was developed by \citet{Imada2005}, \citet{Imada2007} presented a detailed case study using four-spacecraft mission Cluster. They found that the energy spectrum of energetic electrons becomes harder downstream, in the reconnection outflow region where magnetic field intensity increases. In their analysis, the power-law indices ($\delta$) at the magnetic pile-up region were 4.0 -- 5.7. Furthermore, \citet{Imada2011} statistically studied the relationship between energetic electrons and reconnection characteristics, using 10 reconnection events observed by Geotail. They argued that the electrons are `efficiently' accelerated in a thin current sheet during fast reconnection events, and the power-law indices ($\delta$) in and around the reconnection region were 3.0 -- 4.5. A caveat is that they used an integrated flux in the $\gtrsim$38 keV range, combined with an {\it a priori} assumption that electrons exhibit the kappa distribution. 

A similar energization process by gradient and curvature drift has been considered in the collapsing region (Section \ref{sec:collapse}). A notable aspect of the work \cite{Hoshino2001a} is their consideration of  a chaotic process that occurs when the particle gyro-radius become comparable to the curvature of the magnetic field.

Fluxropes have also received considerable attention in various contexts of electron acceleration \citep[see a review by, for example,][]{Birn2012}. Many of these studies consider electron acceleration in magnetic islands, i.e., a 2D picture of fluxropes.  However, the extent to which such  2D pictures can explain observations remains unclear. A recent study of the dayside magnetopause reported the detection of a magnetic fluxrope flanked by two active X-lines, producing colliding plasma jets near the center of the flux rope \citep{Øieroset2011}. While the authors report detection of non-thermal electrons within the flux rope core,  they also described departures from the 2D pictures of electron acceleration. Recent theoretical and simulation studies demonstrated an importance of  3D effects in electron acceleration \citep[e.g.][]{Dahlin2015} and it remains as an interesting topic of research.

In the magnetotail, energetic electrons have been observed within small-scale fluxropes with spatial scales of a few times the ion inertia length $d_{\rm i}$ \citep[e.g.][]{Chen2007,Chen2009, Retino2008, Wang2010b, Huang2012a}. \citet{Chen2007} showed that energetic electron fluxes peak at sites of compressed density within fluxropes. Based on a further analysis, it was reported that the energetic electrons exhibited a power-law energy spectrum inside the fluxropes and that the power-law index $\Gamma$ ranged from 6 to 7.3  (or $\delta$ from 5 to 6.3) \citep{Chen2009}. \citet{Retino2008} also found an enhanced flux of energetic electrons within a small-scale fluxrope during a crossing of reconnecting current sheet. They argued that, while the hardest spectrum was observed for field-aligned electrons at magnetic separatrices, the highest flux was observed within the small-scale fluxrope for perpendicular electrons. After these observational studies, results from 2D PIC simulations have been reported in the context of electron acceleration by the secondary magnetic islands  \citep[e.g.][]{Oka2010b, Wang2017}.

\subsubsection{Reconnection Downstream and Collapsing Region}
\label{sec:collapse}

%XXXXXXXXXXXXXXXXXXXXXXXXXXX
\begin{figure}[t]
\includegraphics[width=\textwidth]{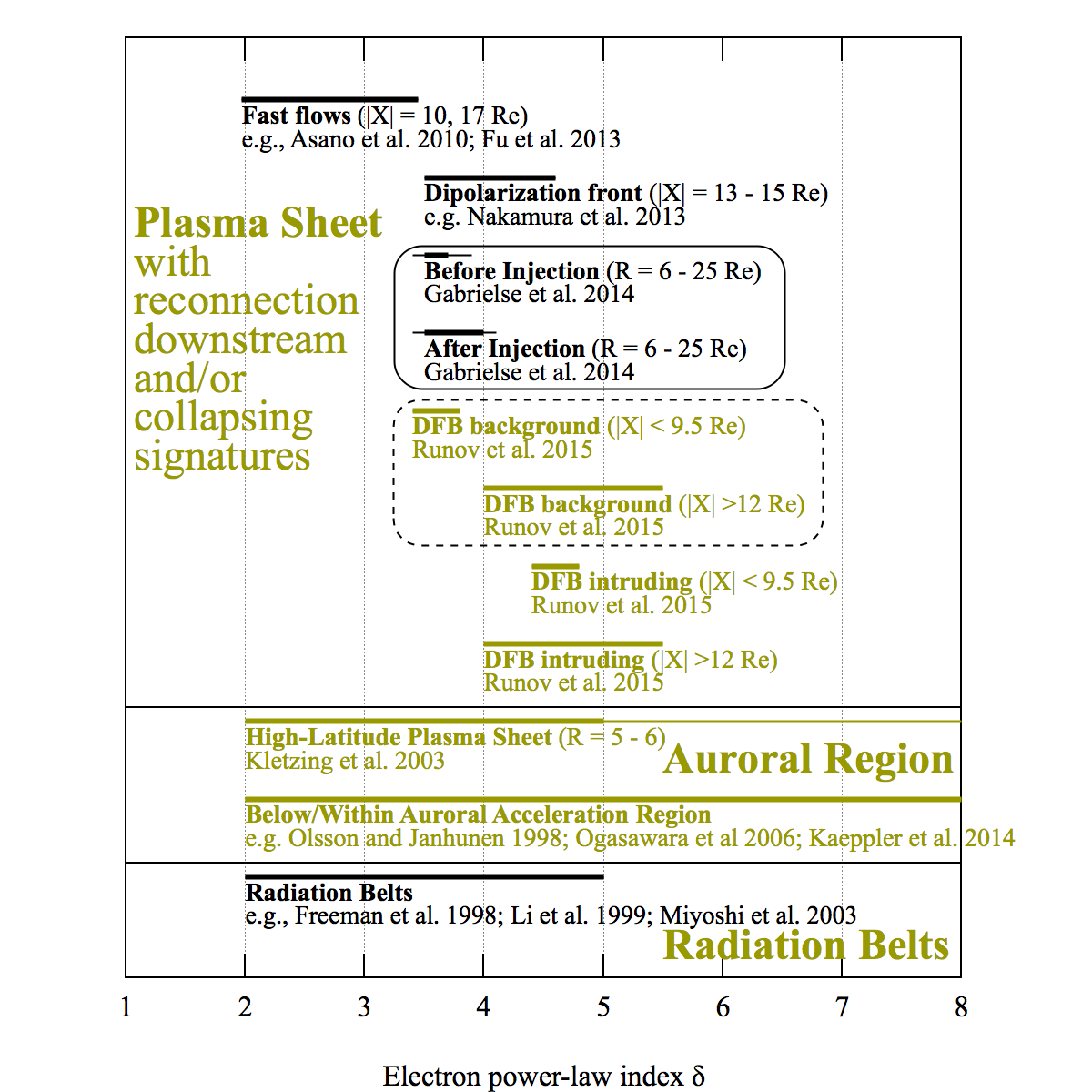}
\caption{An overview of electron power-law index $\delta$ reported by different studies of electron energy spectra in Earth's plasma sheet and auroral region (See also Fig. \ref{fig:emt_index_a}). The category `collapsing region' is for studies of specific features in the reconnection region downstream.  For statistical studies, typical values are shown by thick lines. For non-statistical studies, only  thick lines are used to display reported values. Studies (spectral analysis) over an energy range larger than two-orders of magnitudes are shown in ocher color. The other items with narrower energy ranges are shown in black. Note that different studies may be using not only different energy ranges but also different methodologies and instrumentations. The round-corner rectangles highlight the studies that compare power-law indices in different conditions but with the same methodology and instrumentation. The $\kappa$ values in \cite{Gabrielse2014} and \cite{Runov2015} were defined in differential energy flux and are converted to the $\delta$ values as in differential particle flux in this figure. }
\label{fig:emt_index_b}
\end{figure}
%XXXXXXXXXXXXXXXXXXXXXXXXXXX

\begin{figure}
\includegraphics[width=0.7\textwidth]{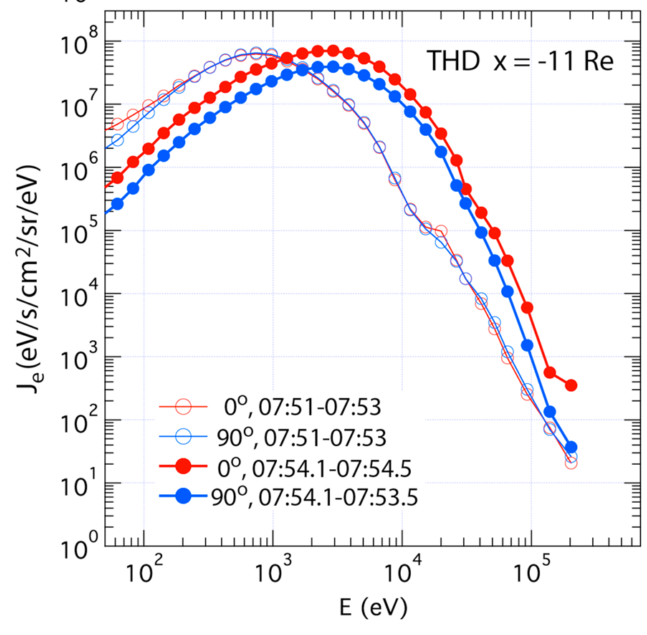}
\caption{Electron energy spectrum before (open circles) and after (filled circles) a jet arrival in the collapsing region, adapted from \cite{Birn2014a}. This observation was made by  THEMIS-D (P3) on 2009 February 27.  The spectrum is taken by the Electro-Static Analyzer (ESA) and  Solid State Telescope (SST) below and above $\sim$ 28 keV, respectively. In general, the pre-existing plasma is cold and dense whereas intruding plasma is hot and tenuous. }
\label{fig:emt_spectra}
\end{figure}

During magnetospheric disturbances, enhancements of energetic electrons are observed at and around the geosynchronous orbit (i.e., the geocentric distance of 6.6R$_{\rm E}$) \citep[e.g.][]{Lezniak1970a, Swanson1978, Erickson1979, Birn1997b}. How energetic electrons can be `injected' into the inner magnetosphere and contribute toward the radiation-belt formation has long been a subject of debate \citep[e.g.][and references therein]{Birn2012, Lui2012, Lui2013}.  Energetic electrons are also observed during an inward motion of dipolarization fronts (jet fronts) and the associated features such as plasma flows and magnetic fluxes \citep[e.g.][]{Deng2010a, Asano2010, Fu2011, Fu2013a, Runov2011b, Runov2015,  Nakamura2013, Duan2014, Gabrielse2014, Gabrielse2016, Turner2016, Liu2016}. Example energy spectra from the collapsing region are shown in Fig. \ref{fig:emt_spectra}.  A general argument is that electrons are accelerated while bouncing along the collapsing magnetic field lines \citep[e.g.][]{Birn2012}. Based on analyses of pitch angle distributions, \cite{Fu2011} argued that Fermi acceleration dominates inside a decaying flux pileup region (FPR), while betatron acceleration  dominates inside a growing FPR. Here, Fermi acceleration refers to Fermi's Type B interaction with an evolving magnetic field line (or the `slingshot' effect; see Appendix \ref{sec:fermitheory}) and is not necessarily a stochastic acceleration. Recently, more detailed picture of the behaviors of energetic electrons have been developed through statistical analyses \citep[e.g.][]{Duan2014, Gabrielse2014, Runov2015}. 

%XXXXXXXXXXXXXXXXXXXXXXXXXXX
%XXXXXXXXXXXXXXXXXXXXXXXXXXX

As for power laws,  \cite{Nakamura2013} reported $\delta$ = 4 -- 6 in a plasma sheet with multiple dipolarization fronts (jet fronts). In a statistical study focused on “dispersionless injection”, it was reported that the average electron power-law index of superposed spectra in the background population ($\delta$ = 3.4 -- 3.9; converted from their values in differential energy flux) was already very similar to that of the injected electrons ($\delta$ = 3.4 -- 4.1; converted from their values in differential energy flux) \citep[e.g.][]{Gabrielse2014}. \cite{Runov2015} investigated particle energy spectra by taking an average of many cases of dipolarizing flux bundle (DFB) events in both background populations (i.e., before arrival of dipolarization fronts, also known as jet fronts) and intruding populations (i.e., inside the DFBs or bulk flow plasmas). Then they divided the magnetotail into four regions, i.e., $r<9.5R_{\rm E}$, $9.5<r<12 R_{\rm E}$, $12<r<15.5 R_{\rm E}$ and $15.5<r<25 R_{\rm E}$ where $r$ is the geocentric radial distance, and measured the power-law index of the average spectra. They found that the power-law index $\delta$ is $\sim$5 in the outermost region and the spectra become harder with the decreasing radial distance (Fig. \ref{fig:emt_index_b}).  

For ions, \cite{Gabrielse2014} and \cite{Runov2015} found much softer spectra with $\delta \sim$ 7 -- 8 and 5 -- 6, respectively, in the magnetotail. In the inner magnetosphere and around geosynchronous orbit, flux increases are often limited in energy range and/or do not exhibit a clear power-law \citep{Baker1981, Birn1997b, Birn2012}. 

The observations of energetic particles in the collapsing region have been complemented by a large number of theoretical studies, primarily on the basis of test particle simulations in assumed field configurations (e.g., Li et al., 1998; Zaharia et al., 2000; Gabrielse et al., 2012, 2016) or using the dynamic fields of MHD simulations of magnetotail dynamics and reconnection (e.g., Birn et al., 1994, 1997b, 1998, 2004, 2013; Ashour-Abdalla et al., 2011; Pan et al., 2014a,b). The combination, and sometimes direct comparison, of observations and simulations have provided a consistent picture of the particle energization in the near-Earth  magnetotail outside the immediate vicinity of the reconnection site (for a summary, see, also, Birn et al., 2012). The most important element is the motional ($v \times B$) electric field associated with earthward flow bursts.

Charged particles can enter the acceleration region of dipolarizing flux bundles (DFBs) (i.e., the region of intensified motional electric field associated with collapsing) in two ways, either from low latitudes via cross-tail drift or from higher latitudes when the field line on which the particle resides becomes reconnected and the particle trapped in the section earthward of the reconnection site (Birn et al., 2015). 

Unlike ions, electrons are more adiabatic, except for the vicinity of the X-line, and exhibit either many gyrations (at pitch angles near 90$^{\rm o}$) or many bounces (at low pitch angles) between mirror points closer to Earth. The energy gain thus occurs in multiple steps, either by betatron or by first-order Fermi Type B acceleration (Northrop, 1963). At the highest energies (hundreds of keV), however, they also drift across the acceleration region in a simple fashion. This limits the maximum energy gain by the integrated cross-tail voltage difference, associated with the electric field of the DFB, similarly for ions and electrons. Taking the speed of a DFB at 1000 km/s, a magnetic field magnitude of 20 nT, and a typical width of, $\sim$20,000 km, this yields a maximum energy gain of 400 keV, close to typical values of injections observed at geosynchronous distance. 

A caveat is that some of previous simulations assumed that particles already exhibit a power-law energy spectrum at the boundary of the simulation box. While this is a reasonable assumption based on the fact that power-law tails can exist even during quiet times (See Section \ref{sec:activequiet}),  it remains unclear when or in what conditions a power-law forms. It also remains unclear if or how the power-law spectra can be modulated by the energization process in the collapsing region.

%----------------------------------------
\subsection{Auroral Region}
%----------------------------------------
\label{sec:aurora}
The electron distributions described above comprise the source population for auroral primaries \citep{Paschmann2003}. These electrons, which stimulate auroral emission, are driven from the magnetospheric source along converging geomagnetic field lines toward the ionosphere with a flux sufficient to satisfy the current  or voltage requirements of the magnetospheric ‘generator’  \citep{Lysak1990}. Consequently, the recognition that the hot tenuous source electron distributions have extended power-law-like supra-thermal tails, often characterized as kappa distributions, is a potentially important development in advancing understanding of auroral arc formation \citep{Pierrard1996, Dors1999, Janhunen1998a}. In the following we briefly review pertinent observations and theoretical results, and in the light of these contributions indicate where further advances could be made.

\begin{figure}
\includegraphics[width=1.\textwidth]{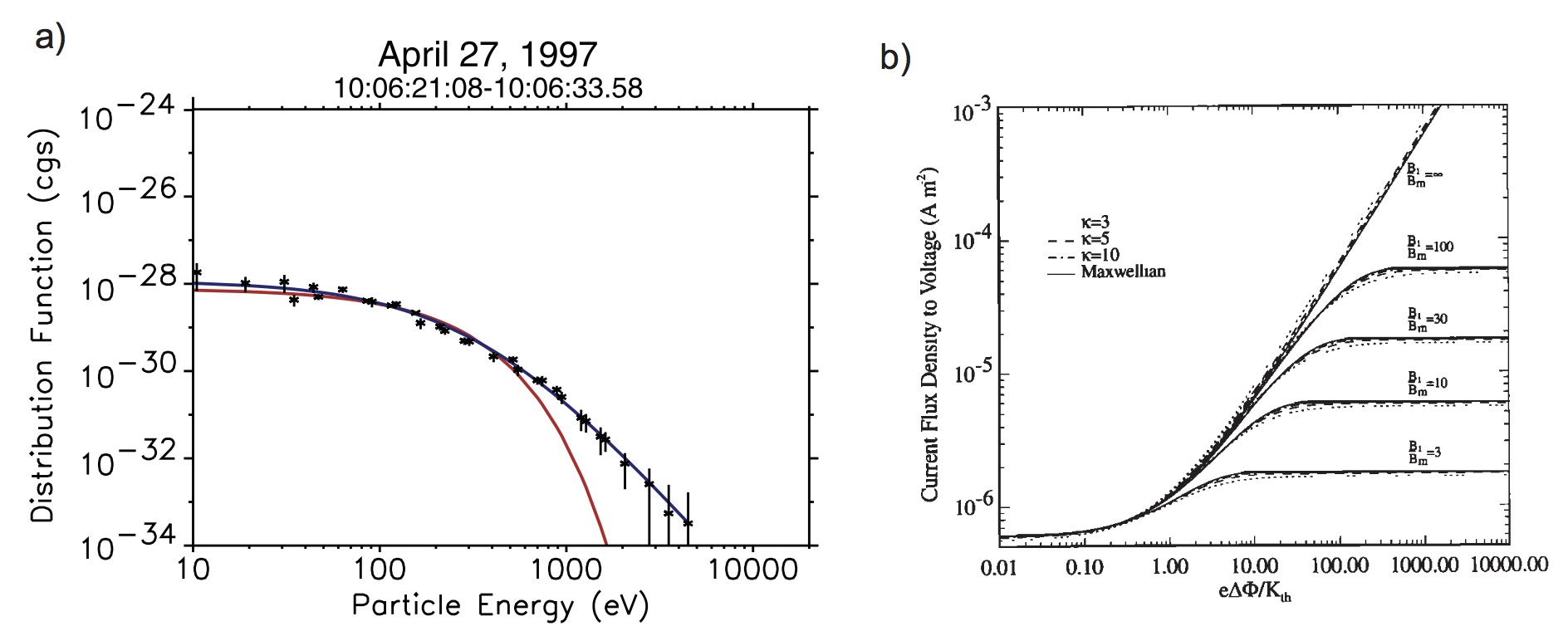}
\caption{Electron supra-thermal tails in auroral arcs. (a) Example of Maxwellian (red curve) and kappa (blue curve) distribution fits to electron spectra in an auroral source region observed from the Polar spacecraft in the high latitude plasma sheet. The kappa fit is within the 99\% confidence interval, but the Maxwellian is not. The fit finds = 3.9 $\pm$ 0.6 or $f \propto E^{-4.9}$ for $E > $1 keV (adapted from Figure 6 of \cite{Kletzing2003}).  (b) Current density  $J$ versus dimensionless acceleration potential $\phi$  normalized by electron temperature ($K_{\rm th}$) for a generalized kappa current-voltage relationship. Line-styles show a variety of $\kappa$ indices for different magnetic field ratios between the source region ($B_M$) and the ionosphere ($B_I$). The Maxwellian result is shown by the black line.  Density is 1 cm$^{\rm-3}$ and $K_{\rm th}$ = 500 eV (adapted from Figure 2 of \cite{Dors1999}).}
\label{fig:au_kletzing}
\end{figure}

The recognition of enhanced supra-thermal tails as a prevailing feature of the source distribution for auroral primaries was implicit from the analysis of ISEE particle measurements reported by  \cite{Christon1988, Christon1989a, Christon1991a}. These authors demonstrated that the observed electron spectra over an energy extending from $\sim$100 eV to hundreds of keV is well described as a kappa distribution with index $\kappa$ (= $\delta$) = 4 -- 6. At energies above the average energy ($E_0$) of the electron distribution these fits indicate supra-thermal tails having power-law form with $\kappa$ = 5 -- 7 (i.e. for $E \gg E_0, f \sim E^{-(\kappa+1)}$).  Observed distributions such as these in the plasma sheet were explicitly connected to the auroral acceleration process \citep{Kletzing2003}. An example electron spectrum as reported by these authors from the Polar spacecraft in the high latitude nightside plasma sheet is shown in Figure \ref{fig:au_kletzing}a \citep[][Figure 6]{Kletzing2003}. This plot demonstrates a marked deviation from Maxwellian form well described by a kappa distribution with $\kappa \sim$ 3.9 providing a power-law spectrum  above 1 keV varying as $E^{-4.9}$.  These measurements in the auroral source regions have demonstrated that a Maxwellian is a perhaps a poor starting point for building a kinetic model for the auroral acceleration process.

%XXXXXXXXXXXXXXXXXXXXXXXXXXX
%XXXXXXXXXXXXXXXXXXXXXXXXXXX

The realization that auroral primaries have distinct power-law-like supra-thermal tails has also become apparent from measurements performed at altitudes below the auroral acceleration region.  \cite{Olsson1998} used observations from the Freja spacecraft to demonstrate that a kappa distribution generally provided a better fit to the observed distribution irrespective of activity level with $\kappa$ = 4 -- 7 for field-aligned primaries. \cite{McIntosh2014} performed a statistical study of DMSP spacecraft observations to show that $\sim$30\% of measured precipitating electron spectra require a kappa description. \cite{Ogasawara2006a, Ogasawara2017} reported $E$-region observations from two separate rocket campaigns over 90 -- 140 km and over 90 -- 336 km respectively showing $\kappa$ = 4 -- 9 and $\kappa$ = 3 -- 30 through discrete auroral arcs.  Similarly, \cite{Kaeppler2014} from the ACES sounding rocket reported $\kappa$ indices increasing from 2 to 11 as the rocket travelled from the equatorward to poleward side of an auroral arc. 

These observations have been supported by the development of theoretical models which include the contribution of the supra-thermal tail in a time independent relationship between the field-aligned current ($J$) and accelerating potential ($\phi$) for quasi-static auroral arcs \citep{Pierrard1996, Dors1999, Janhunen1998a}.   In general these models are extensions of those derived earlier for Maxwellians \citep{Knight1973, Lemaire1974, Lundin1978, Fridman1980} which seek to determine the effective field-line conductivity ($K$) through the auroral acceleration region from a consideration of the source electron distribution and the converging geometry of the geomagnetic field. Over much of the observed energy range of auroral primaries ($\sim$100 eV - 10s keV) these models predict a simple Ohmic-like `current-voltage' relationship where $K$  is independent of $J$ and $\phi$ expressed as
\begin{equation}
J = K\phi
\label{eq:au_j}
\end{equation}
Both $J$ and $\phi$ can be measured by spacecraft to derive the observed value of $K$ \citep[e.g.][]{Lyons1979, Sakanoi1995a, Elphic1998}. Intuitively, the energy dependence of the electron spectrum enters the evaluation of $K$ because the energy gain required to shift electrons into the loss cone, where they can contribute to the net current, increases with energy. Consequently, $K$  may be expected to be reduced as the spectrum becomes harder. On the other hand more energetic electrons can carry a larger current for a lower density.  These competing influences result in a non-obvious variation of $K$ with changes in spectral index (or $\kappa$), density, and temperature.  Comparisons between results expected for a Maxwellian versus a kappa distribution indicate that the linear relationship suggested by Eq. \ref{eq:au_j} remains valid up to larger values of $\phi$ for a kappa distribution before saturating. This allows greater energy deposition. It has also been shown that for the same density and temperature of the plasma sheet source the expected total energization or potential drop along the geomagnetic field to the ionosphere increases by $>$5\% for \cite{Dors1999}. This rather modest change is apparent in Figure \ref{fig:au_kletzing}b where the current-voltage relation is presented for a range of $\kappa$ values. However, by combining the current voltage-relation with current continuity through the ionosphere \cite{Dors1999} demonstrated that kappa distributions with $\kappa$ = 5 provide up to double the precipitating energy flux and drive precipitation over a broader region than the equivalent Maxwellian. \cite{Olsson1998} performed an observational test of the effect of a supra-thermal tail by using Freja spacecraft observations in a generalized kappa model derived for MHD simulations \citep{Janhunen1998a}. Using this approach they derived  $K$ based on Maxwellian and kappa fits to the observed electron distribution. Less than 20\% variation between functional forms was found, however some measurements suggested a variation of up to 40\%.

While these models have shown how the presence of a supra-thermal tail alters the relationship  between the field-aligned current and the acceleration potential they are limited by the omission of electron populations occupying classically forbidden regions of phase space in the acceleration region. Observations through the acceleration region show distributions where significant fractions of the total density reside in trapped orbits inaccessible to electrons conserving the first adiabatic invariant. While these electrons may not contribute appreciably to current they alter the manner which quasi-neutrality is maintained through the acceleration region and hence the distribution of density and potential along the field line \citep{Chiu1978, Sakanoi1995a, Ergun2000}. This lacuna is reflected in the surprising fact that none of the observational studies identified above were performed using measurements taken within the acceleration region itself. There are significant advantages for performing these analyses using measurements made within the acceleration region. These include avoiding the degradation of the spectrum at low altitudes traversed by rockets and removing the contribution from backscatter and secondary electrons at low altitudes to the primary electron spectrum. Measurements performed within in the acceleration region also reveal all the components of the electron and ion distributions which self-consistently carry $J$ and support $\phi$. Therefore significant advances could be made in subsequent studies by exploiting direct measurements from within the auroral acceleration such as those readily available from the FAST mission.

%----------------------------
\subsection{Radiation Belts}
\label{sec:belts}
%----------------------------

The radiation belts are the highest energy populations of charged particles in Earth's magnetosphere. The electron outer radiation belt is highly variable associated with magnetic storms\footnote{See the first footnote in Section \ref{sec:intro} for `storm'.}, and disappearance and subsequent flux enhancements occur associated with magnetic storms \citep[e.g.][]{Miyoshi2005}. 

Two different acceleration processes contribute to the energization of the outer belt electrons \citep[e.g.][]{Ebihara2011}. One process is the radial diffusion. Through the conservation of the first adiabatic invariant, the electron energy increases when electrons move toward the Earth. The drift resonance with the MHD-fast mode waves, i.e. Ultra Low Frequency (ULF) pulsations is an elementary process for this acceleration \citep[e.g.][]{Elkington1999}. If this process is dominant without significant source and loss process during the transport, the slope of the power-law distributions should be conserved. Another process is the wave-particle interactions. The cyclotron resonance with whistler mode chorus causes the non-adiabatic acceleration through violation of all adiabatic invariants \citep[e.g.][]{Summers1998, Miyoshi2003, Horne2005}. The resultant energy spectrum depends on the resonance conditions. If the resonance occurs with a wideband wave-number spectrum, the energy spectrum of the accelerated electrons should be a power law \citep{Ma1998a}, while non-linear wave-particle interactions cause different energy spectrum \citep{Furuya2008}.

Various data indicated that radiation belt electrons from 100 keV to 1.5 MeV can be modeled as a power-law spectrum in the outer radiation belt \citep[e.g.][]{Freeman1998, DesRoziers2006} although the spectrum can be more complex than a simple power-law \citep{Baker1998}. \cite{Freeman1998} derived time variations of the slope of the power-law spectrum. Typically the power-law index $\delta$ (as measured in differential flux) at geosynchrnous orbit ($R = 6.6$ R$_{\rm E}$) is about 2. During the main phase of a storm, the outer belt flux decreases and the energy spectrum becomes soft. The power law index $\delta$ is 4 -- 5 during the main phase. During the recovery phase, the flux of sub-relativistic and relativistic energy electrons gradually increases. During the period, the energy spectrum becomes hard and the power-law index recovers to the pre-storm level.

Similar variations of the power-law index are observed at different locations. \cite{Li1999} derived the time variation of energy spectrum using the GPS satellite at $R$ = 4.2. Assuming a power-law form, they measured the slope between data points in the 0.4 -- 0.8 MeV and 0.8 -- 1.6 MeV ranges. The power-law index $\delta$ was 4.6 during the main phase and then became as small as 1.4, associated with the flux enhancement during the recovery phase. \cite{Miyoshi2003} derived spatial-temporal evolutions of the energy spectrum in the 30 -- 300 keV range during the storm time. They found that the spectrum hardening takes place first in the inner part of the outer belt, and then the hard energy spectrum can be seen at the outer part of the outer belt. This indicates the non-adiabatic acceleration process occurs in the inner part of the outer belt, which are different from the radial diffusion.

% ==== Just a memo ===
%As for comparisons between energetic electrons in the plasma sheet and those of radiation belts, it was reported that the intensity of the energetic electrons at the outer boundary of the radiation belts (8 - 11 R$_{\rm E}$) can be well above the fluxes of relativistic electrons inside the outer radiation belt (even after adibatic energization is taken into account) \citep{Lui2012, Lui2013}. Based on the results, the authors argued that the inward radial transport can be a potential source of the relativistic electrons within the radiation belts, even in the strong pitch angle diffusion limit.

%\cite{Dai2015} reported near-earth injection of MeV electrons associated with intense dipolarization electric fields. 

%%%%%%%%%%%%%%%%%%%%%%%%%%%%%%%%%%%%%%%%%%%%%%%%%%%%%%%%%%%%%%%%%%%%%%
%%%%%%%%%%%%%%%%%%%%%%%%%%%%%%%%%%%%%%%%%%%%%%%%%%%%%%%%%%%%%%%%%%%%%%
\section{Power Laws in Other Environment}
\label{sec:oth}
\addtocontents{toc}{\setcounter{tocdepth}{3}}% Allow \section in ToC

%%%%%%%%%%%%%%%%%%%%%%%%%%%%%%%%%%%%%%%%%%%%%%%%%%%%%%%%%%%%%%%%%%%%%%
%%%%%%%%%%%%%%%%%%%%%%%%%%%%%%%%%%%%%%%%%%%%%%%%%%%%%%%%%%%%%%%%%%%%%%

\begin{figure}[t]
\includegraphics[width=\textwidth]{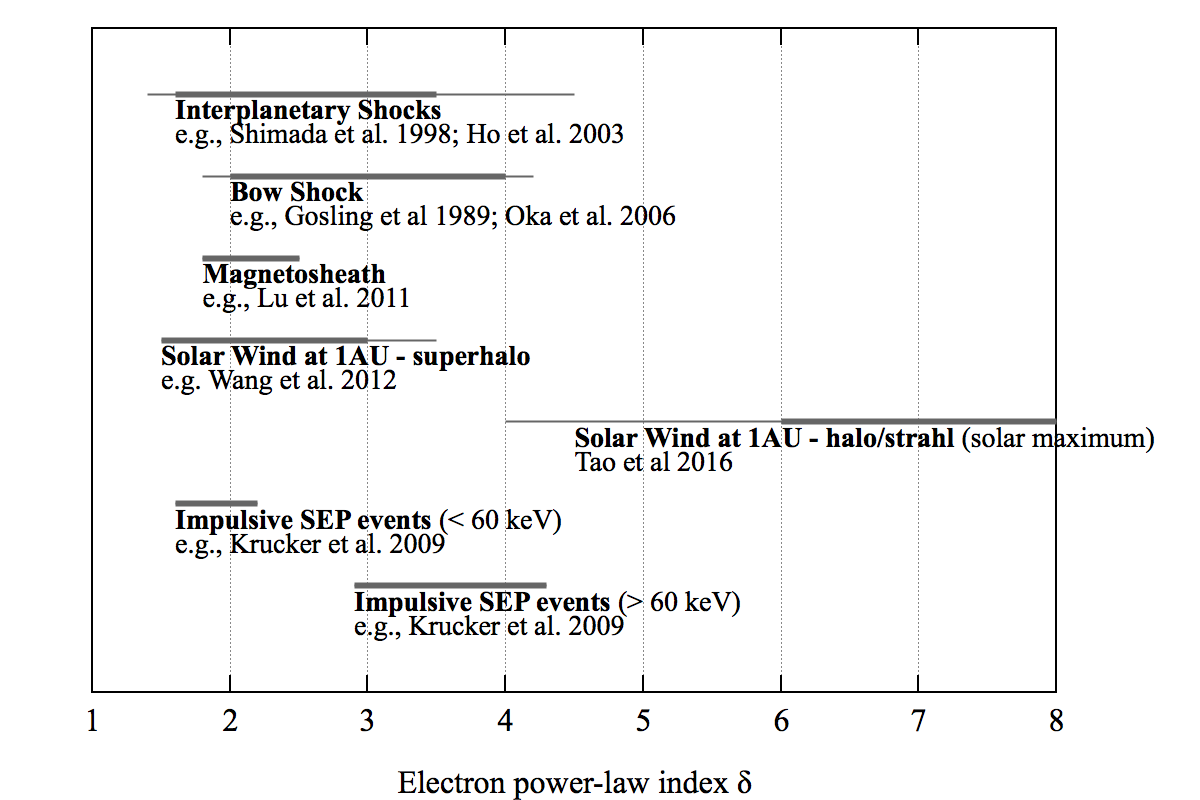}
\caption{An overview of electron power-law index $\delta$ in other environment such as shocks, solarwind and solar energetic particles (SEPs).  The solar wind halo/strahl components exhibit softer ($\delta > $8) spectra during solar minimum and is not shown.}
\label{fig:oth_index}
\end{figure}

In this section, we review power-law spectra in other regions such as solar energetic particles, shocks, magnetosheath, and the solar wind. Figure \ref{fig:oth_index} summarizes the reported values of $\delta$.

%XXXXXXXXXXXXXXXXXXXXXXXXXXX
%XXXXXXXXXXXXXXXXXXXXXXXXXXX

%----------------------------------------
\subsection{Solar Energetic Particles (SEPs)}
\label{sec:sep}
%----------------------------------------

\begin{figure}[t]
\includegraphics[width=\textwidth]{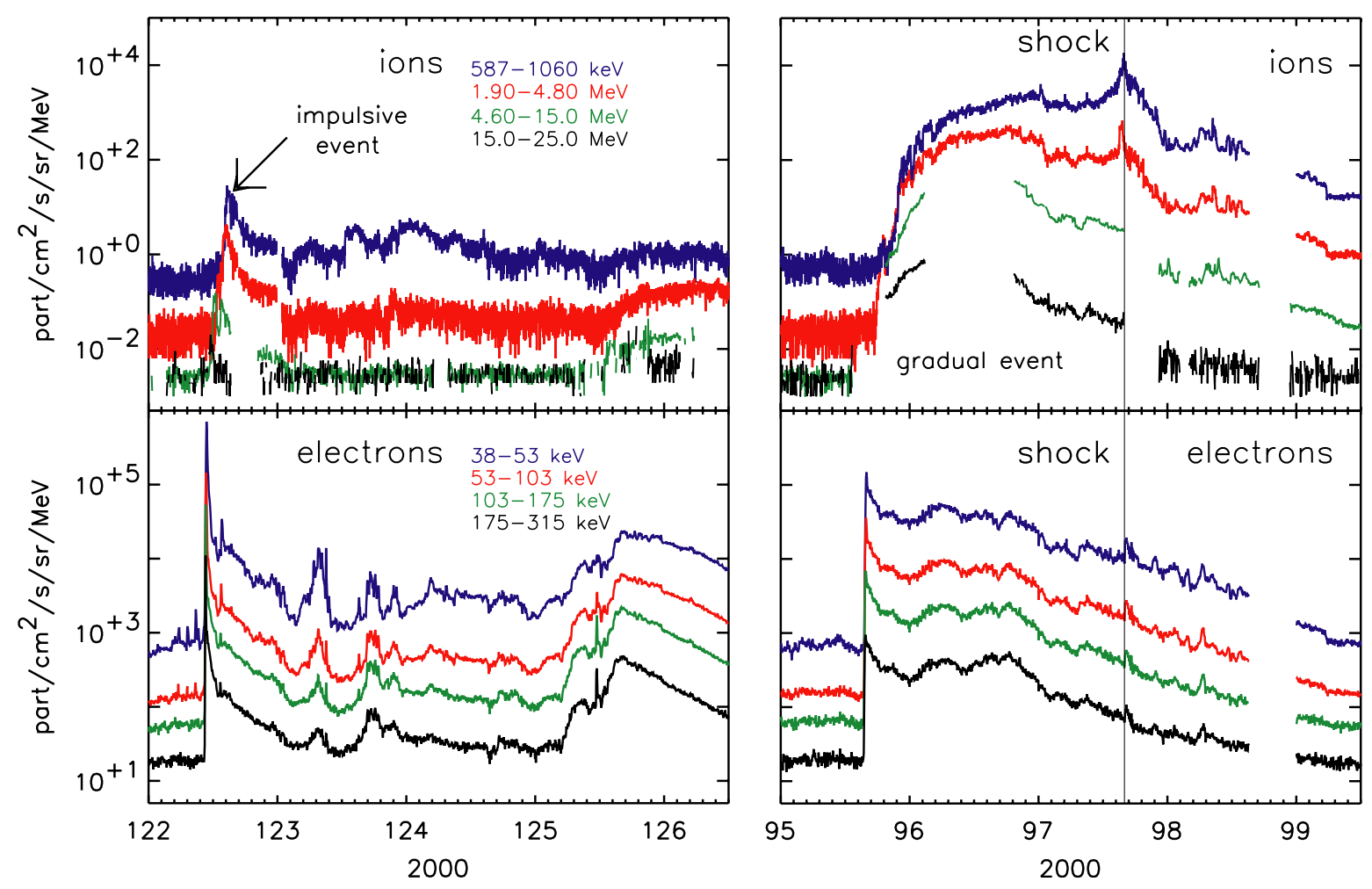}
\caption{Typical profiles of impulsive (left) and gradual (right) Solar Energetic Particle (SEP) events for ions (upper) and electrons (lower) as measured by ACE/EPAM, adapted from \cite{Lario2005} \textcopyright COSPAR. Reproduced with permission. The horizontal axes show Day-of-Year (DOY). }
\label{fig:oth_sep}
\end{figure}

Solar Energetic Particles (SEPs) are the particles detected in interplanetary space in association with solar eruptive events such as flares and coronal mass ejections (CMEs). Typically, SEP events are classified into `gradual' and `impulsive' events depending on the observed time profile of the SEP fluxes (Figure \ref{fig:oth_sep}) \citep[e.g.][]{Reames1999}. While gradual and impulsive events are generally attributed to particle acceleration by CME-driven shocks and solar flares, respectively, the more precise and detailed origins of both types of SEPs (including the possibility of mixed sources) are still debated. In fact, there have been numerous studies of SEPs (including protons, heavy ions, and electrons) and it is beyond the scope of this paper to review further details of SEPs. Readers are referred to recent reviews \citep[e.g.][]{Verkhoglyadova2015, Desai2016, Kahler2017,Dalla2017} and references therein. In this section, we will focus only on electrons during SEP events.

%XXXXXXXXXXXXXXXXXXXXXXXXXXX
%XXXXXXXXXXXXXXXXXXXXXXXXXXX

During a gradual SEP event (the typical duration of which is several days), the particle flux can enhance locally at and around the passage of an interplanetary shock (CME-driven shock) with a typical duration of hours (See, for example, the flux enhancements at the vertical lines in Figure \ref{fig:oth_sep}  right panels). Such a local enhancement is historically referred to as Energetic Storm Particles (ESPs) and indicates particle acceleration and trapping by the traveling interplanetary shock, although all particles in the intensity-time profiles of an SEP event (not just ESPs) could be accelerated by the shock.  We will describe energetic electron events that are clearly associated with shocks (including ESP events) in Section \ref{sec:shock} where both interplanetary shocks and Earth's bow shock are discussed.  

Energetic electrons (that are not necessarily focused on ESPs and shocks)  have provided important information for a better understanding of SEPs \citep[e.g.][ and references therein]{Desai2016, Dalla2017}. For example, it has been reported that the relative peak intensity of energetic ($>$ 0.5 MeV) electrons (i.e., electron-proton ratio) is substantially higher in relatively smaller events of gradual SEPs (with the proton intensity $<$ 3 cm$^{-2}$s$^{-1}$sr$^{-1}$), unless the SEP events are poorly connected to the observer \citep[e.g.][]{Cliver2007}. Therefore, it was argued that the gradual SEP events with relatively high intensities of electrons are attributed to acceleration in solar flares rather than acceleration in CME-driven shocks. In fact, at least in the specific example in Fig. \ref{fig:oth_sep}, energetic electrons in a gradual SEP event show an impulsive time-profile in the early phase. SEP electrons have also been studied in terms of anisotropy \citep[e.g.][]{Dresing2014}, spatial distribution in the heliospheric longitude direction \citep[e.g.][]{Klassen2016}, and mean free path \citep[e.g.][]{Agueda2014}.

The energy spectra of SEP electrons have also been studied and compared with hard X-ray spectra at flares \citep[e.g.][]{Krucker2007,Krucker2008d,Krucker2009}. A statistical study of impulsive SEP events reported that electron energy spectra (measured at 1 AU by Wind) typically exhibit a double power-law with a break around 60 keV \citep{Krucker2009}, although the origin of the double-power-law form is unclear. They reported that the average power-law indices below and above the break energy are $\delta_{\rm low} = 1.9 \pm 0.3$ and $\delta_{\rm high} = 3.6 \pm 0.7$, respectively. Both components are flatter/harder than those of hard X-ray emission from flares (See Figures \ref{fig:flr_index} and \ref{fig:oth_index}). They also reported that the number of electrons measured during the impulsive SEP event (presumably escaping from the solar origin) is much lower than the number of electrons at the flaring site, deduced from the hard X-ray measurements. Thus, they argued that the electron population at the sun and near Earth are different and might be produced by different acceleration mechanisms. They also suggested that, alternatively, the escape from the sun is energy-dependent. Note, however, that \cite{Krucker2007} find a strong correlation of $\gamma$ (flare) and $\delta$ (SEP) for impulsive events with close temporal correlation between the hard X-ray burst and the {\it in situ} electron events, the so-called prompt events. This is an important argument to keep in mind, since it hints at a common origin of both populations for the subset of prompt events \citep{Krucker2009}.

%----------------------------
\subsection{Shocks and Sheaths}
\label{sec:shock}
%----------------------------

Particles are accelerated to high energies by shocks. In the standard `diffusive shock acceleration' (DSA) model, the power-law index $s$ of shock-accelerated particles $f(p) \propto p^{-s}$ would be
\begin{equation}
s = \frac{3r}{r-1}
\label{eq:dsa}
\end{equation}
where $r$ is the shock compression ratio. In sub-relativistic shocks in the heliosphere, $r$ is no larger than $\sim$4 and so $s$ is 4 or larger or $\delta$ is 1 or larger. Even if the shock were weak, say $r$ = 1.4, the spectrum is still hard (flat) with $\delta \sim$4. This already implies that shocks in the heliosphere produce  hard (flat) spectra. 

In fact, {\it in situ} measurements of proton energy spectra at interplanetary shocks showed $\delta$ = 1 -- 4 in the energy range 40 -- 1000 keV \citep[e.g.][]{Scholer1983, VanNes1984}. The particle and magnetic/electric field data at interplanetary shocks were used to test the diffusive shock acceleration (DSA) theory \citep[e.g.][]{Kennel1984}. However, not all of interplanetary shocks show the features predicted by DSA such as a power-law energy spectrum and an exponential increase of particle flux prior to the shock arrival \citep[e.g.][]{VanNes1984, Lario2003, Lario2005a, Fisk2012}. Even when they show a power-law energy spectrum, the power-law index often deviates from the prediction by Equation \ref{eq:dsa}.  It is not clear exactly when or in which conditions ions show significant flux increase at and around shock fronts and how the power-law index could be understood. However, when shocks exhibit a power law with substantially enhanced fluxes, the power-law index $\delta$ seems to fall somewhere between 1 and 4. 

Unlike ions, electrons show the predicted, exponential flux increase on the upstream only in limited cases. Using Geotail, \cite{Shimada1999a} reported  the power-law index $\delta$ = 1.4 -- 1.5  at an interplanetary shock and \cite{Oka2009} reported $\delta \sim$ 3.3 at Earth's bow shock.  Focusing on the power-law index, \cite{Ho2003} reported that, most of their 28 interplanetary shock events showed a power law with $\delta$ between 1.5 and 4.5 in the $>$38 keV range (See also \cite{Ho2008}). Just like the ion case, \cite{Ho2003} argued that the power-law indices are not well reproduced by the formula Eq. (\ref{eq:dsa}). In the lower energy range ($>$ 2 keV), electrons show a localized `spiky' enhancement or almost no enhancement at interplanetary shocks \citep[e.g.][]{Tsurutani1985, Lario2003, Lario2005a}. Electrons at Earth's quasi-perpendicular bow shock also show a spiky enhancement localized at the shock transition layer. Despite the profile inconsistent with DSA, power-law energy spectra do form within the transition layer, indicating electrons can be accelerated by a mechanism very different from the classical DSA scenario \citep{Gosling1989a, Oka2006}. In case studies by \cite{Gosling1989a}, $\delta$ was reported to be between 2 -- 3. In a statistical study by \cite{Oka2006}, $\delta$ ranged from 2 to 4.

The similar power-law spectra have been observed on the downstream side of the shock front or Earth's magnetosheath. A caveat is that the non-thermal electrons tend to exhibit perpendicular anisotropy, which could be due to escape of field-aligned electrons from the acceleration region (i.e., the shock transition layer) \citep{Feldman1983a, Gosling1989a}. In a long duration observation of Earth's magnetosheath, \cite{Lu2011a} analyzed pitch-angle averaged electron energy spectra and obtained electron power-law indices $\Gamma$ = 2.8 -- 3.5 ($\delta$ = 1.8 -- 2.5). They argued, based on simulations, that a soft power law (with large $\delta$) would evolve into harder spectra through excitation of whistler waves. \citep{Lu2010a}. 

To summarize, collisionless shocks in the heliosphere produce a power-law energy spectrum with $\delta <$ 4. Similar power-law distributions have been observed in Earth's magnetosheath.

%----------------------------------------
\subsection{Quiet-Time Solar Wind}
\label{sec:sw}
%----------------------------------------

It has been reported that the solar wind ions in interplanetary space exhibit a non-thermal tail  even when there is no significant disturbance such as shocks, magnetic clouds, counter-streaming electrons and magnetic holes \citep[e.g.][]{Gloeckler2000, Fisk2006}. The non-thermal tail can often be modeled as a power  law with an exponential roll-off, indicating there is a limit to the maximum energy particles can acquire. More importantly, the power-law index $\delta$ seems to be $\sim$1.5 in many cases. This is the smallest power-law index a kappa distribution can take (Section \ref{sec:definition}). It was argued that such a hard and universal power-law index can be explained by a `pump mechanism' in which particles are accelerated by a local compression in a turbulent medium but escape by spatial diffusion before the local compression turns into an expansion, which would decelerate the accelerated particles \cite[e.g.][]{Fisk2006,Fisk2014}. 

%XXXXXXXXXXXXXXXXXXXXXXXXXXX
\begin{figure}
\includegraphics[width=0.6\textwidth]{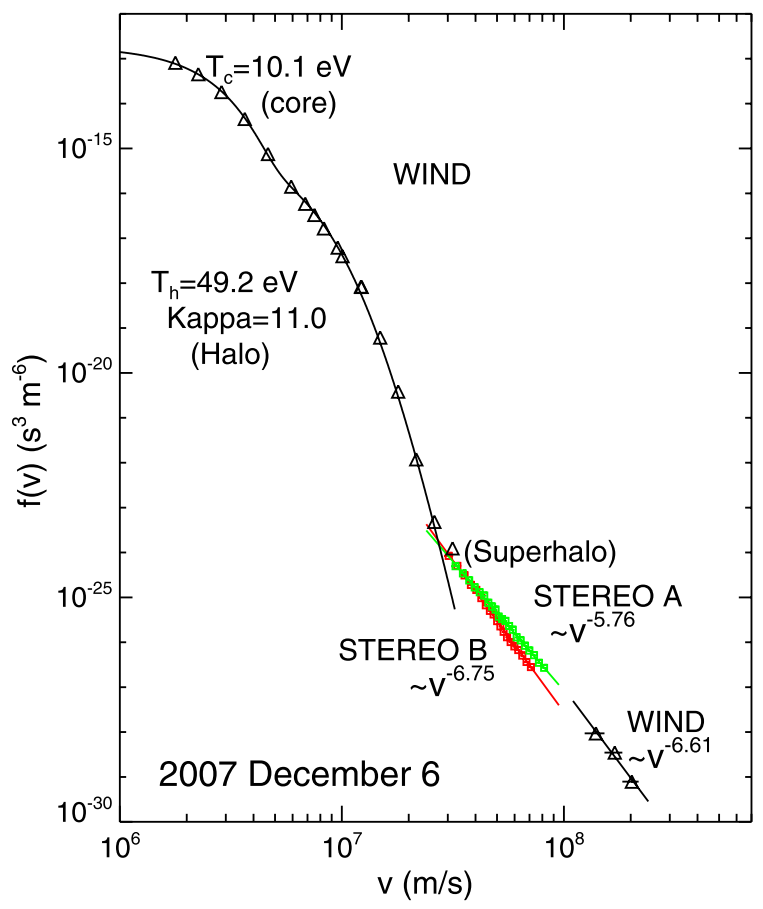}
\caption{Electron velocity distribution in a time period of quiet solar wind, adapted from \cite{Wang2012a}. The solid curves represent the fit by a combined Maxwellian and kappa distribution (for core-halo components) and a power-law (for super-halo component). }
\label{fig:oth_superhalo}
\end{figure}
%XXXXXXXXXXXXXXXXXXXXXXXXXXX

Non-thermal electrons have also been observed  in the quiet-time solar wind  (Figure \ref{fig:oth_superhalo}). In addition to the `core' component with the temperature of a few to $\sim$10 eV, there is a hotter `halo' component with the temperature of several tens of eV \citep[e.g.][]{Marsch2006}. This halo component is often isotropic and may have a non-thermal tail, which could be fitted by the kappa distribution. However, \cite{Tao2016} reported that the tail (at 1 AU) is generally steep and the power-law index $\delta$ (or $\kappa$) is typically between 4 and 16. They also reported that the spectral slope depends on the solar activity and that the tails are substantially harder (flatter) during the solar maximum. The `strahl' electrons are in an energy range similar to those of the halo electrons but they are streaming anti-sunward along the interplanetary magnetic field. The power-law spectra of the strahl electrons look similar to those of the isotropic halo component \citep{Tao2016}.   

Both the halo and strahl components can evolve during the propagation of the solar wind. The relative number of stahl electrons is decreasing with radial distance from the sun, whereas the relative number of halo electrons is increasing \citep[e.g.][]{Maksimovic2005, Stverak2009}. It is often argued that the strahl component represents  the escaping heat flux from the solar corona and that it eventually evolves to become the halo component via pitch angle scattering in the solar wind \citep[e.g.][]{Feldman1975, Stverak2009}. In terms of the power-law index, \cite{Stverak2009} reported that, for both halo and strahl components in the slow solar wind, $\delta$ is large, $\sim$ 9, at $\sim$ 0.3 AU and the spectra gradually become hard $\delta \sim$ 2 -- 4 beyond 1 AU \citep[See also][]{Pierrard2016}. 

The super-halo component (at 1 AU) \citep[e.g.][]{Lin1998, Wang2012a} more generally exhibits a harder power law. This component is within  the energy range of order tens to hundreds of  keV, similar to the energy range of those electrons we have reviewed in Sections \ref{sec:flares} and \ref{sec:emt}. For the super-halo component, the typical values of the power-law index $s$ of $f(v) \propto v^{-\gamma}$ are 5 -- 9 \citep{Wang2012a} so that $\delta$ = 1.5 -- 3.5.

%%%%%%%%%%%%%%%%%%%%%%%%%%%%%%%%%%%%%%%%%%%%%%%%%%%%%%%%%%%%%%%%%%%%%%
%%%%%%%%%%%%%%%%%%%%%%%%%%%%%%%%%%%%%%%%%%%%%%%%%%%%%%%%%%%%%%%%%%%%%%
\section{Summary and Discussion}
\label{sec:dsc}
\addtocontents{toc}{\setcounter{tocdepth}{3}}% Allow \section in ToC

%%%%%%%%%%%%%%%%%%%%%%%%%%%%%%%%%%%%%%%%%%%%%%%%%%%%%%%%%%%%%%%%%%%%%%
%%%%%%%%%%%%%%%%%%%%%%%%%%%%%%%%%%%%%%%%%%%%%%%%%%%%%%%%%%%%%%%%%%%%%%

\subsection{Solar Flares vs Earth's Magnetotail}

There have already been many comparative studies of solar flares and terrestrial substorms from various points of view \citep[e.g.][]{Obayashi1975, Bratenahl1976, Terasawa2000b, Bhattacharjee2004a, Lin2008, Birn2009, Tsurutani2009}. These studies have discussed the basic similarities and differences in terms of (1) plasma parameters/environments, (2) morphologies of the phenomena and (3) the physics of magnetic reconnection. This review is an extension of the comparative studies to the realm of particle acceleration and non-thermal features in the possible key regions of solar flares (Section \ref{sec:flares}) and Earth's magnetotail (Section \ref{sec:emt}). 

As compiled in Figures \ref{fig:flr_index}, \ref{fig:emt_index_a} and \ref{fig:emt_index_b}, the power-law index $\delta$ takes on a wide variety of values  in solar flares and Earth's magnetotail. Nevertheless, we find a notable resemblance between the $\delta$ values in solar flares and those in Earth's magnetotail when $\delta$ is measured close to the reconnection site. 

%XXXXXXXXXXXXXXXXXXXXXXXXXXX
\begin{figure}[t]
\includegraphics[width=\textwidth]{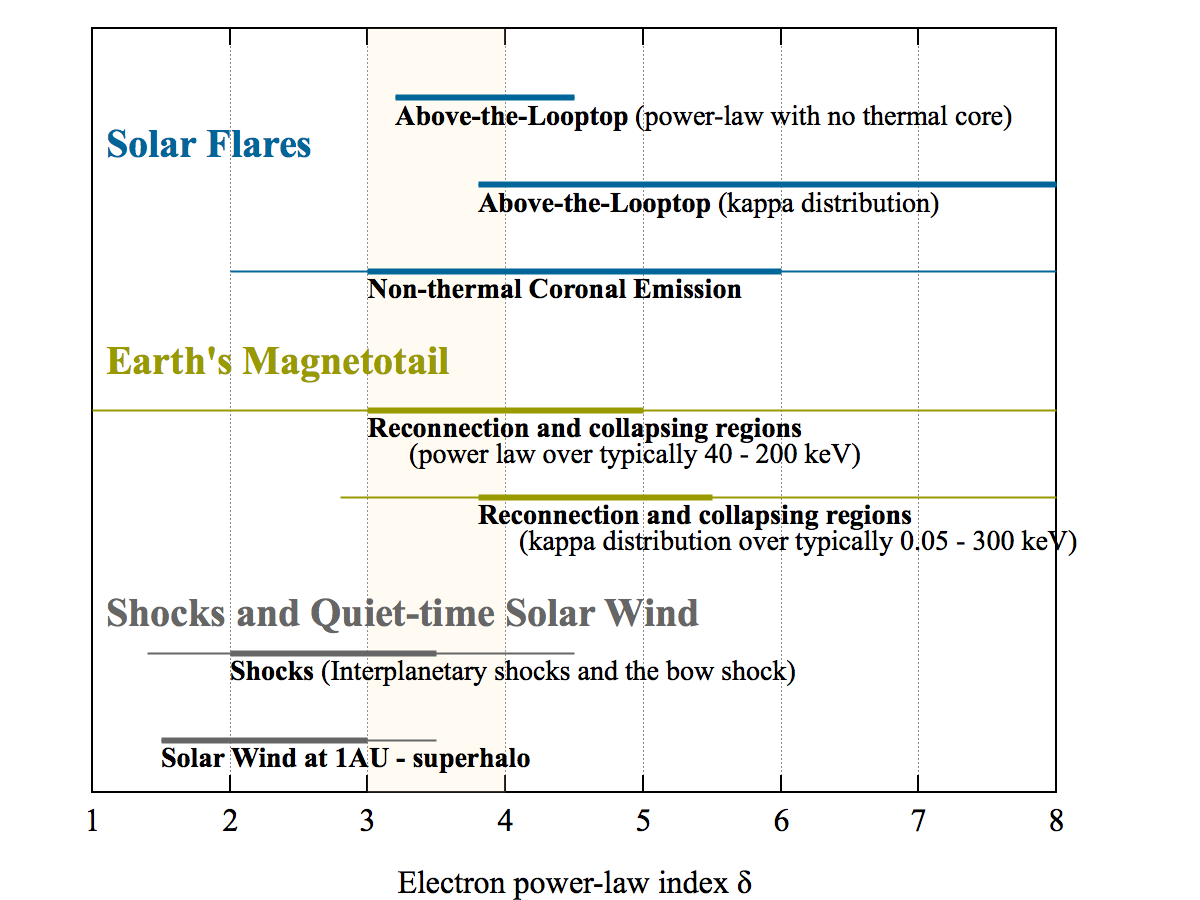}
\caption{An overview of electron power-law index $\delta$ in the regions close to magnetic reconnection and in shocks and quiet-time solar wind. The orange shade in the $3 < \delta < 4$ range highlights the result that, in many cases, the power-law spectra in solar flares and Earth's magnetotail plasma sheet are softer (steeper) while those of shocks and the solar wind are relatively harder (flatter). As described in Section \ref{sec:flares}, the `above-the-looptop' source is a special case of non-thermal coronal emission in which the higher energy source is located clearly anti-sunward from the lower energy source.}
\label{fig:dsc_index}
\end{figure}
%XXXXXXXXXXXXXXXXXXXXXXXXXXX

\subsubsection{Regions close to Reconnection Site}
\label{sec:close}

In order to examine the possible consequence of magnetic reconnection, we examine data from regions/structures that are closest to magnetic reconnection.  For solar flares, the above-the-looptop (ALT) sources are the closest X-ray sources to the reconnection site (as expected in the standard model).  Due to the limited energy coverage (typically 15 -- 80 keV), different spectral models can be used for the ALT source, as shown in Figure \ref{fig:dsc_index}.  Figure \ref{fig:dsc_index} does not show $\delta$ values of the double coronal source events because, as far as we are aware, there have been only two studies that estimated the power-law index; and those two studies reported rather dissimilar ranges of values. Future studies will need to establish the typical $\delta$ values of the secondary, upper coronal source. Another caveat is that many cases of hard X-ray emission from the corona show a wide range of values of the power-law index $\delta$ and it can be as small as $\sim$2 (Section \ref{sec:flares}). Only in rare cases, the non-thermal source exhibits a clear `above-the-looptop' configuration at roughly 20 -- 60 Mm from the solar surface and is well separated from the lower-energy source by 10 -- 20 Mm. The typical distance $d$ between the lower and higher energy sources are smaller, $|d|< 2$, and the source height is also closer to the solar surface, 4 -- 20 Mm.

For Earth's magnetotail, the collapsing region appears to be the counterpart of the solar hard X-ray above-the-looptop (ALT) source. We also note that reconnection outflows typically experience braking (and bouncing) at radial distances between $\sim$8 and $\sim$12 R$_{\rm E}$ in the collapsing region. (Thus, this specific region in the collapsing region is sometimes referred to as the `flow-braking region'.) % This flow-braking region may be the counterpart of the solar hard X-ray above-the-looptop (ALT) source). However, plasma signatures (except the significant bulk-flow decrease) are not making this flow-braking region to stand out from other parts of the magnetotail plasma sheet. In fact,
During magnetotail reconnection, various features such as `fast flows', `dipolarizing flux bundles', and `dipolarization fronts' are observed not only in the collapsing region (between $\sim$8 and $\sim$12 R$_{\rm E}$) but also throughout the plasma sheet including the region very close to (or even within) the Hall region (or ion-scale diffusion region). The power-law index $\delta$ also appears to show the similar lower-limit 3 -- 4 in both the collapsing and reconnection regions (Figures \ref{fig:emt_index_a} and \ref{fig:emt_index_b}), as summarized in Fig. \ref{fig:dsc_index}. Some studies show that the power-law index $\delta$ can be substantially smaller in the $<$ 10 R$_{\rm E}$ range \citep[e.g.][]{Stepanova2015, Runov2015}. However, such an inner region resembles more like the looptop region rather than the above-the-looptop (ALT) region. Also, the inner region is where the radiation-belt electrons are generated, while an analogous region to the radiation belts does not seem to exist in solar flares. Thus, it appears reasonable to avoid values from the inner magnetosphere when comparing with solar flare ALT source. We also note that the there may be a fast-mode shock in the ALT source and it may be fundamentally different from the collapsing (or flow-braking) region we see in the magnetotail (as a fast-mode shock have not been detected in the magnetotail). 

With all these factors taken into account, it is evident in Figure \ref{fig:dsc_index} that electron power-law spectra in the regions close to reconnection are relatively soft and $\delta$ is larger than $\sim$ 4 in many cases (or $\sim$3 depending on the spectral model and/or energy range). The roughly similar lower limit at 3 -- 4 is interesting, despite the orders of magnitudes difference of plasma parameters (See Appendix \ref{sec:param}).

\subsubsection{Regions away from Reconnection Site}
\label{sec:away}

For solar flares, the footpoint sources are in the chromosphere and are away from the reconnection site in the standard model. The typical $\delta$ values of the footpoint sources are 3.5 -- 4.7 (Figure \ref{fig:flr_index}). This is comparable to the typical $\delta$ values of 3 -- 5 of the non-thermal emission in the corona. 

The magnetospheric analogue of the footpoint sources is the top-side ionosphere. These are the regions which emit the brightest extreme-ultraviolet (EUV) and X-rays \citep[e.g.][ and references therein]{Borg2007}. As reviewed in Section \ref{sec:aurora}, power-law tails do exist in and around the auroral region (in the altitude of less  than $\sim$ 1000 km). Unlike the power-law indices in solar footpoint sources, the power-law indices in the aurora region takes a wide range of values. The variation is so large that the distribution can be very close to Maxwellian $\delta > $8. In some cases, the power-law spectra can be very hard with $\delta$ as small as $\sim$2. The spectra can be hard even in the high-latitude plasma sheet, geometrically corresponding to the location along the loop, somewhere between the footpoint and looptop regions in a solar flare loop.

\subsection{Shocks vs Reconnection}

Figure \ref{fig:dsc_index} also shows the $\delta$ values from shocks and quiet-time solar wind (as described in Section \ref{sec:oth} and illustrated in Figure \ref{fig:oth_index}).   It is evident  that shocks produce electron energy spectra  harder than those in solar flares and Earth's magnetotail where magnetic reconnection plays an important role in their dynamics. Although there can be an overlap in the $\delta = $ 3 -- 4 range, the difference suggests that electron acceleration at shocks leads to a larger non-thermal fraction of electron energies when compared to magnetic reconnection (and associated processes). 

This difference between the shocks (including the downstream regions) and reconnection-related region may provide a hint for solving the problem of electron acceleration mechanism in the above-the-looptop (ALT) source. It has been suggested that a fast-mode termination shock plays an important role for accelerating electrons in the ALT source (as described in Section \ref{sec:alt}), although the power-law spectrum  is soft (as mentioned above).  Even in the recent study of fast-mode termination shock by \cite{Chen2015a}, the power-law index as measured by RHESSI was $\delta \sim$ 4.5 - 6.2 (using the thin-target assumption; the spectra would be even softer for a thick-target assumption). 

In contrast, {\it in situ} measurements in space have shown that a fast-mode shock produces a hard power-law ($\delta \sim$ 2 -- 4) even in a weak shock condition similar to that of the above-the-looptop source, i.e., low Alfv\'{e}n Mach number ($\sim$ 2) and large shock angle ($>$ 80$^{\rm o}$) \citep{Oka2006}. However, the spectrum can be softer ($\delta < 4$) if the shock angle were smaller. Thus, the fact that solar flare observations show soft spectra in the ALT source suggests that the shock angle is sufficiently small ($<$ 80$^{\rm o}$) or the shock acceleration scenario does not apply. It is also very possible that the spatial resolution of the X-ray measurement is insufficiently low and that a (possibly hard) power-law spectrum at the shock may have been averaged out. Therefore, a significantly higher spatial resolution is crucial for future studies of the shock acceleration scenario. 

\subsection{Origin of Power-Law Tails in Earth's Magnetotail}

We point out a possibility that electrons form a power-law tail outside of the magnetotail and that such a non-thermal component is transported into the magnetotail without a significant change of the spectral form. As reviewed in Section \ref{sec:activequiet}, power-law distributions are present in the magnetotail plasma sheet even during quiet times \citep[e.g.][]{Anderson1965, Christon1989a} and the power-law index $\delta$ does not necessarily change by activity or locations \citep[e.g.][]{Asnes2008, Gabrielse2014, WuMingyu2015}. These observations indicate that the existence of power-law high-energy tails of particle distributions cannot be taken as evidence for their creation inside the magnetotail. It should be noted that non-thermal tails are already present in the solar wind and in the magnetosheath (as reviewed in Section \ref{sec:oth}) in the energy range above a few hundreds of eV and these tails may enter into the magnetotail.  

In fact, it has been suggested the bulk of the solar wind (and magnetosheath) plasma is transported into the plasma sheet from the distant tail \citep[e.g.][]{Gosling1984}, through dayside reconnection \citep[e.g.][]{Baker1996} and the flanks of the magnetotail \citep[e.g.][]{Terasawa1997}, although there is a substantial contribution from the ionosphere to the near-Earth plasma sheet. Magnetic reconnection in the distant tail could be important for the plasma transport from the distant tail to the near-Earth tail, although the precise mechanism of transport remains an open question \citep[e.g.][]{Hultqvist1999}. We also note that the average (kinetic) temperatures increase from the magnetosheath to the magnetotail by an order of magnitude, but the typical ion to electron temperature ratio remains almost constant (5 -- 10) \citep[e.g.][]{WangCP2012}. This can be taken as evidence that the major entry mechanism (which is still debated) is consistent with adiabatic heating, which would keep the shape of the energy distributions unchanged, so that the quiet time populations in the tail may be considered as heated magnetosheath populations. 

Such an external origin scenario of electron power-law tails does not contradict with the observations that electrons are energized significantly in the magnetotail. Non-thermal electrons that enter from the magnetosheath can further acquire energies in the magnetotail as indicated by particle flux increases, well documented in the magnetotail over a wide range of distances (as reviewed in Section \ref{sec:emt}). And the energization process may, but need not, involve a change in the power law index. 

However, the precise origin of the electron power-law tails and their evolution in the magnetotail remain inconclusive, and the external origin scenario of electron power-law tails needs to be studied further. This is because, we have shown, through a review of various studies, that electron energy spectra in the plasma sheet are relatively softer ($\delta \sim$ 3 -- 5) than those in the solar wind and magnetosheath ($\delta \sim$ 2 --4), suggesting different sources of the power-law tails. In fact, a statistical analysis reported that there is no clear correlation between energetic plasma sheet electrons ($>$38 keV) and solar wind electrons of comparable energies \citep{BurindesRoziers2009}.  Similarly, \cite{Luo2012} reported that, for most of their magnetopause crossing events ($>$70\%), the fluxes and phase-space-densities of energetic ($>$ 38 keV) electrons in the magnetosheath were less than those in the magnetosphere. They suggested that the energetic electrons in the magnetosheath cannot be a direct source sufficient for the energetic electrons inside the magnetosphere. Also, detailed case studies of the reconnection diffusion region (including the Hall region and electron diffusion region) report that electrons are energized across the diffusion region (as reviewed Section \ref{sec:mrx}), suggesting that the lobe electrons rather than magnetosheath electrons entering through the tail flanks are accelerated. 

\subsection{Possible Future Studies}

Our review led us to suggest similar values of $\delta$ (around 3 -- 5) in the regions close to magnetic reconnection in solar flares and Earth's magnetotail. However, it is desired to have a larger number of studies of $\delta$ (from both observational and theoretical points of view) to examine the generality or validity of this idea. 

An issue we have encountered in this review is that different studies use different models, methodologies or techniques to derive the power-law index, possibly leading to inconsistent results or larger variations of the $\delta$ values. For example, while some observations have a limited coverage of the energy range or a limited number of data points in the energy range (ending up using only a single functional form), other observations have  a wide energy-range coverage or many data points (ending up using a combination of different spectral models). Thus, it is desirable to carry out a more systematic and coordinated studies of power-law distributions using consistent methodologies and conditions for comparing particle spectra in different regions.  For studies of solar flares, we propose more frequent use of the relativistic Bethe-Heitler bremsstrahlung cross-section as implemented in the RHESSI/OSPEX software for more convenient comparisions with electron distributions measured {\it in-situ}. We also support development of new instrumentations or techniques. In particular, future solar  X-ray missions with  focusing optics (currently being proposed in Unites States and Japan) would achieve higher sensitivities and larger dynamic ranges.  For studies of the magnetotail,  MMS completed its first main tail season (`Maha Phase') very recently (2017 fall). With the state-of-the-art measurements achieved by MMS (for example, large energy coverages combined with unprecedentedly high time resolution), we envision a substantial advancement of our understanding of the power-law tails (for both ions and electrons) in the magnetotail in the coming years.  

From a theoretical point of view, we already have a variety of different particle acceleration mechanisms (as briefly summarized in Section \ref{sec:gen} and thoroughly reviewed in other review papers). However, in order to quantitatively predict the power-law index $\delta$, we need a more generalized model for the `escape' process via, for example, diffusion, because it would make the energy spectrum softer than the hardest possible spectrum (i.e., $\delta = 1.5$) as described in Section \ref{sec:gen}. Because particles need to be trapped in a specific region to be accelerated, the modeling of the `escape' process can be equivalent to the modeling of particle trapping in acceleration region. In flares, the fact that the non-thermal coronal source is generally isolated rather than spatially extended along the loop already indicates that a certain degree of trapping is at work in the (above-the-)looptop region (See also Section \ref{sec:relation}). Thus, there have been many theoretical studies of trapping and related transport of flare particles from different approaches \citep[e.g.][]{Minoshima2011, Li2012, Chen2013a, Kontar2014}. For Earth's magnetotail, the basic behaviors of energetic electrons seem to be explained by the drift motion in and around the dipolarizing flux bundles \citep[e.g.][]{Birn2004, Birn2013, Gabrielse2016} (See also Section \ref{sec:collapse}). However, some observations suggest a need for taking into account particle diffusion processes \citep[e.g.][and references therein]{Imada2008a, Stepanova2015}.

\appendix
\section{Plasma Parameter Regimes}
\label{sec:param}
\addtocontents{toc}{\setcounter{tocdepth}{3}}% Allow \section in ToC
Plasma parameter regimes in the solar corona and Earth's magnetotail generally do not overlap (Table \ref{tab:parameters}). For example, magnetic field magnitudes in the corona (2 -- 200 G) are $>$ 4 orders of magnitude larger than those in the lobe region of Earth’s magnetotail ($\sim$ 30 nT). Also, the number densities in the corona ($\sim$10$^{8}$ cm$^{-3}$) are roughly 10 orders of magnitude larger than those in the magnetotail (0.01 cm$^{-3}$), and the corona can be somewhat collisional depending on particle energies especially in a flaring region.

\begin{table}
\caption{Typical plasma parameters in different regions. 
Note 1 MK $\sim$ 86 eV and 1 G = 10$^5$ nT. The solar flare looptop parameters are for the above-the-looptop events 
that showed large loop sizes $>$ 40 Mm.  Many flares are smaller 
$\lesssim$20 Mm and the parameters may be different by orders of magnitude. For comparison, typical parameters at $H \sim$ 50 Mm in the quiet-time corona
are shown in the first column (Taken from \cite{Aschwanden1999} and references therein). 
The `collapsing region' in this table refers to the current sheet center $B_{x, \rm GSM} \sim$ 0 around $|X_{\rm GSM}|$ = 8 -- 11 R$_{\rm E}$ after arrival of dipolarization fronts. The ion (proton) gyro radius is based on the temperature and magnetic field magnitude that are also listed in the table. }
\label{tab:parameters}       % Give a unique label
\begin{tabular}{llllll}
\hline\noalign{\smallskip}
Physical quantities & Solar Corona & Solar Corona & Magnetotail & Magnetotail   & Solar Wind \\
					& (in quiet time &(large flare         & (lobe)      & (collapsing & (at 1 AU) \\
                    & at $H \sim$ 50 Mm) &         looptop)&            &   region) & \\

\noalign{\smallskip}\hline\noalign{\smallskip}
density (cm$^{-3}$)    & $\sim 10^{8}$   & $10^{10}$ -- $10^{11}$ & $<$ 0.01 & 0.1 -- 1   & 1 -- 10              \\
temperature (MK)       & $\sim$2         &  10 -- 30 & $<$ 1     & 10 -- 100  & 0.1           \\
magnetic field (G)     & 2 -- 200        & $\gtrsim$100  & $10^{-4}$ & $10^{-4}$  & $10^{-5} - 10^{-4}$ \\
  & & & & & \\
Alfv\'{e}n speed (km/s)& 400 -- 40000 & 500 -- 2000 & $>$ 2000  & 100 -- 1000& 20 - 100            \\
plasma $\beta$         & $<$ 0.1        & $\lesssim$ 1 & $<$ 0.1   & 0.1 -- 100 & 0.1 - 50 \\
ion gyro  radius (km)  &$10^{-4}-10^{-2}$& $< 10^{-3}$ & $<$ 100   & 100 -- 1000 & 10 -- 1000 \\
electron plasma & $10^8$ & $\sim 10^9$  & $10^3$ & $10^{-4}-10^{-3}$ & $10^{-5}-10^{-4}$ \\
$\quad\quad$frequency (Hz)  & & & & & \\
electron gyro   & $10^6$ -- $10^8$ & $> 10^8$  & $10^{2}-10^{3}$ & $10^{2}-10^{3}$ & $10^{1}-10^{2}$ \\
$\quad\quad$frequency (Hz)  & & & & & \\
%spatial scale (Mm) & --  & 40 &  -- &  60  & -- \\
%magnetic Reynolds number     & $10^{12}$       　& $>10^{12}$  & \\ 
%characteristic size (cm)     & several $\times 10^9$ & several $\times 10^9$ & - \\
%released energy (erg)        & $10^{29} - 10^{32}$ &  $10^{22}-10^{23} $ & - \\
%sound speed (km/s)                  & 200        &  500 - 1000 & 50                  \\
\noalign{\smallskip}\hline
\end{tabular}
\end{table}

Interestingly, the Alfv\'{e}n speed $V_A$ turns out to be similar, of the same order of magnitude, and the plasma beta $\beta$ sometimes differ by only 1 -- 2 orders of magnitude \citep{Terasawa2000b, Lin2008}. A caveat is that the ion gyro radius is $\sim$ 5 orders of magnitude smaller in the solar coronal environment. Thus, if normalized by the ion gyro radius (or the thickness of a reconnecting current sheet \footnote{Our experimence with particle-in-cell (PIC) simulations as well as Earth's magnetotail suggests that a current sheet needs to be as thin as the ion gyro radius before magnetic reconnection to take place.}), the spatial scale of a solar flare is again many orders of magnitude larger than that of the magnetotail. Therefore, plasma conditions remain very different in solar flares and Earth's magnetotail.

Note that solar flares can occur in various spatial scales at different altitude \citep[e.g.][]{Shibata2007}. If magnetic reconnection takes place in the chromosphere with a characteristic spatial scale of $\sim$10$^3$ km or less, the environment would be dominated by partially ionized plasmas with frequent Coulomb collisions.

The guide field $B_{\rm g}$ (the magnetic field component perpendicular to the reconnection plane) is also an important parameter, especially from the reconnection physics point of view. It is zero when two magnetic field lines  are approaching toward each other (to be reconnected) at exactly 180$^{\rm o}$ angle. It is non-zero when the shear angle is less than 180$^{\rm o}$.  In the near-Earth plasma sheet, say R$<$30R$_{\rm E}$, $B_{\rm g}$ is often small and can be $\sim$0.1 B$_{\rm 0}$ where B$_{\rm 0}$ is the asymptotic value in the magnetotail lobe region. However, $B_{\rm g}$/B$_{\rm 0}$ can be as large $\sim$1 in the  distant magnetotail as well as in the solar corona.

\section{Non-Extensive Statistical Mechanics}
\label{sec:nonextensive}

Classical particle systems in thermal equilibrium have their phase-space distribution stabilized into a Maxwell-Boltzmann function. These systems are characterized by limited or no correlations among their particles' energies or individual phase-space. In contrast, space plasmas are particle systems frequently described by stationary states out of thermal equilibrium, namely, their distribution is stabilized into a function that is not given by the Maxwell-Boltzmann formulation and is typically described by a kappa distribution (See Section \ref{sec:definition}, Eq. (\ref{eq:kappa})). These systems are characterized by long-range interactions that induce `correlations' resulting to a collective behavior among particles.\footnote{Correlations are represented by a collective behavior of plasma particles via electrostatic and/or electromagnetic fields and can keep space plasmas away from the thermal equilibrium \citep[e.g.][]{Livadiotis2015d}. Thus, correlations can be present at different spatial scales. The smallest correlation length is represented by the Debye length \citep[e.g.][]{Livadiotis2014} and the largest correlation length is represented by, perhaps, the minium wave number $k_{\rm min}$ of power spectral density, often discussed in the context of solar wind turbulence theory. Without any correlation, plasmas would be completely uniform with no characteristic spatial scale. As a result of correlations (and associated collective behaviors), space plasmas would  reach a steady state represented by the kappa distribution. As a result of collisions, on the other hand, space plasmas would reach a steady state represented by the Maxwell distribution as collisions would destry correlations \citep[e.g.][]{Livadiotis2015d}.}

%While the long-range Coulomb force should be shielded in most situations considered in this review paper, particles can still interact with each other through electrostatic or electromagnetic fields and can be `correlated'. The nonextensive statistics may be better applied to a quasi-stationary state of space plasmas filled with wave turbulence.}

Non-extensive statistical mechanics was introduced by \cite{Tsallis1988} as a generalization of  Boltzmann-Gibbs statistical mechanics. It has been applied to power-law distributions in various fields of natural and social science. Inspired by multifractals, Tsallis introduced a free parameter $q$ and postulated that the entropy can be expressed as
% \begin{equation}
% S_{q} \equiv k_B\frac{1-\sum\limits_{E} p^q}{q-1}
% \end{equation}
\begin{equation}
S_{q} \equiv k_B\frac{1-\int f(v)^q dv}{q-1}, 
\end{equation}
where $f(v)$ is the probability function (This is a continuum limit in which intergral is used instead of sum of discretized volume element). In the limit of $q$ $\rightarrow$ 1, it reduces to the traditional Boltzmann-Gibbs entropy, $S_1 = - k_B \int f(v) \ln f(v) dv$.   While Boltzmann-Gibbs entropy is extensive (or additive due to the logarithmic measure), Tsallis entropy is non-extensive, i.e., the entropy of the whole is not equal with the sum of the partial entropies of the parts. Note that the Boltzmann-Gibbs entropy is maximized under the constraints of canonical ensemble (i.e., normalization of the distribution, and fixed energy of the system) leading to the Maxwell-Boltzmann distribution. On the other hand, the maximization of Tsallis entropy under the constraints of the canonical ensemble leads to Tsallis distribution which is, in one-dimensional form, 
\begin{equation}
f(v) \propto \left[1+(q-1)(\beta v^2) \right]^{-\frac{1}{q-1}}, 
\end{equation}
where $\beta$ is a parameter related to the inverse of temperature and the $q$-index \citep[e.g.][]{Tsallis1988, Tsallis1998}. This  distribution recovers the Maxwell-Boltzman distribution in the limit of $q \rightarrow$ 1.

It was soon recognized that, with the transformation $\kappa = 1/(q-1)$, the Tsallis distribution corresponds to the empirically derived kappa distribution (Section \ref{sec:definition}, Eq. (\ref{eq:kappa}))\footnote{As already noted in the footnote of Section \ref{sec:definition}, these slightly different forms of kappa distribution can be equivalent under a certain transformation.}. The connection of kappa distributions with statistical mechanics, and specifically with the concept of non-extensive statistical mechanics, has been established by several authors \citep[e.g.][]{Treumann1997, Milovanov2000, Leubner2002, Livadiotis2009}. \cite{Livadiotis2009} showed that the consistent connection of the theory and formalism of kappa distributions with non-extensive statistical mechanics is based on several fundamental physical notions and concepts, including that of temperature. For further details, see the `historical comments' in Chapter 1 of \cite{Livadiotis2017}.

The $\kappa$ parameter (that appears in the power-law index) is interwoven with the statistical correlation between the energy of the particles. It has been shown that a simple relation exists between the correlation $\rho$ (ranging between 0 and 1) and $\kappa$, i.e., $\rho=(3/2)\kappa$ for 3 degrees of freedom per particle \citep{Livadiotis2011}. The largest value of kappa is infinity, corresponding to the system residing at thermal equilibrium, where the particles are characterized by zero correlation. The smallest kappa is 3/2 and corresponds to the furthest state from thermal equilibrium, called anti-equilibrium, where the particle energies are highly correlated.

Finally, we mention that the relations of kappa with spectral indices may change in the presence of nonzero potential energies \citep{Livadiotis2015a},  anisotropies of the velocity space (Chapter 10 of \cite{Livadiotis2017}),  and generalized formalisms of kappa distributions (e.g., the Lp-normed kappa distribution, \cite{Livadiotis2016, Randol2016}).

\section{Fermi Acceleration}
\label{sec:fermitheory}

Here we summarize very briefly the original theory by Fermi and its variants, following reviews by, for example, \cite{Tsuneta2011}. For a more complete review, readers are referred to literatures by, for example, \cite{Drury1983, Blandford1987,Longair1994} and \cite{Kulsrud2005}.

If we consider a particle with mass $m$ and speed $v$ in an acceleration region full of magnetic `clouds', the particle energy gain by an elastic collision with a cloud (in 1D space) would be $\Delta E_\pm = (1/2)m(v \pm 2V)^2 - (1/2)mv^2 \sim \pm 2mvV$ where $V \ll v$ is the speed of the cloud and the positive and negative signs denote the head-on and head-tail collisions, respectively. Here, the probability of head-on and head-tail collisions can be expressed as
\begin{equation}
p_{\pm} = \frac{v \pm V}{2v},
\end{equation}
where $p_+ + p_- = 1$ and $p_+$ is slightly larger than $p_-$. Thus, the average energy gain by a collision becomes $\Delta E = p_+ \Delta E_+ + p_- \Delta E_- = 2mV^2$. Using the average collision time $\sim l/v$, where $l$ is the typical distance between clouds, the energy gain per unit time becomes
\begin{equation}
\frac{dE}{dt} = \alpha E,
\label{eq:egain}
\end{equation}
where $\alpha \equiv 4V^2/vl$. In a 3D configuration, $\alpha = (4/3)V^2/vl$ as described by, for example, \cite{Longair1994} for the relativistic case. If the particle is relativistic $v \sim c$, $\alpha$ can be regarded as a constant so that solving Eq. \ref{eq:egain} yields  $E \propto \exp{(\alpha t)}$, indicating exponential energy increase. (The non-relativistic case will be described later.)  Because of the dependence on $V^2$, this process is called \textbf{second-order Fermi} acceleration.

Particle thus accelerated will, at some point, escape from the acceleration region. If we define such a time scale $t_{\rm esc}$, the probability of a particle remaining in the acceleration region could be expressed as $P(t) = \exp{(-t/t_{\rm esc})}$. Then, $dN/dE = (dP/dt)(dt/dE)$ becomes a power law, as does the differential density $N(E)$
\begin{equation}
N(E) \propto E^{-\left(1+ \frac{t_{\rm acc}}{t_{\rm esc}} \right)},
\label{eq:fermi2}
\end{equation}
where we have used the acceleration time scale $t_{\rm acc} \equiv \alpha^{-1}$ because of Equation (\ref{eq:egain}). Therefore, the power-law index is a function of both acceleration (energization `kick') and escape time scales (However, the quantity $\tau_{\rm acc}/\tau_{\rm esc}$ is assumed to be energy independent). The same power-law form can be derived if we consider the continuity equation in energy space
\begin{equation}
\frac{\partial N}{\partial t} + \frac{\partial}{\partial E} \left( \frac{d E}{dt} N \right) = q - \frac{N}{t_{\rm esc}}
\label{eq:cont}
\end{equation}
where $q$ is the source term. If we neglect the source term and use Equation (\ref{eq:egain}), then we can readily obtain Equation (\ref{eq:fermi2}). 

This Fermi process would be more efficient if there were head-on collisions only (one can imagine a trapped particle bouncing between two walls approaching toward each other). In this case, the average energy gain by a collision becomes $\Delta E = 2mvV$. The average collision time could be  $\sim 2l/v$ so that we again obtain Equation (\ref{eq:egain}) but with $\alpha=2V/l$, which is a constant in both relativistic and non-relativistic regimes. Because of the dependence on $V$ this is known as  \textbf{first-order Fermi} acceleration.

As for the non-relativistic case of the second-order Fermi acceleration, $\alpha$ in Eq. (\ref{eq:egain}) is not a constant and the equation should be rewritten as 
\begin{equation}
\frac{dE}{dt} = \alpha E^{\frac{1}{2}}
\label{eq:egain2}
\end{equation}
with the constant $\alpha = 2\sqrt{2m}V^2/l$. Using Equation (\ref{eq:cont}),  the differential density becomes
\begin{equation}
N(E) = E^{-0.5}\exp{\left(-\sqrt{\frac{E}{E_0}}\right)}
\end{equation}
where $E_0 \equiv \alpha^2 t_{\rm esc}^2/4$ \citep[e.g.][]{Tsuneta1995}. This is an unrealistically hard power-law with an exponential cutoff at $E_0$. Thus, we expect first-order Fermi acceleration (instead of second-order) to be applicable in many cases of power-law observations reviewed in this paper. When acceleration occurs by a DC electric field (i.e., $m(dv/dt) = -eE$  with electric field $E$ larger than the Dreicer electric field), Eq. (\ref{eq:egain2}) appears again with $\alpha = \sqrt{2/m}eE$ if we allow $E = (1/2)mv^2$ \citep[e.g.][]{Tsuneta1995}. Thus, a DC electric field alone likely cannot  explain the power-law spectra in solar flares as well as Earth's magnetotail because they are much softer (steeper). 

Let us now consider what a `cloud' could be or how particles receive an energization `kick' in the acceleration region. The nature of the `kick' can actually  vary in different situations. The most widely known application of  Fermi acceleration is the diffusive shock acceleration (DSA) whereby particles move back and forth across the shock front and receive energy `kick' multiple times via collisions with magneto-hydrodynamic (MHD) waves, which are assumed to be convecting together with the plasma bulk flow.  While waves are convected toward the same direction (i.e., downstream) with the speed $V_1$ and $V_2$ in the upstream and downstream regions, respectively, we can find a frame in which the upstream and downstream plasma converge toward each other with the speed $(V_1-V_2)/2$, leading to first-order Fermi acceleration. Additional details of diffusive shock acceleration can be found in, for example, \cite{Blandford1987}.

In solar flares and Earth's magnetotail, where the spatial scale of the acceleration region (or more precisely the magnetic field curvature radius) can be larger than the particle gyro-radius, particle motion may be described by a guiding-center approximation \citep[e.g.][]{Northrop1963}. In this approximation the rate of energy gain may be  expressed
\begin{equation}
\frac{dE}{dt} = -\mu V_{\rm ||} \frac{\partial B}{\partial s} + m(v_{\rm ||}-V_{\rm ||})^2 \mathbf{V}_{\rm \perp}\cdot\frac{\partial \mathbf{b}}{\partial s} + \mathcal{O}(l^2),
\end{equation}
where $\mathbf{V}$ is the velocity of the `cloud,' $\mathbf{b}=\mathbf{B}/B$,  $\partial/\partial s= \mathbf{b}\cdot\nabla$, $\mu = m_e v_{\perp}^2/(2B)$, and $l$ represents the smallness parameter, for instance, the ratio between gyro-radius and a macroscopic scale length \citep{Northrop1963}. The first term on the right-hand side represents \textbf{Fermi Type A acceleration}, i.e., a reflection by a moving magnetic mirror,  and includes the effect of betatron acceleration.  The second term is \textbf{Fermi Type B acceleration}, or the `slingshot' effect \citep{Birn2004}, which occurs when the lines of force are curved but the magnitude of $\mathbf{B}$ may be constant along the guiding center trajectory. 

Figure \ref{fig:fermitypeB} illustrates field-line configurations favorable for Fermi Type B acceleration. Fig. \ref{fig:fermitypeB}a shows the curved field line as illustrated in the original paper by \cite{Fermi1949}. Fig. \ref{fig:fermitypeB}b represents a situation in which the two ends of the field-line are fixed ( or `line-tied') (e.g., the collapsing magnetic fields in solar flares and Earth's magnetotail). While particles can undergo mirror reflection at the footpoints, in the observer's frame  $\partial \mathbf{b}/\partial s$ = 0 and Fermi Type B acceleration is dominant. In the frame co-moving with the `looptop' (i.e., the intersection of the field line and the equatorial plane), Fermi Type A acceleration becomes important. Such a frame dependence  has been reproduced in more realistic simulations of  test-particles in dynamically evolving MHD fields of Earth's magnetotail  \citep[e.g.][]{Birn2004, Birn2013}. Fig. \ref{fig:fermitypeB}c shows a magnetic island in which the curved field-line in Fig. \ref{fig:fermitypeB}a is closed by another curved field-line on the opposite side. The dynamical (shrinking) motion of the magnetic island leads to particle energization \citep[e.g.][]{Kliem1994} (as in the analogy of two walls approaching toward each other) and has drawn considerable attentions in recent years \citep[e.g.][]{Drake2006}. Theoretically, there could also be a hybrid in which a magnetic island hits a fast-mode shock so that the gray region in Fig. \ref{fig:fermitypeB}b represents the shock \citep[e.g.][]{Nishizuka2013}.

\begin{acknowledgements}
We thank Vah\'{e} Petrosian for his comments on the power-law index conversions including Table \ref{tab:index}. We thank Alexander Warmuth and Hugh Hudson for drawing our attention to some important papers on flares with unusually hard power-law spectra. We thank Shiyong Huang for drawing our attention to some important papers on energetic electrons in the magnetotail. We thank Christine Gabrielse for a clarification of her analysis. We thank Alfred Mallet for his comment on turbulence. We thank Fan Guo for his helpful comments about this manuscript.

This work was initiated and partly carried out with support from the International Space Science Institute (ISSI) in the framework of an International Team entitled `Particle acceleration in solar flares and terrestrial substorms'. We particularly thank Lyndsay Fletcher for her contributions to this paper as a member of the ISSI team. The authors are indebted to ISSI and its staff in Bern, Switzerland, for the support of this activity.  MO was supported by NASA grants NNX08AO83G at UC Berkeley and NNH16AC60I at Los Alamos National Laboratory (LANL). FE acknowledges partial support by NASA grant NNX17AK25G.
\end{acknowledgements}

\bibliographystyle{aps-nameyear}      % American Physical Society (APS) style, author-year citations
%\bibliography{/Users/moka/Documents/Mendeley/Synchronized}
% name your BibTeX data base
%\nocite{*}
%###########################################################

%\input{ssr2017.bbl}

\end{document}